\documentclass[final]{gji} 

 \usepackage{amssymb,amsmath,graphicx,natbib}
\newcommand{\dfd}{\mathrm{d}}

\graphicspath{{figures_small/}{../figures_small/}}
\usepackage{lineno}

\begin{document}

\title[Torsional waves in geodynamo simulations]{The dynamics and excitation of torsional waves in geodynamo simulations}
\author[R.J. Teed \textrm{et al.}]{R. J. Teed, C. A. Jones and S. M. Tobias \\ Dept. of Applied Mathematics, University of Leeds, Leeds, LS2 9JT, UK. Email: R.J.Teed@leeds.ac.uk}

\maketitle
\begin{abstract}
The predominant force balance in rapidly rotating planetary cores is between Coriolis, pressure, buoyancy and Lorentz forces. This magnetostrophic balance leads to a Taylor state where the spatially averaged azimuthal Lorentz force is compelled to vanish on cylinders aligned with the rotation axis. Any deviation from this state leads to a torsional oscillation, signatures of which have been observed in the Earth's secular variation and are thought to influence length of day variations via angular momentum conservation. In order to investigate the dynamics of torsional oscillations, we perform several three-dimensional dynamo simulations in a spherical shell. We find torsional oscillations, identified by their propagation at the correct Alfv\'{e}n speed, in many of our simulations. We find that the frequency, location and direction of propagation of the waves are influenced by the choice of parameters. Torsional waves are observed within the tangent cylinder and also have the ability to pass through it. Several of our simulations display waves with core travel times of 4 to 6 years. We calculate the driving terms for these waves and find that both the Reynolds force and ageostrophic convection acting through the Lorentz force are important in driving torsional oscillations.
\end{abstract}


 \linenumbers

%%%%%%%%%%%%%%%%%%%%%%%%%%%%%%%%%%%%%%%%%%%%%%%%%%%%%%%%%%%%%%%%%%%%%%%%%%%%%%%%%%%%%%%%%%%%%%%%%%%%%%%%%%%%%%%%%%%%%%%%%%%%%%%%

\section{Introduction}
\label{sec:intro}

Rapidly rotating planetary dynamos, including the geodynamo, are believed to be operating under the magnetostrophic regime,  \citep[see, for example,][]{jonREV11}. In this regime, although the Lorentz force may be locally strong, the averaged azimuthal Lorentz force must vanish on  geostrophic cylinders \citep{tay63}. A dynamo with a magnetic field organised in such a way is said to be in a Taylor state, which provides a severe
constraint for dynamo generated fields. Any violation of the state can be represented as an acceleration of the cylinders and stretches radial magnetic field into azimuthal field. The resultant Lorentz force acts like a torsional spring in an attempt to restore the Taylor state \citep{bra70} and leads to the driving of torsional oscillations (TOs) of the cylinders. These oscillations, which are dependent only on cylindrical radius and time, are a type of Alfv\'{e}n wave \citep{alf42}.

Torsional waves are believed to be continually driven in the Earth's core and are traceable in observational data. However, there has been some ambiguity as to the period for the fundamental modes of the torsional oscillations. Early observational data \citep{bra84} inferred a decadal timescale; however more recent data obtained from core flow models by \cite{gil10} show a much shorter period of approximately 6 years. Previous work \citep{jau88,jac97,zat97,blo02,buf09} has suggested that torsional oscillations may be responsible for various observed features of the Earth's dynamics; these include changes in length-of-day variations \citep{jau88,jac97} and geomagnetic jerks \citep{blo02}. Additionally, it may be possible to infer information about the magnetic field within the core via core flow models \citep{zat97,buf09}. This is useful since geomagnetic data from the Earth's surface can only be reliably transferred down as far as the core-mantle boundary (CMB) \citep{gub85}.

Numerical simulations are an obvious tool to analyse the dynamics of torsional waves; however, difficulties arise owing to the inability to reach appropriate Earth-like parameter values. Previous efforts \citep{dum03,bus05,wic10} to locate torsional waves in simulations have been undertaken with \cite{wic10} providing the most clear evidence yet of their observation in the region outside the tangent cylinder (OTC). A recent study by \cite{sch12} has focused on the reflection of Alfv\'{e}n waves at boundaries. They suggest that simulations run with rigid boundary conditions cannot exhibit wave reflection when the viscosity is too large.

We investigate torsional wave production and dynamics in numerical simulations. We employ a systematic exploration of available parameter space and include analysis of the  region inside the tangent cylinder (ITC) which was omitted in previous studies. This allows us to attempt to observe not only torsional waves ITC but also the propagation of such waves across the tangent cylinder (TC). We estimate core travel times for the oscillations and, by bandpass filtering our data, we are able to determine whether the timescales that identified TOs operate on are correct. We also explore possible excitation mechanisms by calculating the relevant driving terms. In particular, we separate the Lorentz force into its constituent parts: a restoring force and a driving force.

%%%%%%%%%%%%%%%%%%%%%%%%%%%%%%%%%%%%%%%%%%%%%%%%%%%%%%%%%%%%%%%%%%%%%%%%%%%%%%%%%%%%%%%%%%%%%%%%%%%%%%%%%%%%%%%%%%%%%%%%%%%%%%%%

\section{Mathematical formulation}

\label{sec:math}

We adapt the model described by \cite{jon11} to incompressible systems (using the Boussinesq approximation). We shall extend to the compressible parameter space in future work. Our geometry is based on the Earth's core using a spherical polar coordinate system, $(r,\theta,\phi)$. We consider a spherical shell that is radially bounded above at $r=r_o$ by an electrically insulating mantle and below at $r=r_i$ by an electrically insulating inner core. The system rotates about the vertical ($z$-axis) with rotation rate $\Omega$ and gravity acts radially inward so that ${\bf g}=-g{\bf r}$. The fluid is assumed to have constant values of $\rho$, $\nu$, $\kappa$ and $\eta$, the outer core density, kinematic viscosity, thermal diffusivity and magnetic diffusivity respectively.

Several recent papers \citep{sak09,hor10,chr10} have argued that allowing for internal heat sources (or sinks) and imposing fixed heat flux (as opposed to fixed temperature) thermal boundary conditions in models may significantly influence the generation of solutions with Earth-like magnetic field morphologies. Therefore, following the approach of \cite{hor10}, we also introduce a source of internal heating, $\epsilon$, to the temperature equation. The internal heating must satisfy the heat flux equation so that
\begin{eqnarray}
\frac{4\pi}{3}\epsilon(r_o^3 - r_i^3) = 4\pi\kappa r_i^2\left.\frac{\partial T}{\partial r}\right|_{r=r_i} - 4\pi\kappa r_o^2\left.\frac{\partial T}{\partial r}\right|_{r=r_o},
\label{eq:flux}
\end{eqnarray}
where $T$ is the temperature. We nondimensionalize the basic system of equations on the length scale, $D=r_o-r_i$, magnetic timescale, $D^2/\eta$, temperature scale, $\epsilon D^2/\eta$, and magnetic scale, $\sqrt{\rho\mu_0\Omega\eta}$. The relevant system of coupled equations for velocity, ${\bf u}$, magnetic field, ${\bf B}$, temperature, $T$, and pressure, $p$ are:
\begin{eqnarray}
\begin{split}
 \frac{\partial{\bf u}}{\partial t} + ({\bf u\cdot\nabla}){\bf u} = &-\frac{Pm}{E}\left[\nabla p  + 2{\bf \hat{z}}\times{\bf u} - (\nabla\times{\bf B})\times{\bf B}\right] \\ 
&+ \frac{Pm^2Ra}{Pr} T{\bf r} + Pm\nabla^2{\bf u},
\end{split}
\label{eq:mom}
\end{eqnarray}
\begin{gather}
 \frac{\partial{T}}{\partial t} + ({\bf u\cdot\nabla}){T} = \frac{Pm}{Pr}\nabla^2 T + \text{sgn}(\epsilon), \label{eq:en} \\
 \frac{\partial{\bf B}}{\partial t} - \nabla\times({\bf u}\times{\bf B}) = \nabla^2{\bf B}, \label{eq:ind} \\
 \nabla\cdot{\bf u} = 0, \label{eq:usolen} \\
 \nabla\cdot{\bf B} = 0. \label{eq:Bsolen}
\end{gather}

Equations (\ref{eq:mom}) to (\ref{eq:ind}) are the incompressible Navier-Stokes, temperature and induction equations respectively and (\ref{eq:usolen}) and (\ref{eq:Bsolen}) describe the solenoidal conditions for velocity and magnetic field. The nondimensional parameters appearing in our equations are the Rayleigh number, $Ra$, Ekman number, $E$, Prandtl number, $Pr$, and magnetic Prandtl number, $Pm$, defined by:
\begin{eqnarray}
 Ra=\frac{g\alpha|\epsilon| D^5}{\nu\kappa\eta}, \quad E = \frac{\nu}{\Omega D^2}, \quad Pr = \frac{\nu}{\kappa}, \quad Pm = \frac{\nu}{\eta}.
\end{eqnarray}
The radius ratio, $\beta=r_i/r_o$, is an additional parameter but in this work we restrict ourselves to the value appropriate to the Earth's core, namely $\beta=0.35$. Note that under the nondimensionalization chosen, the internal heating term has been scaled to unity. However, in order to maintain a consistent physical problem, via (\ref{eq:flux}), the internal heating may be either a source or a sink resulting in the need for the sgn$(\epsilon)$ function in (\ref{eq:en}). The magnitude of $\epsilon$ appears only in the definition of the Rayleigh number. In this definition of $Ra$ the quantity $|\epsilon|$ occupies the driving role usually taken by the temperature difference across the domain which appears in the classical definition of the Rayleigh number.

%%%%%%%%%%%%%%%%%%%%%%%%%%%%%%%%%%%%%%%%%%%%%%%%%%%%%%%%%%%%%%%%%%%%%%%%%%%%%%%%%%%%%%%%%%%%%%%%%%%%%%%%%%%%%%%%%%%%%%%%%%%%%%%%

\section{Theory and methods}
\label{sec:meth}

\subsection{Taylor's constraint and torsional oscillations}
\label{sec:taylor}

The analysis of torsional oscillations requires consideration of the forces on geostrophic cylinders and hence the introduction of a cylindrical polar coordinate system, $(s,\phi,z)$, is beneficial. Averages over $\phi$ and $z$ are required and hence for any scalar field $A$ we define
\begin{eqnarray}
\bar{A}(t,s,z) = \frac{1}{2\pi}\int_0^{2\pi}A\dfd\phi, \quad \langle A\rangle(t,s,\phi) = \frac{1}{h}\int_{z_-}^{z_+} A \dfd z.
\end{eqnarray}
Here $h(s)=z_+(s)-z_-(s)$ and OTC we simply have that $z_\pm=\pm\sqrt{r_o^2-s^2}$. Within the tangent cylinder the definition of $z_\pm$ may remain the same if an average over the entire $z$ domain is desired. However, ITC we may wish to average over the two hemispheres separately, which we refer to as ITCN and ITCS for north and south of the inner core respectively. For ITCN (ITCS) we then have that $z_+=\sqrt{r_o^2-s^2}$ and $z_-=\sqrt{r_i^2-s^2}$ ($z_+=-\sqrt{r_i^2-s^2}$ and $z_-=-\sqrt{r_o^2-s^2}$).

For later convenience, we also define two further quantities for a scalar, or vector, field $A$. The first of these quantities, $\tilde{A}$, is simply the time average of $A$ over some time period, $\tau$. The second quantity, $A^\prime$, is the fluctuating part of $A$. Therefore we define $\tilde{A}$ and $A^\prime$ by
\begin{eqnarray}
 \tilde{A}(s,\phi,z) = \frac{1}{\tau} \int_0^\tau A\dfd t \quad \text{ and } \quad A^\prime(t,s,\phi,z)=A-\tilde{A},
\label{eq:meanfluc}
\end{eqnarray}
respectively. $A^\prime$ is useful because it removes from $A$ the mean background state which only varies on a long timescale. Standard torsional oscillation theory relies on the ability to separate the timescales in this way successfully.

The $\phi$ and $z$ averages of the $\phi$-component of (\ref{eq:mom}) illustrate the forces that can accelerate geostrophic cylinders. Three such forces can be identified \citep{wic10}; namely the Reynolds force, Lorentz force and viscous force leading to the equation
\begin{eqnarray}
 \frac{\partial\langle\overline{u_\phi}\rangle}{\partial t} &=& -\langle\overline{\boldsymbol{\hat{\phi}}\cdot(\nabla\cdot{\bf uu})}\rangle + PmE^{-1}\langle\overline{\boldsymbol{\hat{\phi}}\cdot((\nabla\times{\bf B})\times{\bf B})}\rangle \notag \\
 & &+ Pm\langle\overline{\boldsymbol{\hat{\phi}}\cdot\nabla^2{\bf u}}\rangle \notag \\ 
 &\equiv& F_R + F_L + F_V. \label{eq:geocyl} 
\end{eqnarray}
 The Coriolis and buoyancy forces have vanished during the integration process since in the former there is no net flow across the cylinder and no $\phi$-component in the latter. This has consequences in the core where the fluid is believed, at leading order, to be in magnetostrophic balance (between Lorentz, Coriolis and Archimedean forces). \cite{tay63} noted that in systems where the force balance is magnetostrophic the constraint
\begin{eqnarray}
 F_L =  0, \label{eq:taycon}
\end{eqnarray}
arises. 

The Lorentz force can be partially integrated (see, for example, \cite{wic10}) to give
\begin{eqnarray}
 F_L = \frac{Pm}{E}\frac{1}{hs^2}\frac{\partial}{\partial s}s^2h\langle\overline{B_sB_\phi}\rangle + \frac{Pm}{E}\frac{1}{h}\left[\frac{s}{z}\overline{B_sB_\phi}+\overline{B_zB_\phi}\right]_{z_-}^{z_+}.
\label{eq:lorw&c}
\end{eqnarray}
We are able to neglect the magnetic coupling terms in this expression at this stage due to our use of insulating boundary conditions at both the CMB and the inner core boundary (ICB) \citep{jon11}. However, if one were to allow for a conducting inner core (or mantle), the contribution from these surface terms would be nonzero resulting in an additional forcing in the system that is not discussed further here. For discussion of how this coupling term arises see \cite{rob12}.

Upon consideration of the time derivative of the expression for $F_L$ in (\ref{eq:lorw&c}) we find that we require expressions for the time derivatives of components of the magnetic field. We substitute from the induction equation and retain \emph{all terms} on the right-hand-side of (\ref{eq:ind}), to determine that
\begin{align}
 \dot{F}_L\ =&\ \frac{Pm}{E}\frac{1}{hs^2}\frac{\partial}{\partial s}s^2h\langle\overline{\dot{B}_sB_\phi + B_s\dot{B}_\phi}\rangle \\
=&\ \frac{Pm}{E}\frac{1}{hs^2}\frac{\partial}{\partial s}s^2h\left\{\left\langle\overline{sB_s({\bf B}\cdot\nabla)\frac{u_\phi}{s}}\right\rangle\right. \notag \\
& \qquad\qquad\qquad\quad + \left\langle\overline{\frac{B_\phi}{s}({\bf B}\cdot\nabla)(su_s)}\right\rangle \notag \\
& \qquad\qquad\qquad\quad - \left\langle\overline{\left({\bf u}\cdot\nabla +\frac{2}{s^2}\right)(B_sB_\phi)}\right\rangle \notag \\
& \qquad\qquad\qquad\quad + \langle\overline{B_s\nabla^2B_\phi + B_\phi\nabla^2 B_s}\rangle\left.\right\}.
\label{eq:lorfull}
\end{align}
In order to make further progress we use the definitions of (\ref{eq:meanfluc}) to split the velocity and magnetic field into mean and fluctuating parts. Previous studies \citep{wic10,rob12} have essentially assumed that the mean quantities, $\tilde{\bf u}$ and $\tilde{\bf B}$, are the principal parts of the Taylor state and that the fluctuating quantities, ${\bf u}^\prime$ and ${\bf B}^\prime$, are perturbations associated with the TOs. However, this is not the full picture since it requires the assumption that ${\bf u}^\prime$ is purely geostrophic as explicitly stated by \cite{tay63}. In reality the convection will be operating, to some degree, on all timescales and this phenomenon is likely to be an important driving mechanism. Hence rather than assuming a geostrophic form for our velocity fluctuation we instead split it into geostrophic ($s\zeta'$) and ageostrophic parts (${\bf u}_A'$) so that
\begin{eqnarray}
 {\bf u} = \tilde{\bf u} + {\bf u}' = \tilde{\bf u} + s\zeta'(s,t)\boldsymbol{\hat{\phi}} + {\bf u}_A', \qquad {\bf B} = \tilde{\bf B} + {\bf B}'.
\label{eq:uBmeanfluc}
\end{eqnarray}
Upon substitution of these forms into our expression for $\dot{F}_L$, we find that $\zeta'$ only appears in the first term on the right-hand-side of (\ref{eq:lorfull}). Considering only the mean magnetic field parts of this term and calling it $\dot{F}_{LR}$ gives
\begin{eqnarray}
 \dot{F}_{LR} = \frac{1}{hs^2}\frac{\partial}{\partial s}\left(s^3h U_A^2 \frac{\partial\zeta}{\partial s}\right), \qquad U_A=\sqrt{\frac{Pm}{E}\langle\overline{\tilde{B}_s^2}\rangle},
\label{eq:FLRdot}
\end{eqnarray}
where we have defined the Alfv\'{e}n speed, $U_A$. Equation (\ref{eq:lorfull}) can then be written as
\begin{eqnarray}
 \dot{F}_L = \dot{F}_{LR} + \dot{F}_{LD},
\label{eq:FLsplit}
\end{eqnarray}
where $\dot{F}_{LD}$ is a complicated expression made up of the remaining terms on the right-hand-side of (\ref{eq:lorfull}). Thus it involves terms containing the components of $\tilde{\bf B}$, ${\bf B}'$, $\tilde{\bf u}$, ${\bf u}_A'$, as well as $\zeta'$.

If we now take the time derivative of (\ref{eq:geocyl}) and use the result of (\ref{eq:FLsplit}) we find that
\begin{eqnarray}
 s\ddot{\zeta}' = \dot{F}_{LR} + \dot{F}_{LD} + \dot{F}_R + \dot{F}_V,
\label{eq:foringtder}
\end{eqnarray}
noting that $\langle\overline{\boldsymbol{\hat{\phi}}\cdot {\bf u}_A'}\rangle=0$ by definition. By writing the expression for $\ddot{\zeta}'$ in this way we have been able to separate the term involved in the balance of the torsional wave equation from the remaining terms. The standard canonical wave equation as found in previous work \citep[see, for example,][]{bra70} is represented by $s\ddot{\zeta}' = \dot{F}_{LR}$. Consequently, if we time integrate (\ref{eq:foringtder}) to acquire
\begin{eqnarray}
 s\dot{\zeta}' - F_{LR} = F_{LD} + F_R + F_V,
\label{eq:acceqsforce}
\end{eqnarray}
we find that $F_{LR}$ is the restoring force whereas $F_{LD}$, $F_R$ and $F_V$ are driving forces.

Torsional waves in the core must be driven and dissipated by some mechanism(s) and hence the terms on the right-hand-side of (\ref{eq:acceqsforce}), namely $F_R$, $F_V$ and $F_{LD}$, fulfil this role. They are driving (and dissipative) forces which are able to create, destroy and alter the nature of propagating torsional waves. When performing diagnostics on our simulations, one of our interests will be analysing the terms on the right-hand-side of (\ref{eq:acceqsforce}). This will allow us to identify which forces are able to act as excitation mechanisms at various points in the domain. We look at this in section \ref{sec:fb}.

%%%%%%%%%%%%%%%%%%%%%%%%%%%%%%%%%%%%%%%%%%%%%%%%%%%%%%%%%%%%%%%%%%%%%%%%%%%%%%%%%%%%%%%%%%%%%%%%%%%%%%%%%%%%%%%%%%%%%%%%%%%%%%%%

\subsection{Output parameters}

In addition to quantities described in subsection \ref{sec:taylor} we also output several other parameters from our simulations. The magnetic Reynolds number, Elsasser number, Rossby number and dipole moment are defined by
\begin{gather}
 Rm = \frac{UD}{\eta}, \label{eq:Rm}\\
\Lambda = \frac{|B|^2}{\rho\mu\eta\Omega},\\
 Ro = \frac{U}{\Omega D},\\
 f_\text{dip} = \left(\frac{E_M^{(1,0)}(r_o)}{\sum_{l=1}^{12}\sum_{m=0}^{l}E_M^{(l,m)}(r_o)}\right)^{1/2},\label{eq:fdip}
\end{gather}
respectively. Here $E_M^{(l,m)}(r)$ represents the magnetic energy in the $(l,m)$ harmonic at radius $r$. Owing to our choice of nondimensionalization, the magnetic Reynolds and Elsasser numbers can be identified with the nondimensional velocity and square of the magnetic field respectively. The parameters defined in equations (\ref{eq:Rm}) to (\ref{eq:fdip}) give an indication of the sort regime that the dynamo is in, a point we address in section \ref{sec:field}.

%%%%%%%%%%%%%%%%%%%%%%%%%%%%%%%%%%%%%%%%%%%%%%%%%%%%%%%%%%%%%%%%%%%%%%%%%%%%%%%%%%%%%%%%%%%%%%%%%%%%%%%%%%%%%%%%%%%%%%%%%%%%%%%%

\subsection{Methods}

We perform several simulations, using the Leeds spherical dynamo code \citep{jon11} which uses a pseudo-spectral numerical scheme with finite differences in the radial direction. We run the code at parameter regimes and with boundary conditions that facilitate the production of Earth-like dynamos. Guided by previous work \citep{hor10} we therefore employ the use of fixed flux thermal boundary conditions for all of our simulations. Specifically, we set zero flux on the CMB and the flux entering at the ICB is then balanced by a sink term in the temperature equation; that is, sgn$(\epsilon)=-1$. This mathematical setup is, in a physical sense, representative of a model for compositional convection. Rigid kinematic boundary conditions are primarily used, although one set of simulations is repeated with stress-free boundaries as way of comparison. 

In parameter space we perform simulations at a range of Ekman numbers since the existence of torsional oscillations requires the dynamo to be near magnetostrophic balance, which in turn is dependent on a small Ekman number. Thus, by decreasing the Ekman number over the range $10^{-4}$ to $10^{-6}$ torsional oscillations should become more apparent. We focus on $Pr=1$ and each simulation is at the same value of criticality; that is $Ra/Ra_c\simeq8.32$ for all runs. However, we do vary the magnetic Prandtl number, $Pm\in[1,5]$, in order to allow for a range in the magnetic field strength. The values of $Ra_c$ used are for the onset of non-magnetic convection \citep[see, for example,][]{dor04}. Table \ref{tab:runs}  displays the input parameters for the set of runs performed as well as the kinetic boundary conditions employed.

\begin{table}
\centering
\begin{tabular}{ccccccc}
\hline
Run & $E$ & $Ra$ & $Pr$ & $Pm$ & BCs & $\tau$ \\
\hline\hline
4R1 & $10^{-4}$ & $4.937\times10^6$ & $1$ & $1$ & NS & $0.02$ \\
4R2 & $10^{-4}$ & $4.937\times10^6$ & $1$ & $2$ & NS & $0.02$ \\
4R3 & $10^{-4}$ & $4.937\times10^6$ & $1$ & $3$ & NS & $0.02$ \\
4R4 & $10^{-4}$ & $4.937\times10^6$ & $1$ & $4$ & NS & $0.014$ \\
4R5 & $10^{-4}$ & $4.937\times10^6$ & $1$ & $5$ & NS & $0.014$ \\
\hline
5R1 & $10^{-5}$ & $1\times10^8$ & $1$ & $1$ & NS & $0.006$ \\
5R2 & $10^{-5}$ & $1\times10^8$ & $1$ & $2$ & NS & $0.006$ \\
5R3 & $10^{-5}$ & $1\times10^8$ & $1$ & $3$ & NS & $0.006$ \\
5R4 & $10^{-5}$ & $1\times10^8$ & $1$ & $4$ & NS & $0.003$ \\
5R5 & $10^{-5}$ & $1\times10^8$ & $1$ & $5$ & NS & $0.003$ \\
\hline
6.5R1 & $5\times10^{-6}$ & $2.493\times10^8$ & $1$ & $1$ & NS & $0.004$ \\
6.5R2 & $5\times10^{-6}$ & $2.493\times10^8$ & $1$ & $2$  & NS & $0.004$ \\
6.5R3 & $5\times10^{-6}$ & $2.493\times10^8$ & $1$ & $3$  & NS & $0.004$ \\
6.5R4 & $5\times10^{-6}$ & $2.493\times10^8$ & $1$ & $4$  & NS & $0.002$ \\
6.5R5 & $5\times10^{-6}$ & $2.493\times10^8$ & $1$ & $5$  & NS & $0.002$ \\
\hline
6R1 & $10^{-6}$ & $2.132\times10^9$ & $1$ & $1$ & NS & $0.002$ \\
\hline
5F1 & $10^{-5}$ & $1.265\times10^8$ & $1$ & $1$ & SF & $0.008$ \\
5F2 & $10^{-5}$ & $1.265\times10^8$ & $1$ & $2$ & SF & $0.005$ \\
5F3 & $10^{-5}$ & $1.265\times10^8$ & $1$ & $3$ & SF & $0.003$ \\
5F4 & $10^{-5}$ & $1.265\times10^8$ & $1$ & $4$ & SF & $0.003$ \\
5F5 & $10^{-5}$ & $1.265\times10^8$ & $1$ & $5$ & SF & $0.002$ \\
\hline
\end{tabular}
\caption{Table displaying the parameter sets used for the various simulations. Note that all runs have fixed flux thermal boundary conditions with zero flux on the outer boundary and an internal heat sink.}
\label{tab:runs}
\end{table}

Each run is initially time integrated from a random state for at least one tenth of a magnetic diffusion time apart from run 6R1 which is run for a shorter period due to resolution constraints. In order to search for torsional oscillations we then analyse a period of time, $\tau$, of every run. The value of $\tau$ for each run, indicated in Table \ref{tab:runs}, is run dependent and varies between 0.002 and 0.02 of a diffusion time. 

By including the region ITC in our analysis we present ourselves with a complication since it is not obvious how to deal with the regions north and south of the inner core. For example, when performing averages over $z$ do we average over the entire vertical from pole to pole or instead retain the distinction between the hemispheres? Consequently, there is also the issue of how to treat waves propagating across the tangent cylinder since they may originate (or terminate) in either hemisphere. These issues were not present in the previous work on torsional wave analysis in dynamo simulations \citep{wic10} where the region ITC was omitted. We choose to allow for both scenarios by performing both sets of averages. Therefore in our analysis we average over the entire region ITC, but also perform averages over each hemisphere separately (that is over ITCN and ITCS). For the region OTC, averages are always performed across all $z$-space.

%%%%%%%%%%%%%%%%%%%%%%%%%%%%%%%%%%%%%%%%%%%%%%%%%%%%%%%%%%%%%%%%%%%%%%%%%%%%%%%%%%%%%%%%%%%%%%%%%%%%%%%%%%%%%%%%%%%%%%%%%%%%%%%%

\section{Numerical results}
\label{sec:res}

\subsection{Field strength and morphology}
\label{sec:field}

The output parameters calculated from our numerical results are displayed in Table \ref{tab:output}. In this table we also indicate, for each run, whether torsional oscillations were identified and if so, also the region(s) of the shell that they were observed. Within our full set of simulations we are able to identify two major magnetohydrodynamic regimes for which the fluid in each run can organise itself. The weak field regime has $\Lambda\sim O(1)$ whereas the strong field regime has a much larger Elsasser number. As one would expect, the latter regime is found at larger values of the magnetic Prandtl number. Velocity structures are larger in the strong field regime. However, it should be noted that even in the weak field regime the convection is not as small scale as one may expect for such a rapidly rotating system. This is due to the employment of fixed flux thermal boundary conditions, which have been found to significantly affect the size of velocity structures \citep{hor10}.

\begin{table*}
\centering
\begin{tabular}{ccccccc}
\hline
Run & $Rm$ & $\Lambda$ & $Ro$ & $f_{\text{dip}}$ & $U_A(s=r_o)$ & TOs \\
\hline\hline
4R1 & $98.118$& $0.896$ & $0.010$ & $0.890$ & $0.067$ & - \\
4R2 & $135.595$ & $1.888$ & $0.007$ & $0.867$ & $1.436$ & - \\
4R3 & $152.387$ & $5.672$ & $0.005$ & $0.847$ & $15.673$ & - \\
4R4 & $183.966$ & $10.358$ & $0.005$ & $0.776$ & $22.262$ & OTC \\
4R5 & $217.046$ & $15.621$ & $0.004$ & $0.741$ & $29.382$ & OTC,ITC  \\
\hline
5R1 & $128.542$& $0.319$ & $0.001$ & $0.924$ & $5.015$ & OTC \\
5R2 & $203.348$ & $1.740$ & $0.001$ & $0.904$ & $14.283$ & OTC,ITC \\
5R3 & $330.519$ & $16.197$ & $0.001$ & $0.722$ & $90.073$ & OTC,ITC \\
5R4 & $355.911$ & $17.433$ & $0.001$ & $0.713$ & $90.267$ & OTC,ITC \\
5R5 & $437.071$ & $19.252$ & $0.001$ & $0.742$ & $123.902$ & OTC \\
\hline
6.5R1 & $155.4277$& $0.325$ & $0.001$ & $0.917$ & $7.774$ & OTC,ITC \\
6.5R2 & $267.719$ & $2.400$ & $0.001$ & $0.955$ & $22.078$ & OTC,ITC \\
6.5R3 & $383.569$ & $3.631$ & $0.001$ & $0.946$ & $29.173$ & OTC,ITC \\
6.5R4 & $575.840$ & $23.637$ & $0.001$ & $0.752$ & $259.222$ & OTC \\
6.5R5 & $598.998$ & $20.080$ & $0.001$ & $0.752$ & $243.473$ & OTC,ITC \\
\hline
6R1 & $372.872$ & $0.561$ & $<0.001$ & $0.918$ & $15.664$ & OTC,ITC  \\
\hline
5F1 & $172.707$& $0.368$ & $0.002$ & $0.918$ & $5.094$ & OTC \\
5F2 & $226.404$ & $2.164$ & $0.001$ & $0.955$ & $16.588$ & OTC,ITC \\
5F3 & $336.970$ & $18.817$ & $0.001$ & $0.676$ & $94.567$ & OTC,ITC \\
5F4 & $402.806$ & $18.578$ & $0.001$ & $0.738$ & $89.943$ & OTC,ITC \\
5F5 & $560.841$ & $23.636$ & $0.001$ & $0.719$ & $109.473$ & OTC,ITC \\
\hline
\end{tabular}
 \centering
\caption{Table displaying the output parameters calculated for the various simulations.}
\label{tab:output}
\end{table*}

With current estimates that $Rm\approx1000$ for the Earth's outer core, Table \ref{tab:output} indicates that only our high $Pm$, low $E$ runs begin to approach Earth-like magnetic Reynolds numbers. However, simulations in the strong field regime produce Elsasser numbers too large for the Earth where $\Lambda\sim O(1)$. The converse is true of the dipolarity, which decreases to near Earth-like values for our larger $Pm$ runs.

\begin{figure}
\begin{minipage}[b]{0.45\linewidth}
\centering
% \figbox*{}{}{%
% \includegraphics[width=1.0\linewidth]{5R2Br.eps}
  \includegraphics[width=1.0\linewidth]{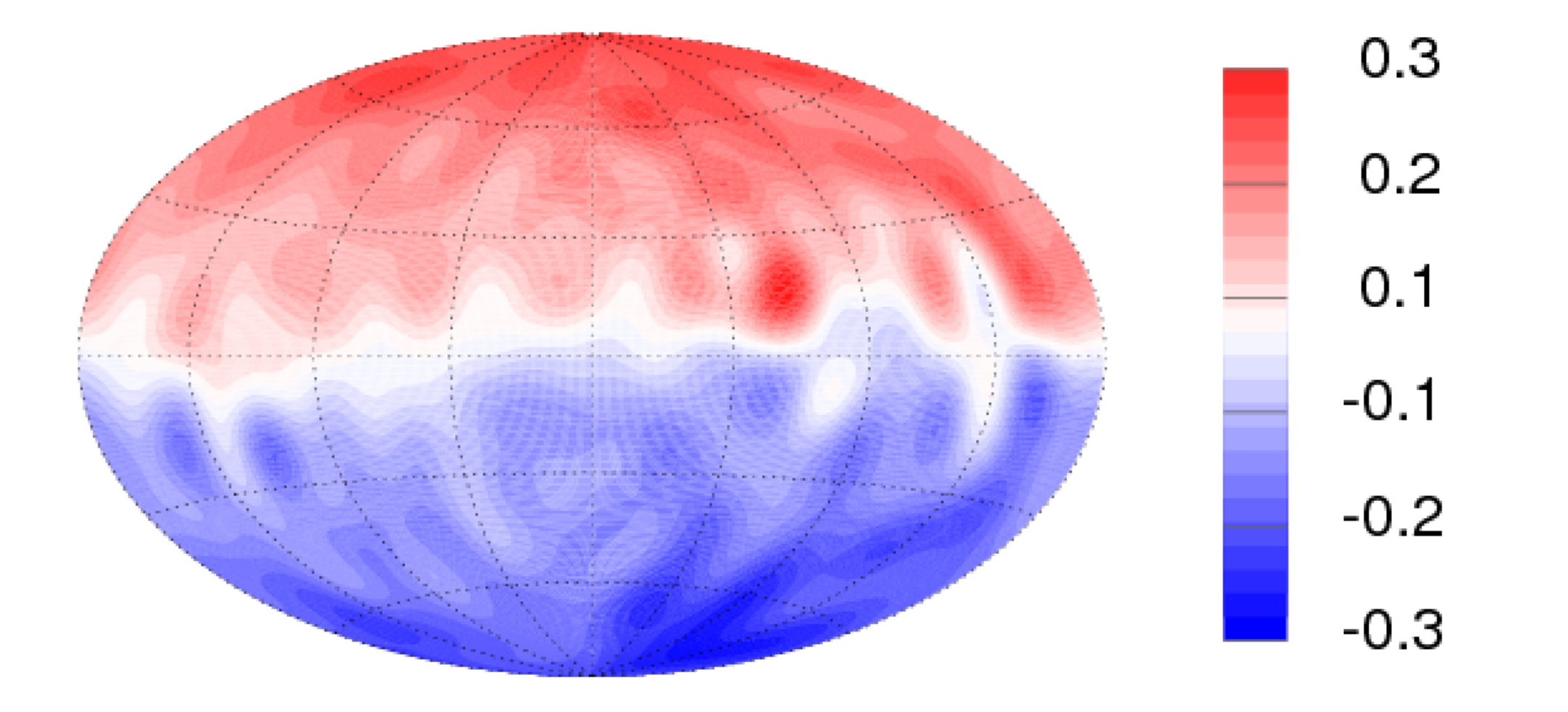}
%}
\caption{The radial magnetic field at the CMB for the run 5R2.}
\label{fig:5R2Br}
 \end{minipage}
\hspace{0.5cm}
 \begin{minipage}[b]{0.45\linewidth}
\centering
%  \figbox*{}{}{%
%   \includegraphics[width=1.0\linewidth]{6.5R5Br.eps}
  \includegraphics[width=1.0\linewidth]{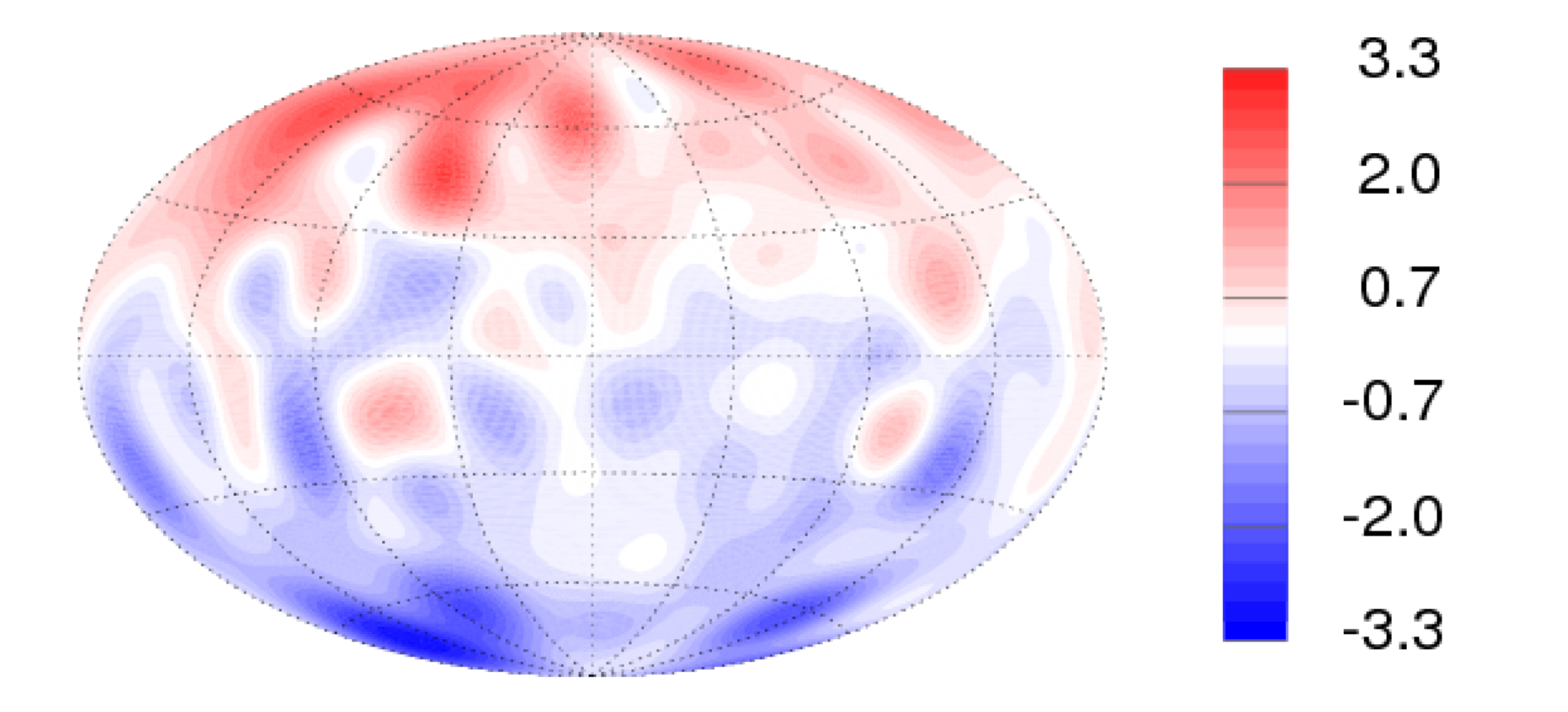}
% }
\caption{The radial magnetic field at the CMB for the run 6.5R5.}
\label{fig:6.5R5Br}
 \end{minipage}
\end{figure}

In Figs \ref{fig:5R2Br} and \ref{fig:6.5R5Br} we plot $B_r$, truncated at harmonic degree 12, at the CMB for runs at two different values of $Pm$. Although both figures show dipolar fields, the dipolarity is visibly stronger in Fig. \ref{fig:5R2Br} than Fig. \ref{fig:6.5R5Br}, which has patches of reversed flux. These plots are representative of the radial magnetic field for the two different regimes seen across all of our runs. As we shall discuss later, the two regimes will also have implications on where and what sort of torsional oscillations can be found.

%%%%%%%%%%%%%%%%%%%%%%%%%%%%%%%%%%%%%%%%%%%%%%%%%%%%%%%%%%%%%%%%%%%%%%%%%%%%%%%%%%%%%%%%%%%%%%%%%%%%%%%%%%%%%%%%%%%%%%%%%%%%%%%%

\subsection{Identification of torsional oscillations}
\label{sec:tos}

In a similar vein to \cite{wic10} we identify torsional oscillations by structures in the azimuthal fields moving radially in $s$ with the local Alfv\'{e}n speed. In order to observe features operating on short timescales we analyse the fields with the time average removed; that is we consider $u_\phi^\prime$ and its spatial average relevant to the problem in hand. For each run we evaluate the quantity $\langle\overline{\tilde{B}_s^2}\rangle$ for use in the definition of $U_A$. 

\begin{figure}
% \begin{minipage}[b]{0.45\linewidth}
\centering
\includegraphics[width=1.0\linewidth]{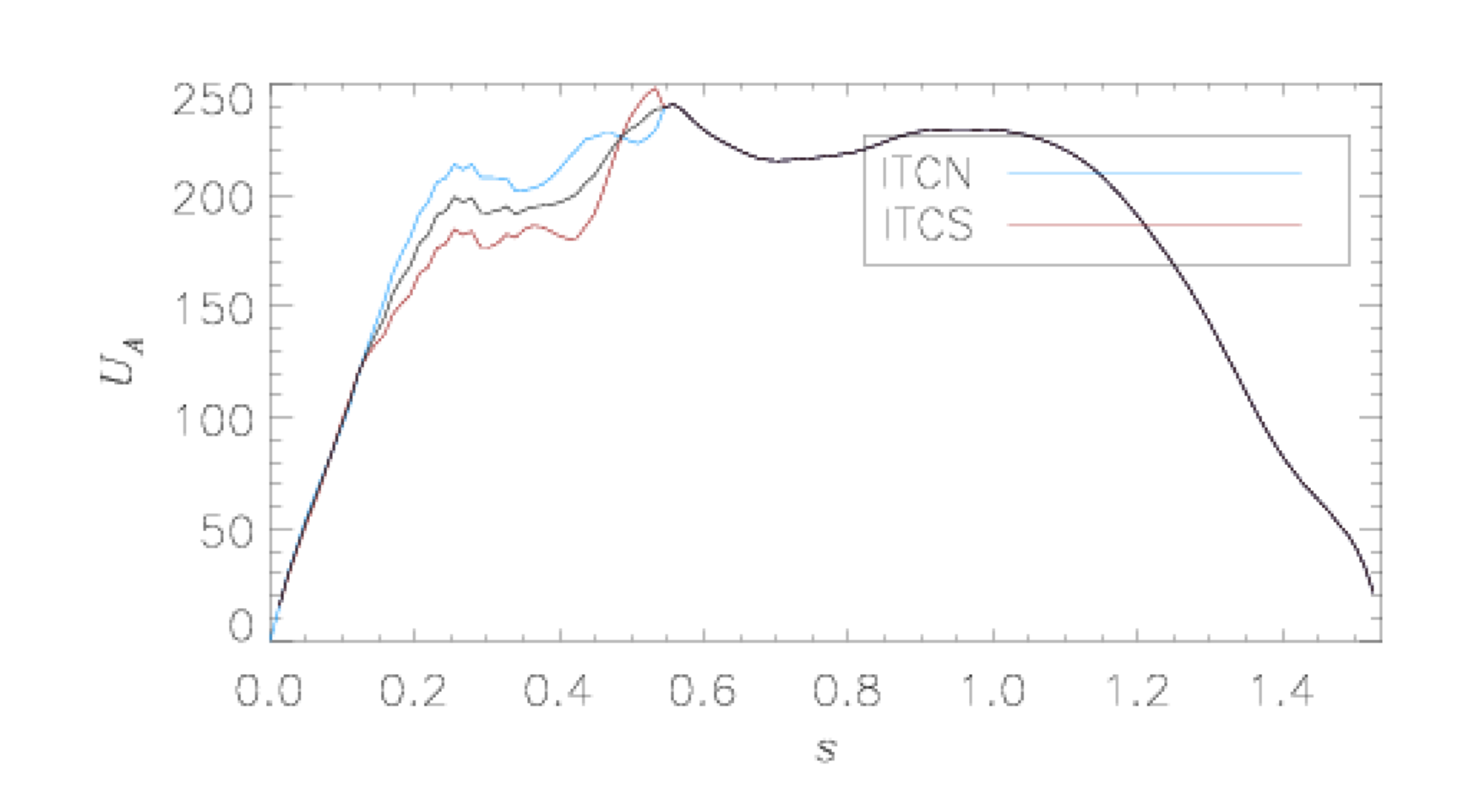}
\caption{Alfv\'{e}n speed, as a function of $s$, for the run 6.5R2.}
\label{fig:6.5R2Alf}
% \end{minipage}
\end{figure}

\begin{figure}
% \begin{minipage}[b]{0.45\linewidth}
\centering
\includegraphics[width=1.0\linewidth]{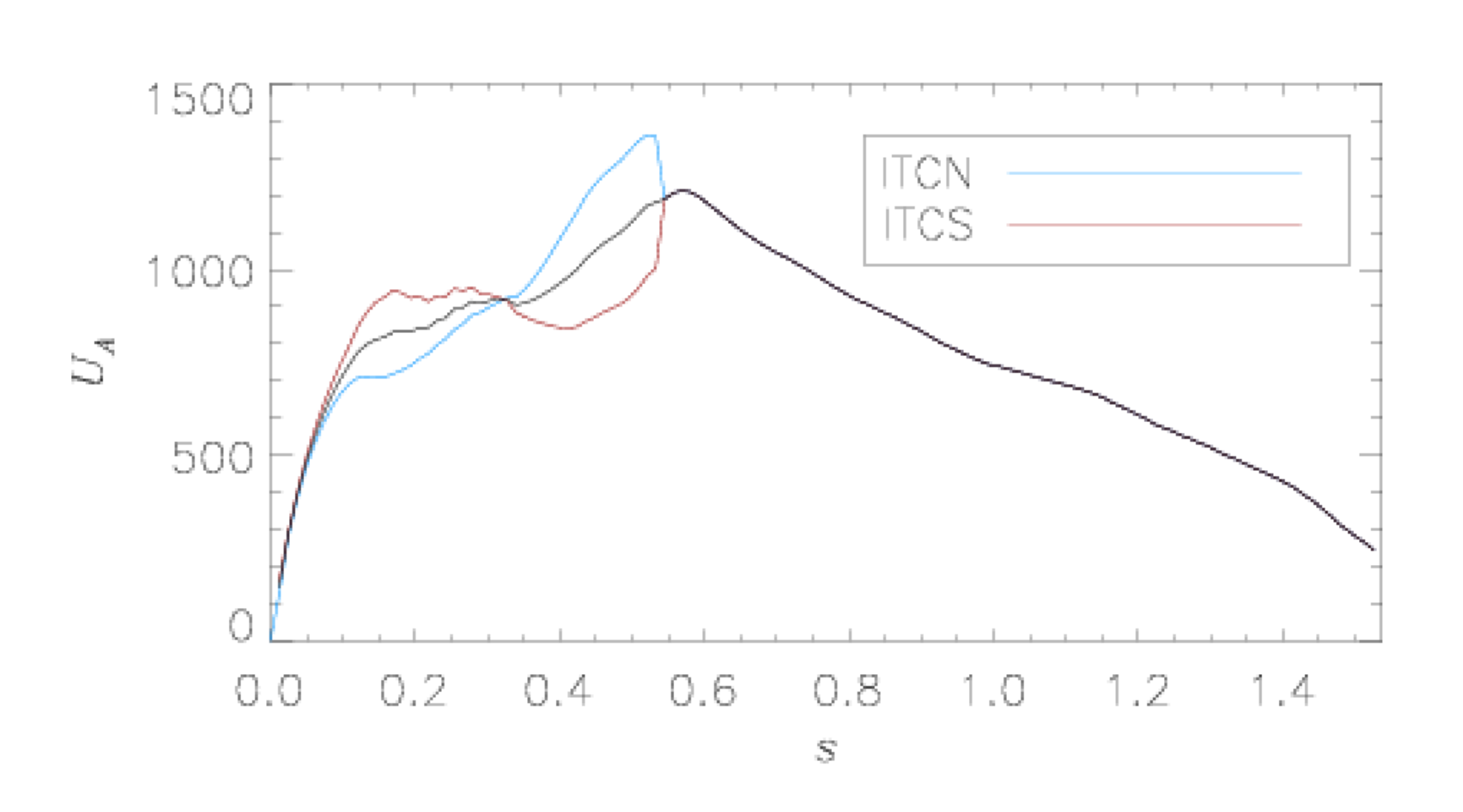}
\caption{Alfv\'{e}n speed, as a function of $s$, for the run 6.5R5.}
\label{fig:6.5R5Alf}
% \end{minipage}
\end{figure}

Figs \ref{fig:6.5R2Alf} and \ref{fig:6.5R5Alf} show $U_A$ as a function of $s$ for the two runs 6.5R2 and 6.5R5 respectively. Blue and red curves indicate a $z$-average over the northern and southern hemisphere respectively whereas the black curve is an average performed over all $z$-space. These plots are typical for all runs with the same values of $Pm$ so we do not present further plots of $U_A$ here. The form of $U_A$ is broadly similar in the two cases: increasing rapidly from the origin (but not identically zero at $s=0$), reaching a peak at the TC (clearly located at $s\approx0.538$) and generally decreasing OTC as the equatorial region at the CMB is approached. The main difference is an increase in the magnitude of the Alfv\'{e}n speed as the magnetic Prandtl number is increased. This is to be expected owing to the dependence of $U_A$ on $Pm$ shown in (\ref{eq:FLRdot}). The only major difference in the form of $U_A$ at different magnetic Prandtl numbers is that runs with lower $Pm$ tend to retain their peak Alfv\'{e}n speed for a significant region OTC. Conversely, at higher $Pm$ the Alfv\'{e}n speed, as a function of $s$, decreases more or less immediately and monotonically from the TC to the CMB at the equator.

% \hspace{0.5cm}

\begin{figure}
%  \begin{minipage}[b]{0.45\linewidth}
\centering
\includegraphics[width=0.8\linewidth]{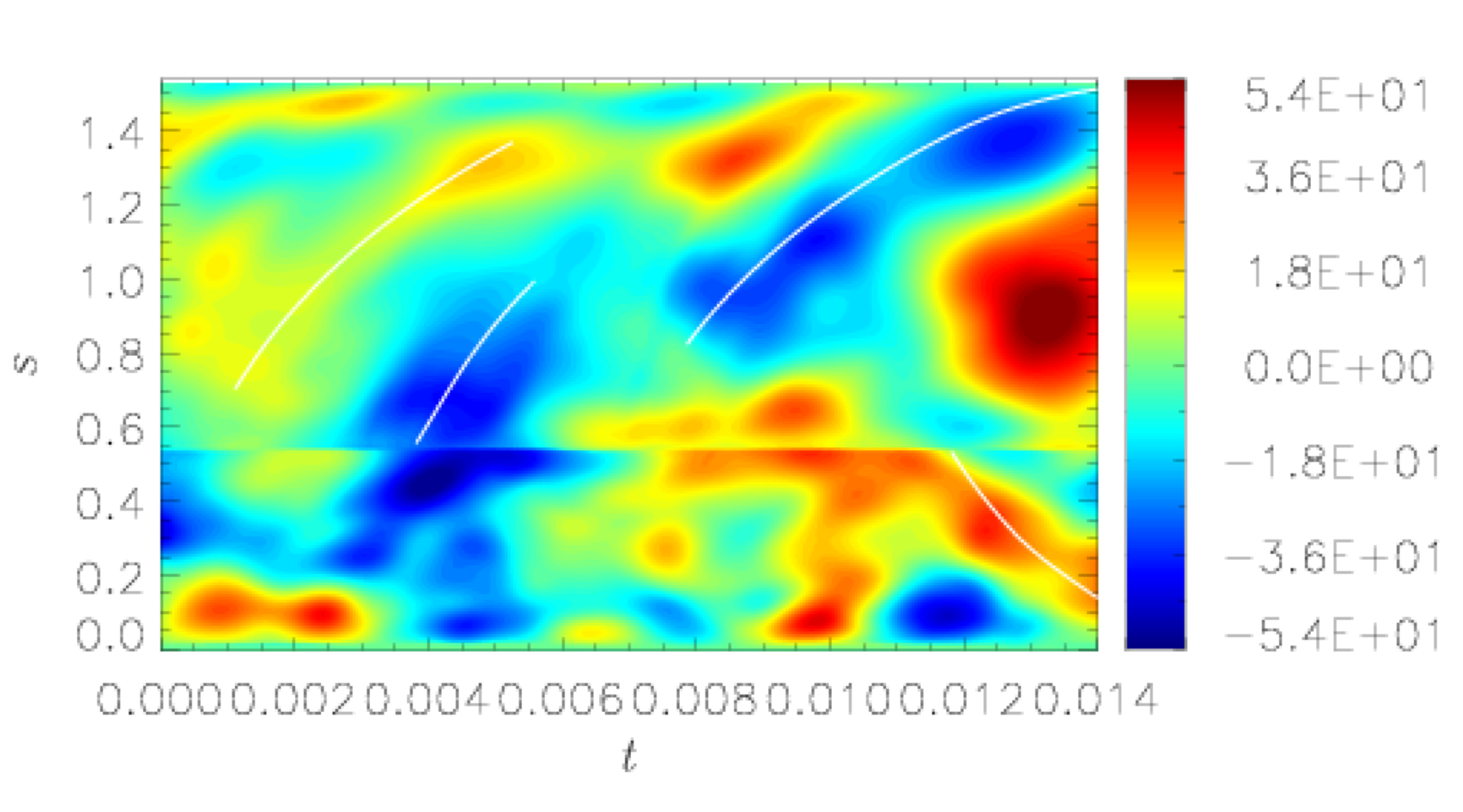} \\
\includegraphics[width=0.8\linewidth]{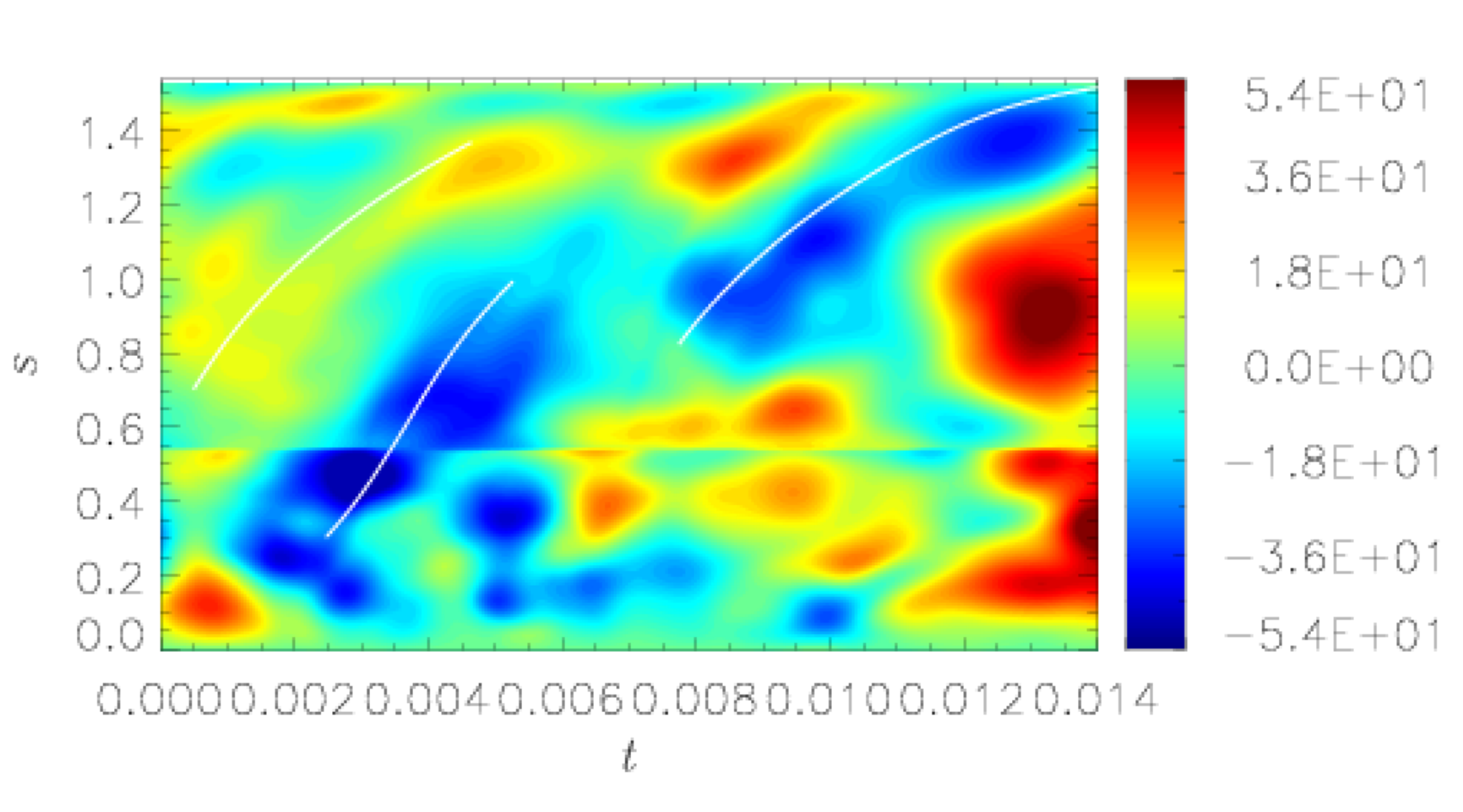} \\
\includegraphics[width=0.8\linewidth]{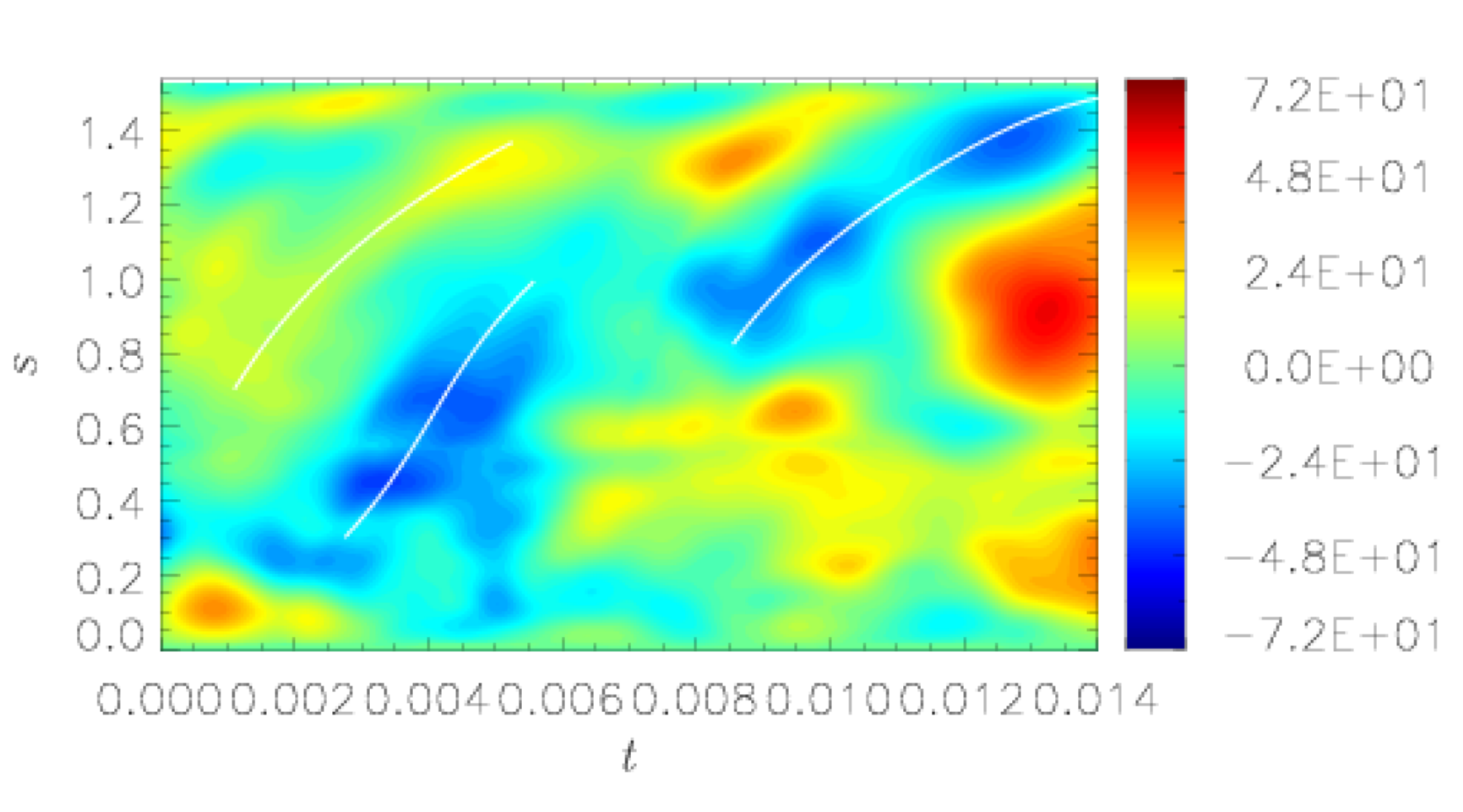}
\caption{Azimuthal velocity, $\langle\overline{u_\phi}\rangle^\prime$, for the run 4R5, as a function of distance, $s$, from the rotation axis and time, $t$, in magnetic diffusion units.}
\label{fig:4R5uphi}
%  \end{minipage}
\end{figure}

\begin{figure}
% \begin{minipage}[b]{0.45\linewidth}
\centering
\includegraphics[width=0.8\linewidth]{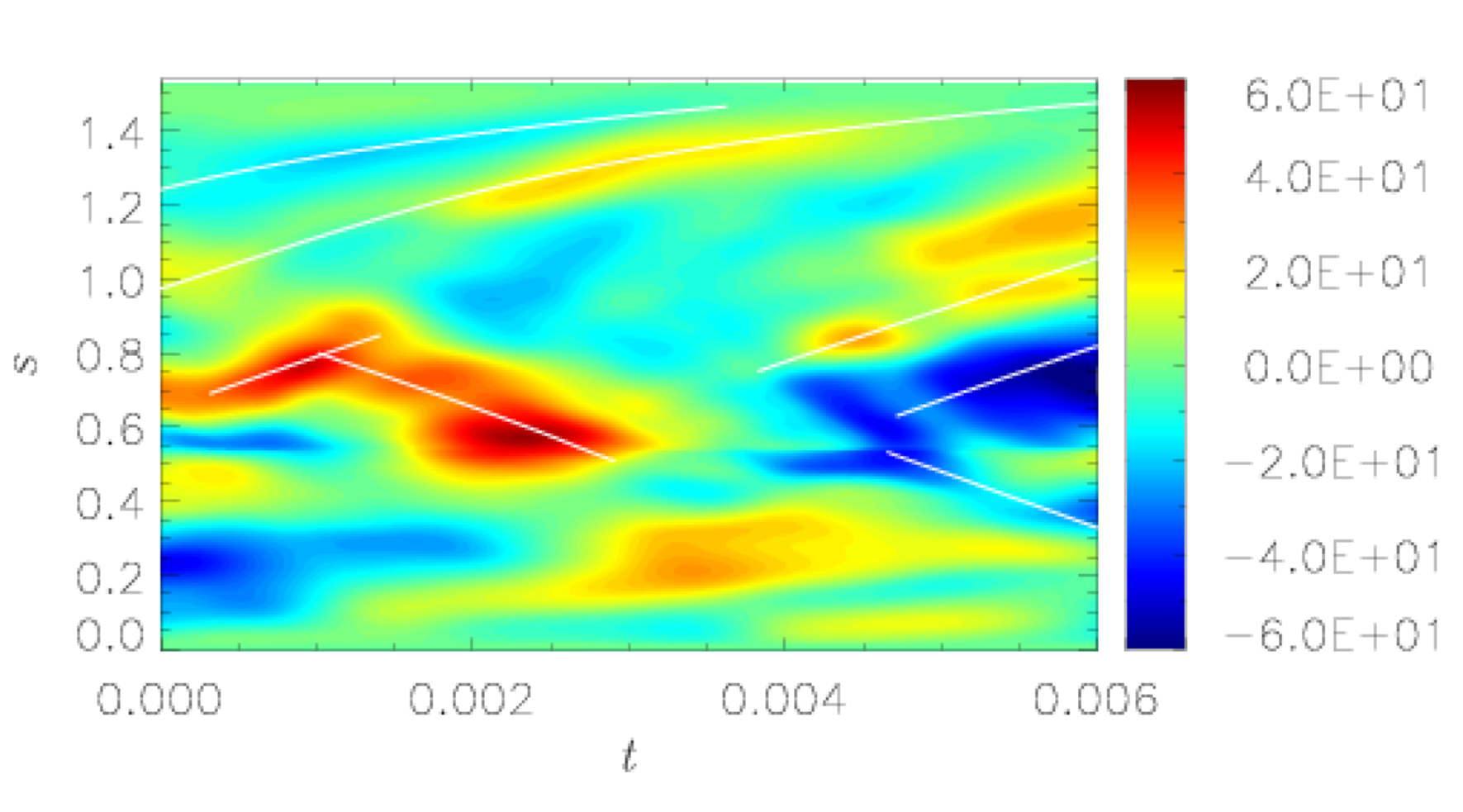}\\
\includegraphics[width=0.8\linewidth]{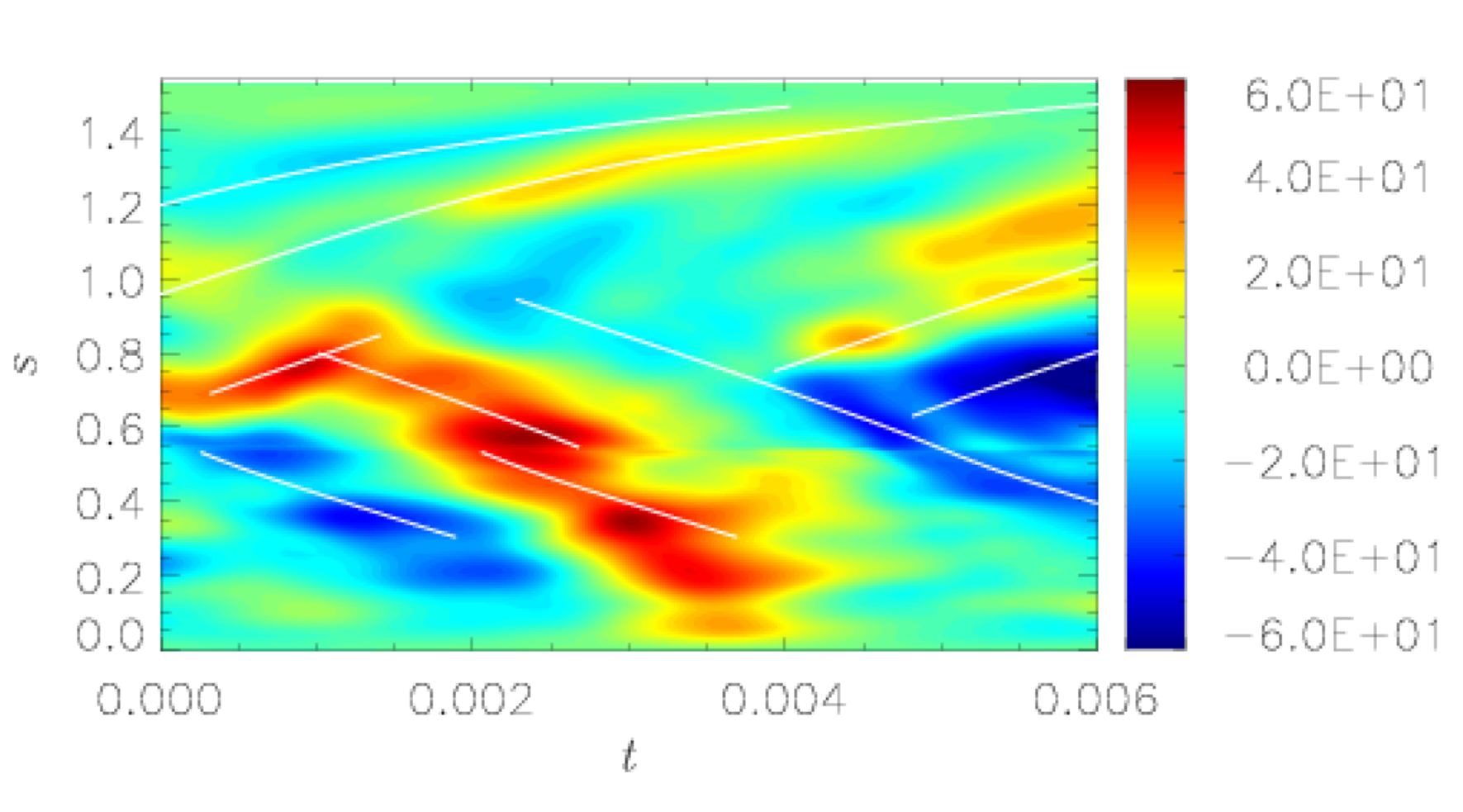}\\
\includegraphics[width=0.8\linewidth]{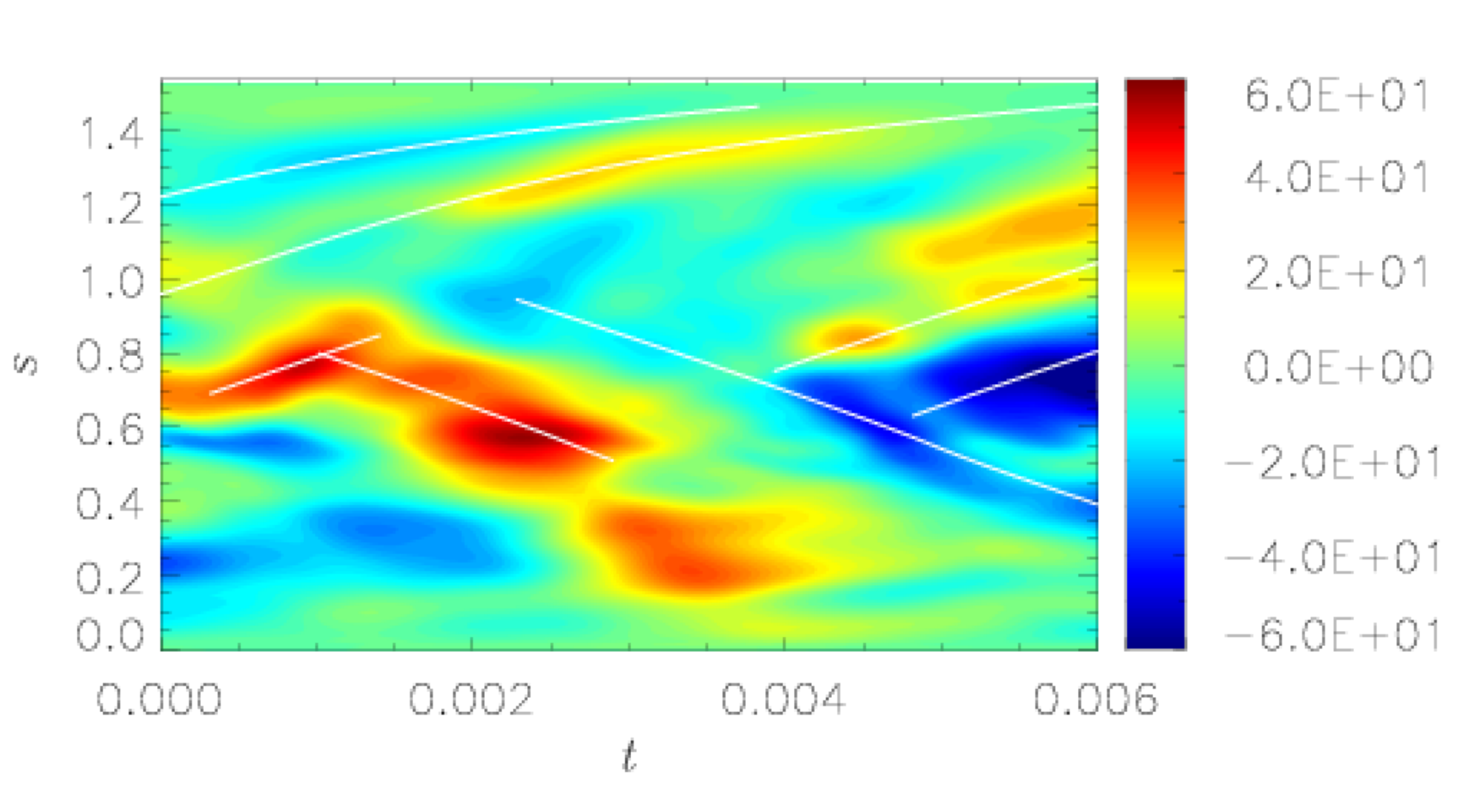}
\caption{Azimuthal velocity, $\langle\overline{u_\phi}\rangle^\prime$, for the run 5R2.}
\label{fig:5R2uphi}
% \end{minipage}
\end{figure}

\begin{figure}
% \begin{minipage}[b]{0.45\linewidth}
\centering
\includegraphics[width=0.8\linewidth]{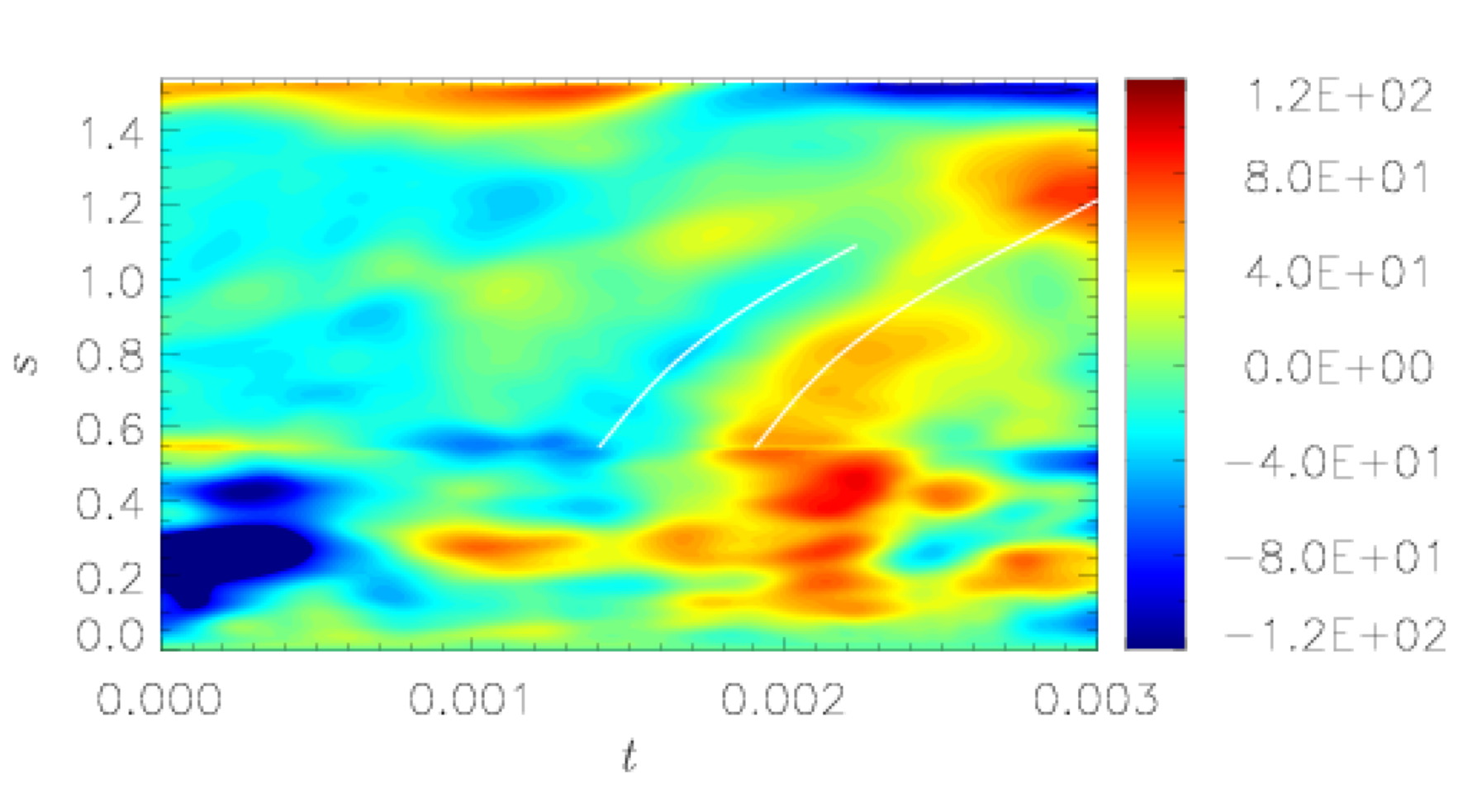} \\
\includegraphics[width=0.8\linewidth]{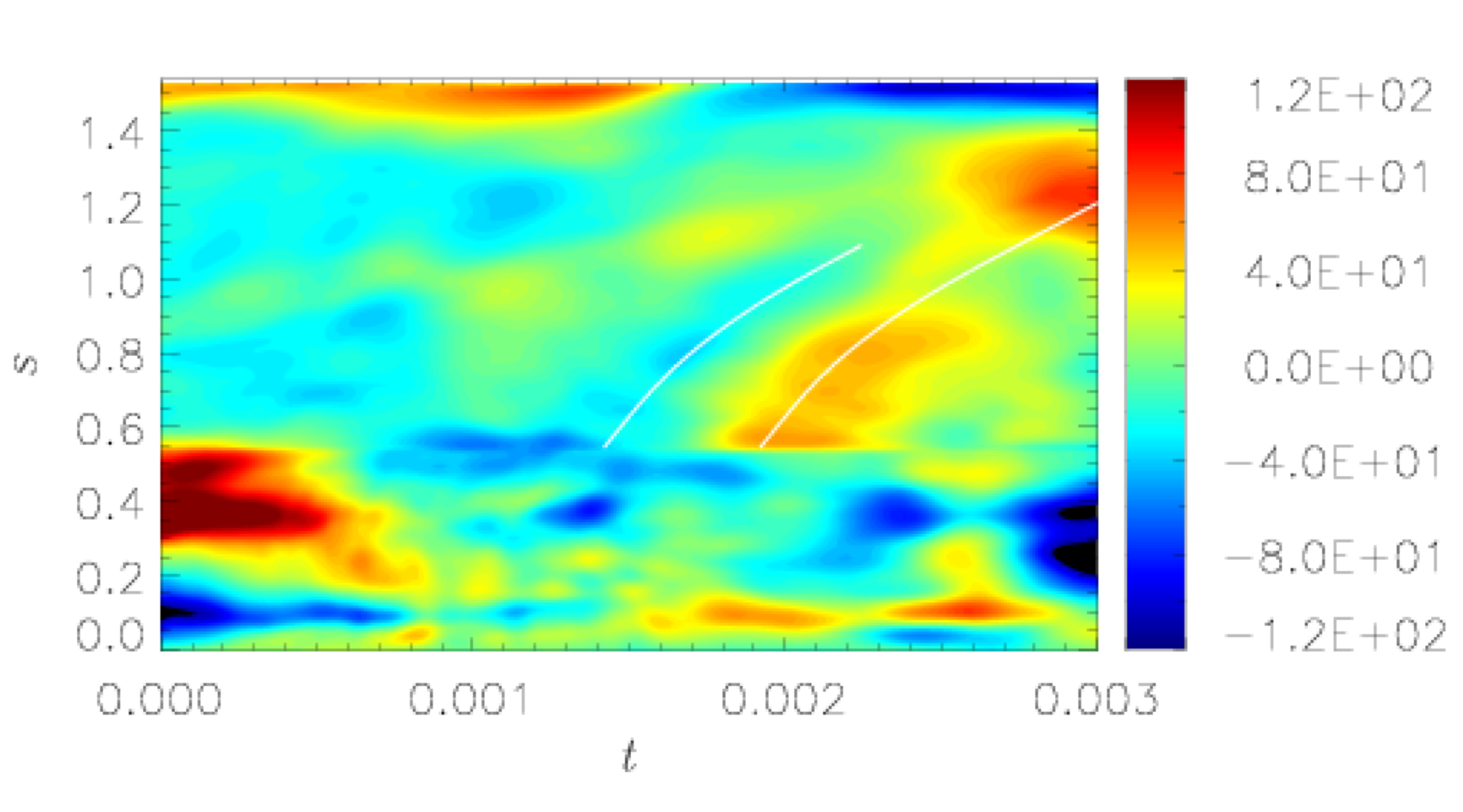} \\
\includegraphics[width=0.8\linewidth]{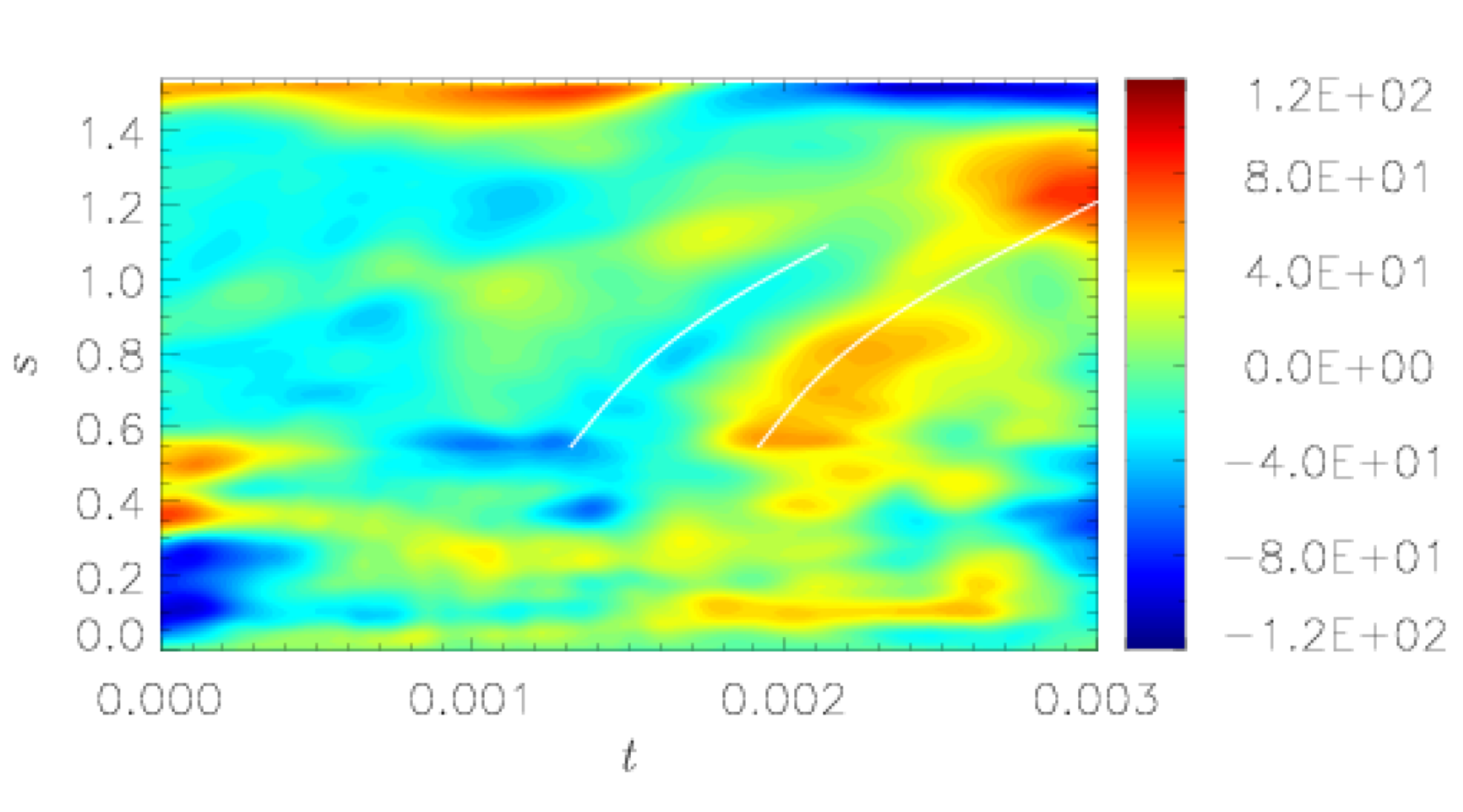}
\caption{Azimuthal velocity, $\langle\overline{u_\phi}\rangle^\prime$, for the run 5R5.}
\label{fig:5R5uphi}
% \end{minipage}
\end{figure}

\begin{figure}
% \begin{minipage}[b]{0.45\linewidth}
\centering
\includegraphics[width=0.8\linewidth]{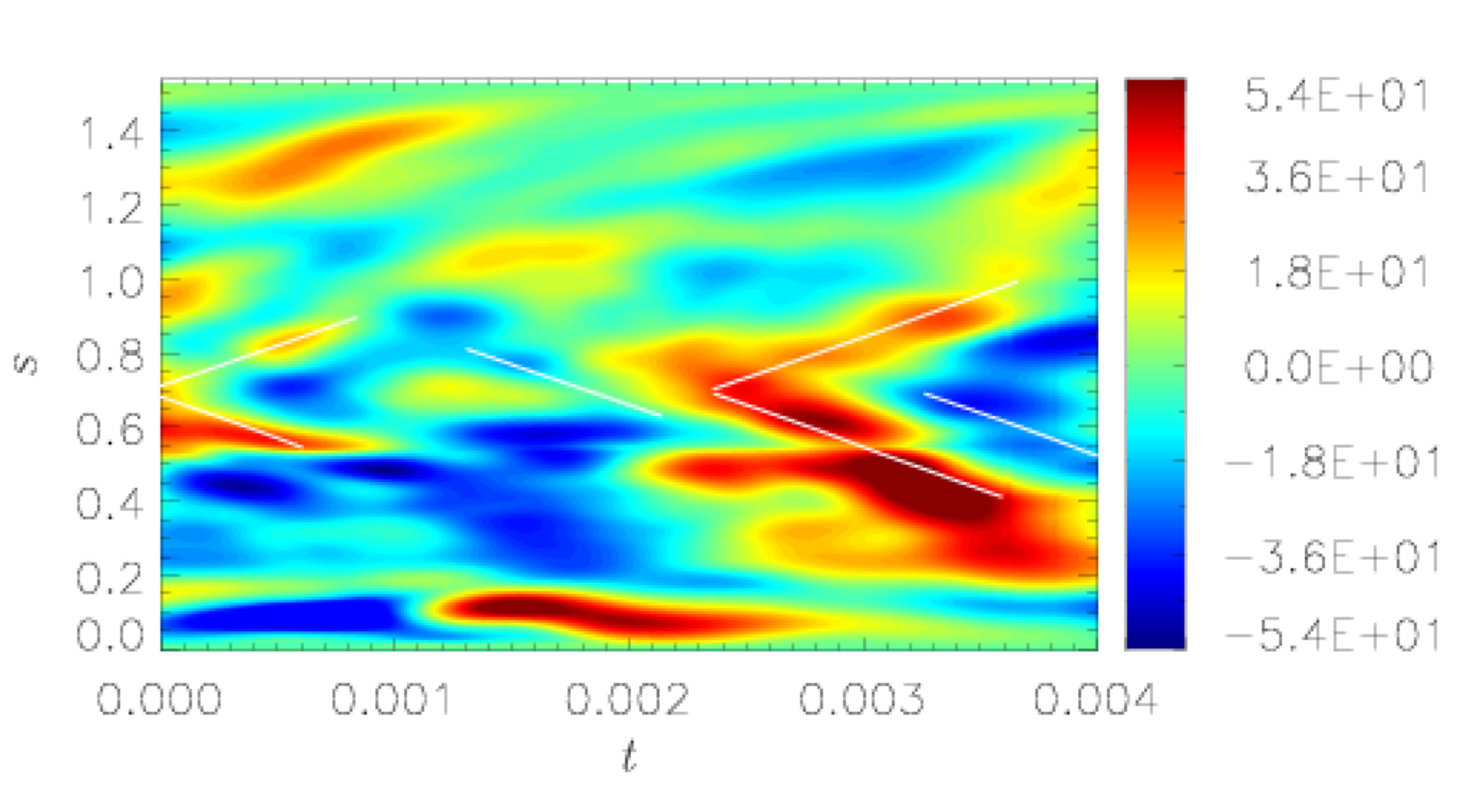}\\
\includegraphics[width=0.8\linewidth]{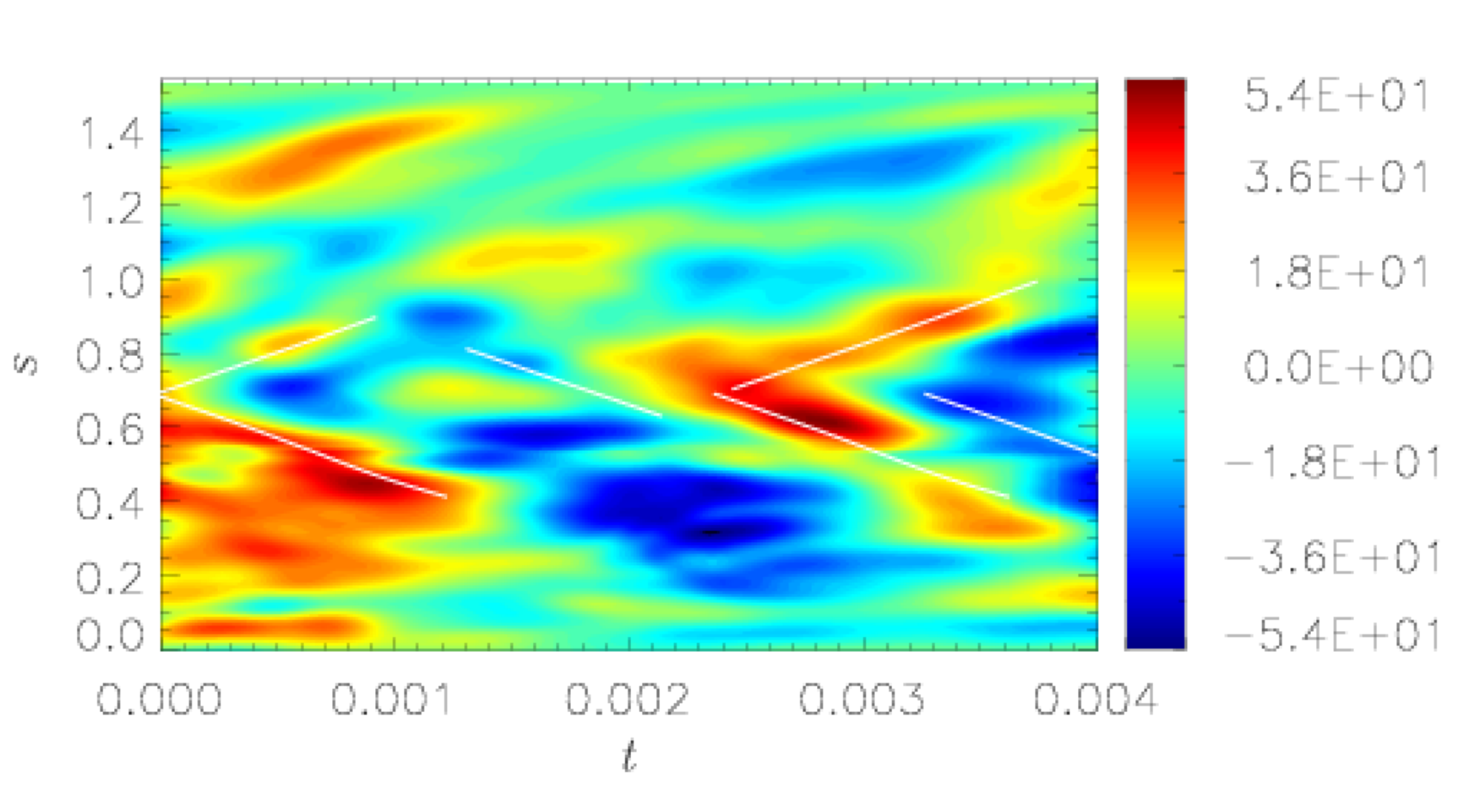}\\
\includegraphics[width=0.8\linewidth]{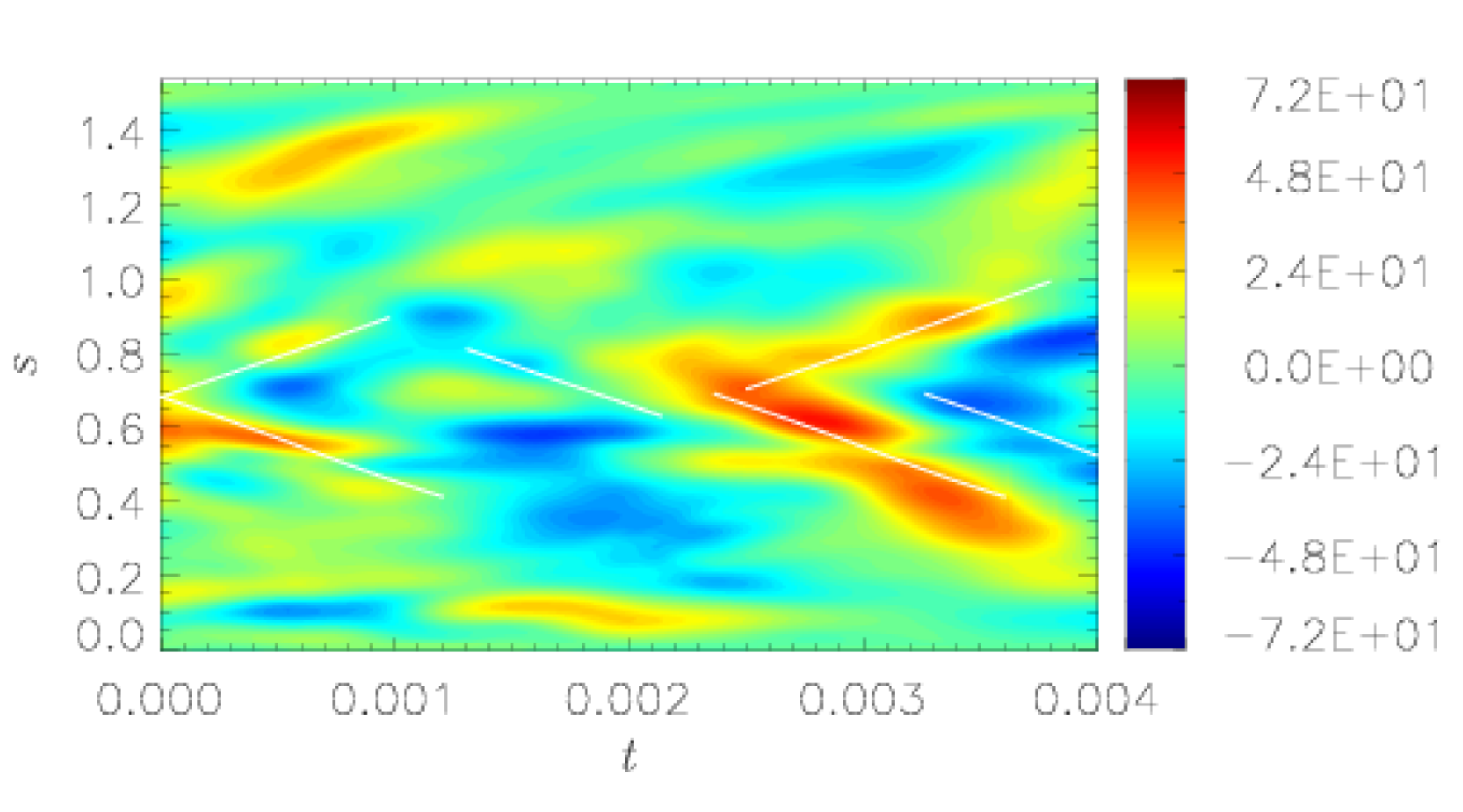}
\caption{Azimuthal velocity, $\langle\overline{u_\phi}\rangle^\prime$, for the run 6.5R2.}
\label{fig:6.5R2uphi}
% \end{minipage}
\end{figure}

\begin{figure}
% \begin{minipage}[b]{0.45\linewidth}
\centering
\includegraphics[width=0.8\linewidth]{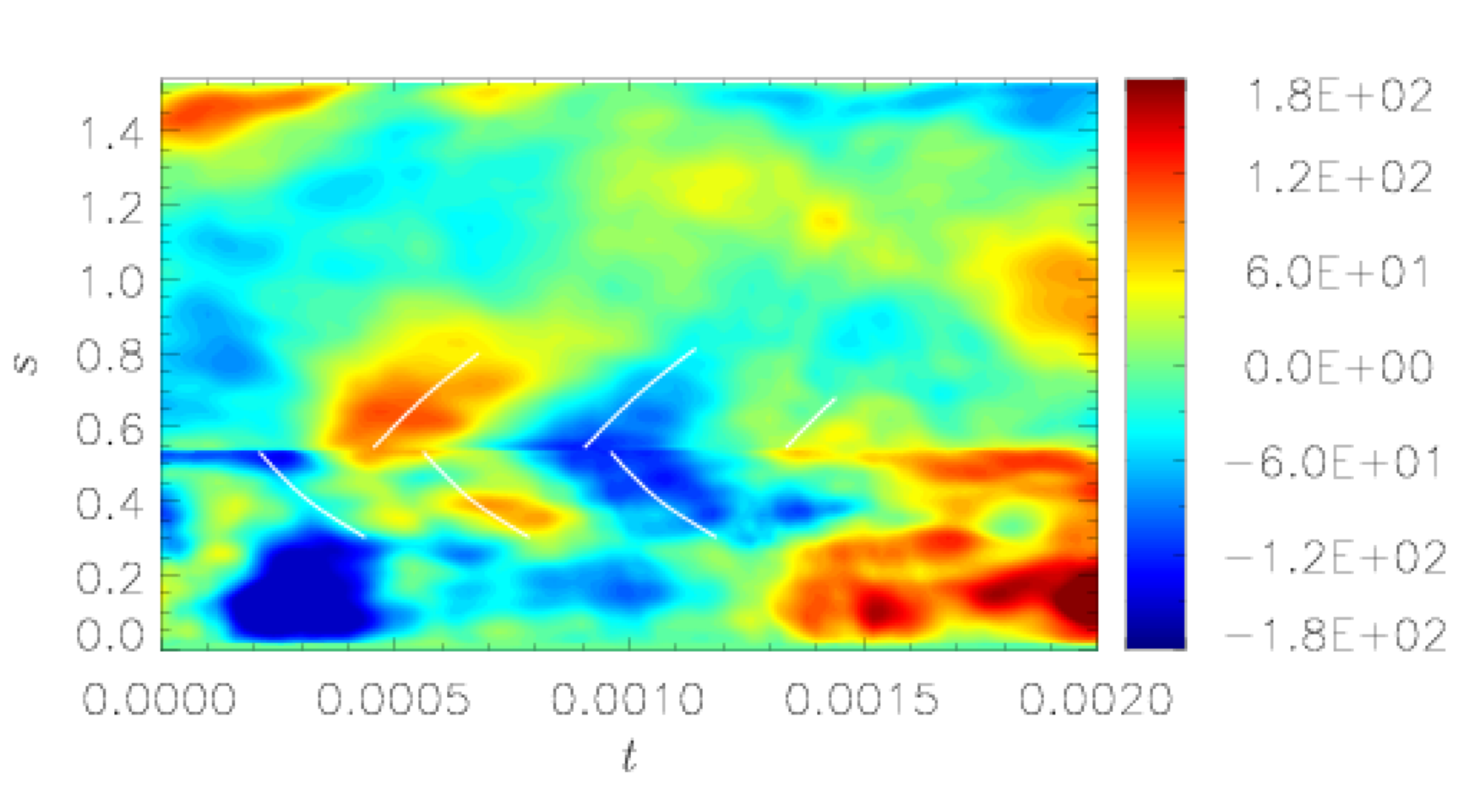}\\
\includegraphics[width=0.8\linewidth]{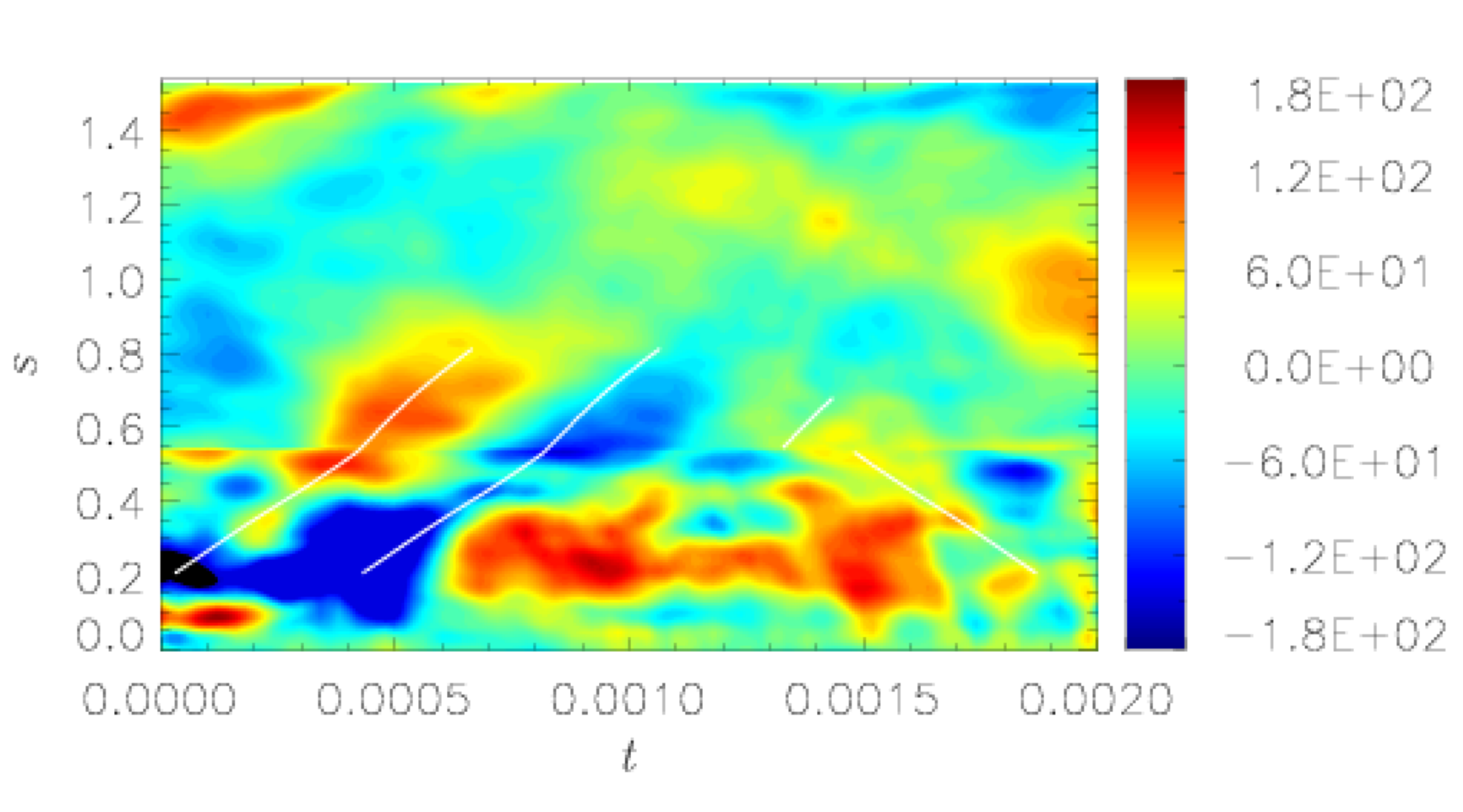}\\
\includegraphics[width=0.8\linewidth]{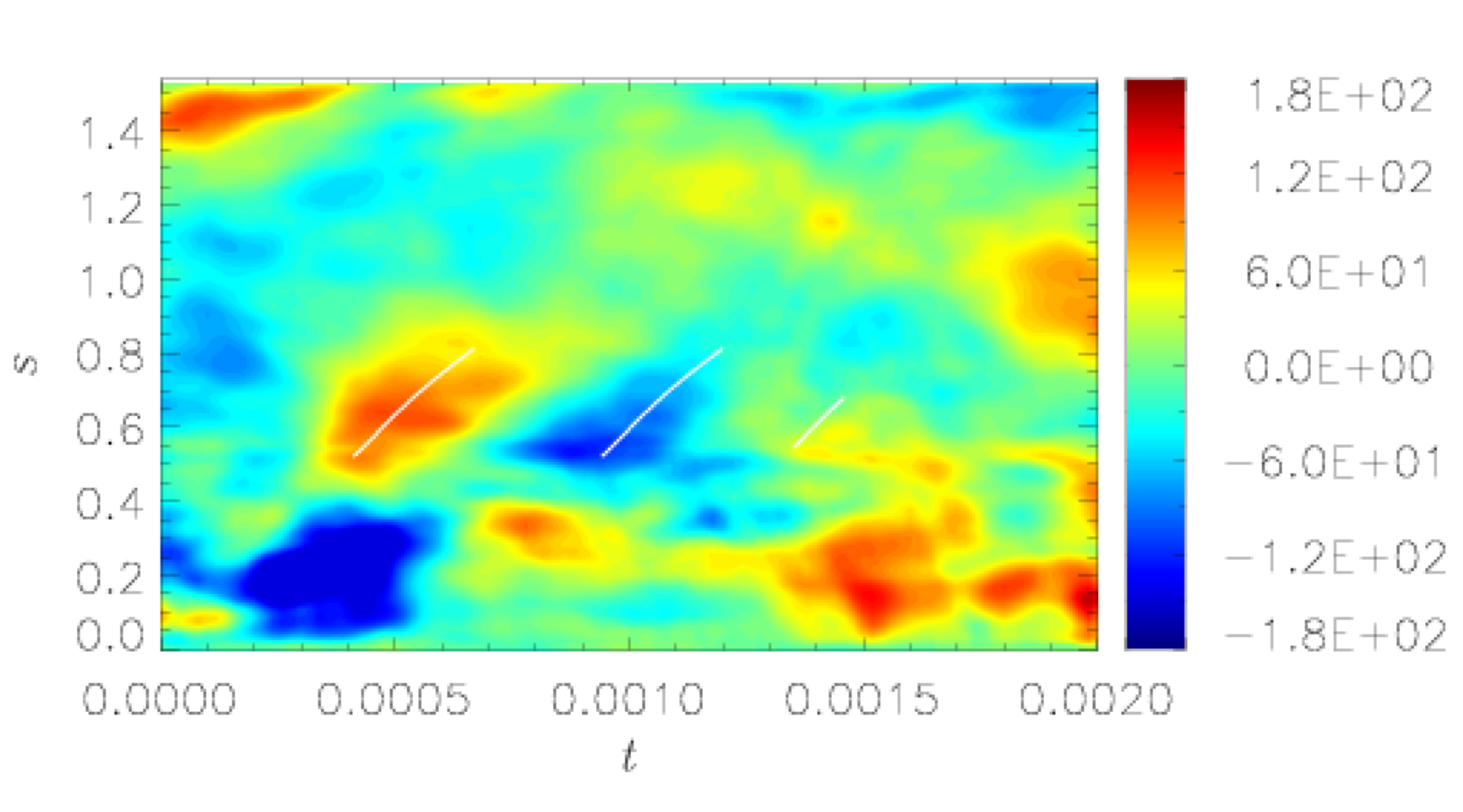}
\caption{Azimuthal velocity, $\langle\overline{u_\phi}\rangle^\prime$, for the run 6.5R5.}
\label{fig:6.5R5uphi}
% \end{minipage}
\end{figure}

In Figs \ref{fig:4R5uphi} to \ref{fig:6.5R5uphi} we display colour-coded density plots of $\langle\overline{u_\phi}\rangle^\prime$ in $ts$-space for several runs. For these figures we have chosen runs from both regimes described in section \ref{sec:field}. Each of the figures contains three plots which display the different possible averaging domains ITC. The top/middle plot is for ITCN/ITCS whereas the bottom plot takes the average over the entire $z$-domain. Each plot contains the same data OTC. Overlaying each plot are several white curves that display trajectories that features take when travelling at the Alfv\'{e}n speed, $U_A$. Note that these curves do not have a constant gradient since the Alfv\'{e}n speed is a function of $s$.

The first run that we display plots for is a run with $Pm=5$ and $E=10^{-4}$, which is the largest value of the Ekman number considered. Runs in the weak field regime were not found to permit TOs at this large an Ekman number. In Fig. \ref{fig:4R5uphi}, for run 4R5, several structures in $\langle\overline{u_\phi}\rangle^\prime$ can be identified as torsional waves since they follow a trajectory predicted by $U_A$. These features appear regularly and can be seen to originate at various locations of the domain indicating that the waves can, but are not obliged to appear from the TC. Within the tangent cylinder a wave propagates inwards from the TC in the northern hemisphere (at $t\simeq0.011$); the only feature to do so in this run.

In Figs \ref{fig:5R2uphi} (for a weak field regime at $Pr=2$) and \ref{fig:5R5uphi} (for a strong field regime at $Pr=5$) the Ekman number has been reduced by an order of magnitude compared with Fig. \ref{fig:4R5uphi}. In both sets of plots several torsional oscillations are again immediately apparent. Features in $\langle\overline{u_\phi}\rangle^\prime$ travel slower in the lower $Pm$ case owing to the smaller magnetic field strength generated at lower magnetic Prandtl number. However, it is certainly noticeable, from the timescale on the plots alone, that waves are propagating significantly faster at lower Ekman number, as expected from (\ref{eq:FLRdot}).

There is evidence of an inward propagating wave passing through the tangent cylinder (at $s\approx0.538$) in Fig. \ref{fig:5R2uphi} shortly after $t=0.002$. It is clear from the top and middle plots that this wave continues to propagate in the southern hemisphere ITC but does not ITCN. At $t\simeq0.005$ a second structure again appears to pass through the TC, this time in both hemispheres. Run 5R2 also has an approximately similar number of inward and outward propagating waves. Conversely, run 5R5 is dominated by two structures originating at the TC and moving radially outwards towards the equator at the CMB. Neither inwards propagating TOs nor TOs within the TC were identified in this run.

% \hspace{0.5cm}

When the Ekman number is reduced further to $E=5\times10^{-6}$, for runs 6.5R2 and 6.5R5, we continue to observe faster moving waves with lower Ekman number. Other than the difference in the speed of the waves, run 6.5R2 is rather similar to run 5R2 since Fig. \ref{fig:6.5R2uphi} displays several oscillations propagating both inwards and outwards as well as persistence through the TC. There are TOs propagating from the TC in run 6.5R5 as well as possible evidence of waves ITC propagating in either direction. However, several of the features highlighted with white curves in Fig. \ref{fig:6.5R5uphi} will become more apparent when we apply bandpass filtering and thus we retain further discussion until section \ref{sec:filter}.

% \hspace{0.5cm}

\begin{figure*}
% \begin{minipage}[b]{0.3\linewidth}
\centering
\includegraphics[width=0.25\linewidth]{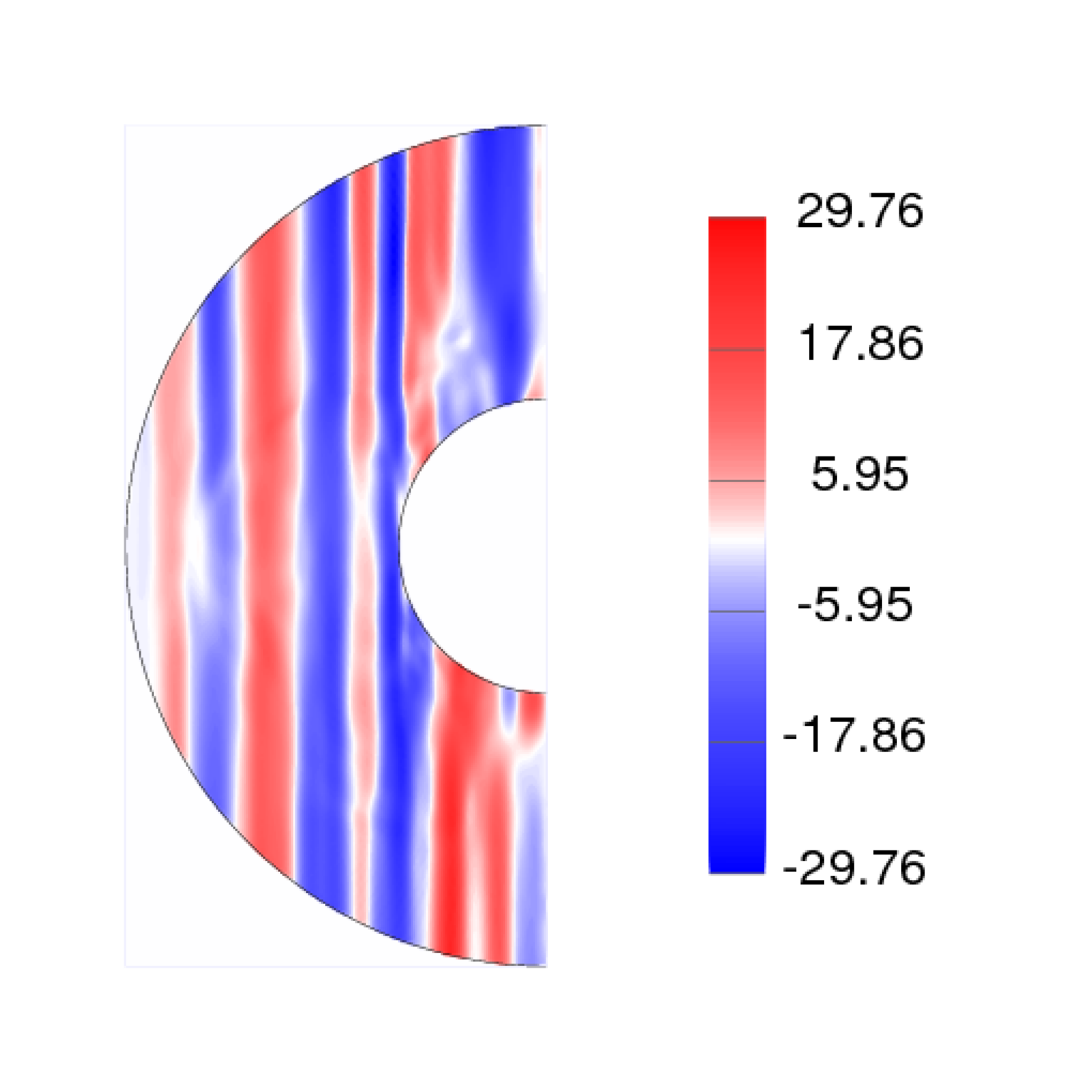}
% \caption{$t=0.0004$}
\label{fig:5R2uphi20}
% \end{minipage}
 \hspace{-0.9cm}
% \begin{minipage}[b]{0.3\linewidth}
\centering
\includegraphics[width=0.25\linewidth]{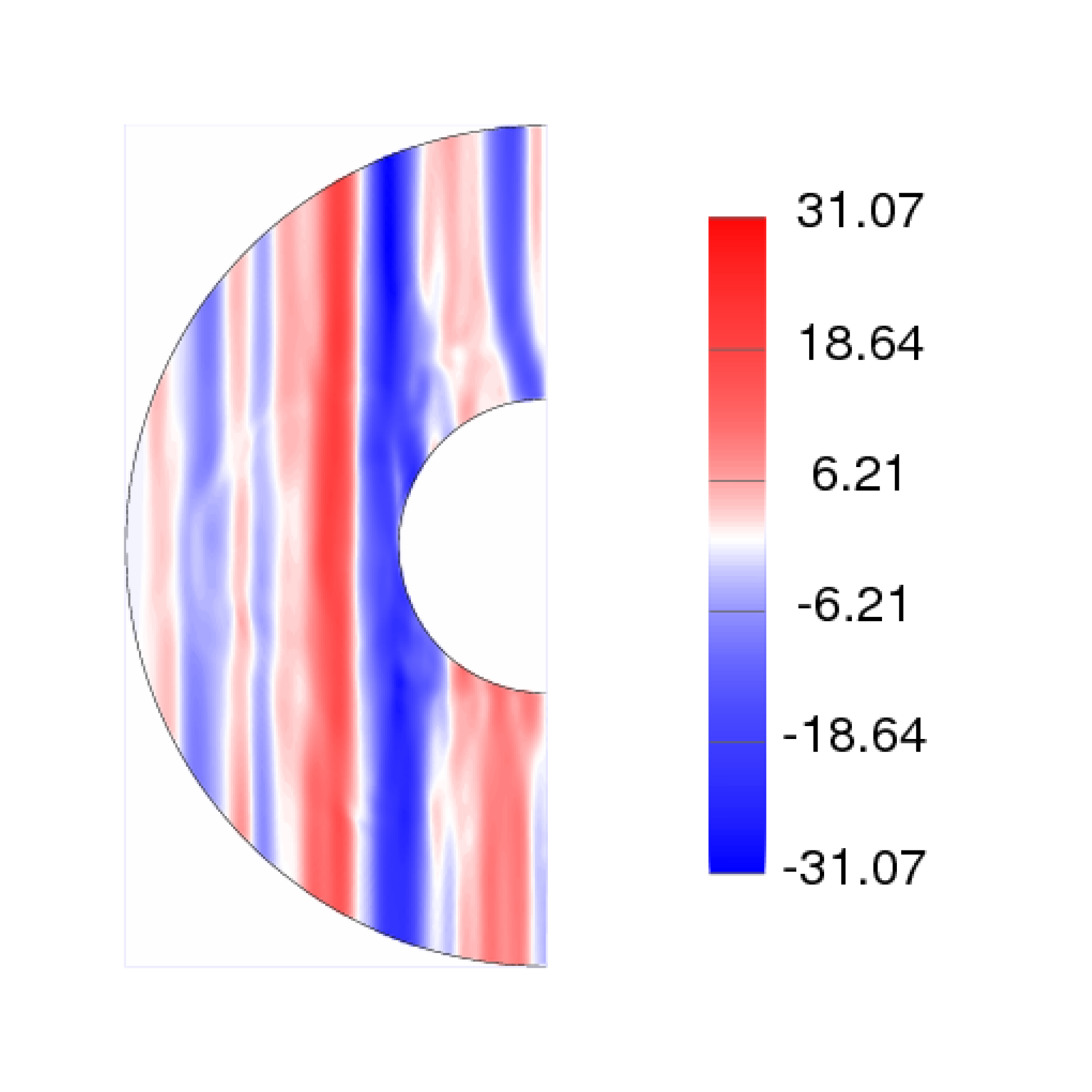}
% \caption{$t=0.0008$}
\label{fig:5R2uphi40}
% \end{minipage}
\hspace{-0.9cm}
% \begin{minipage}[b]{0.3\linewidth}
\centering
\includegraphics[width=0.25\linewidth]{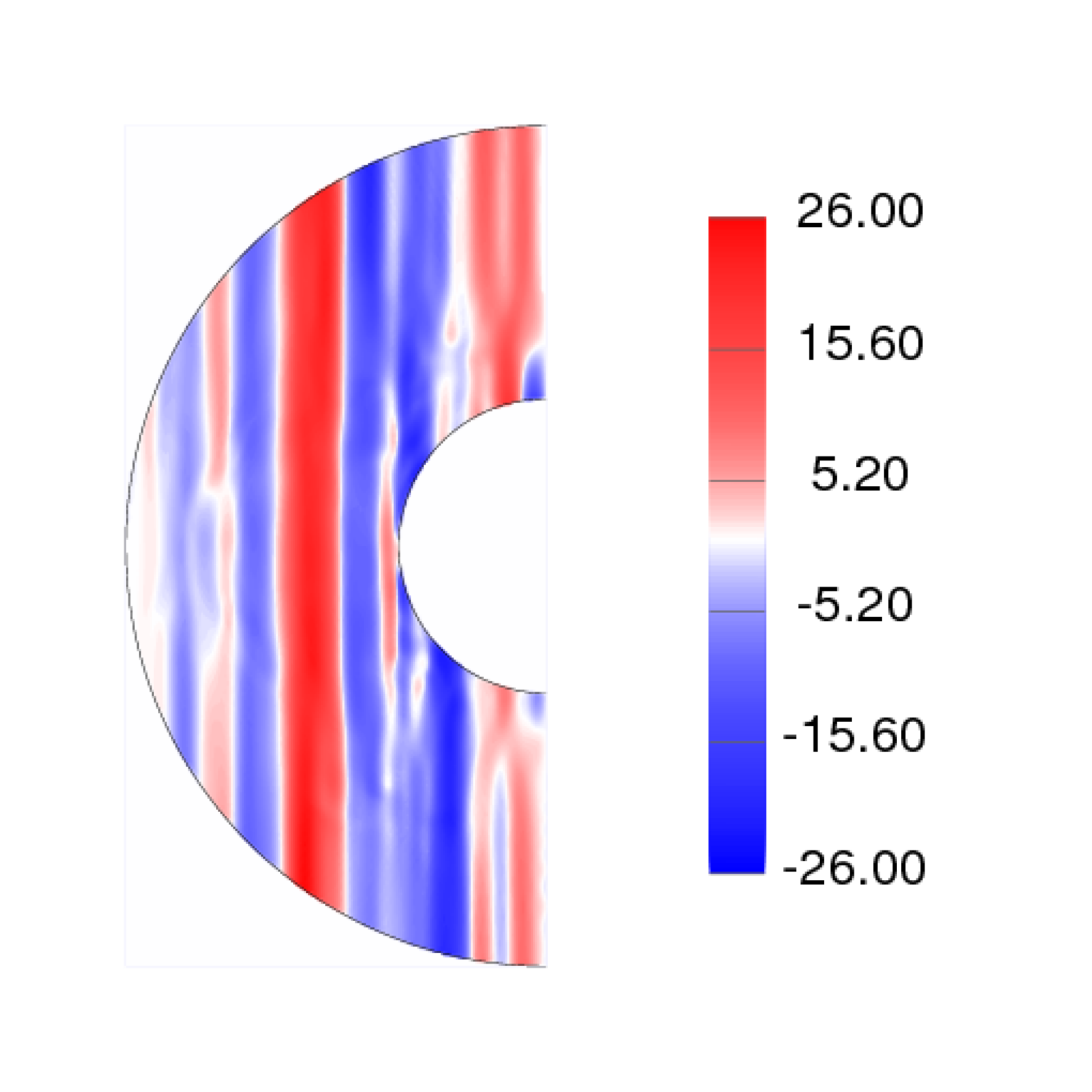}
% \caption{$t=0.0012$}
\label{fig:5R2uphi60}
% \end{minipage}
\hspace{-0.9cm}
% \begin{minipage}[b]{0.3\linewidth}
\centering
\includegraphics[width=0.25\linewidth]{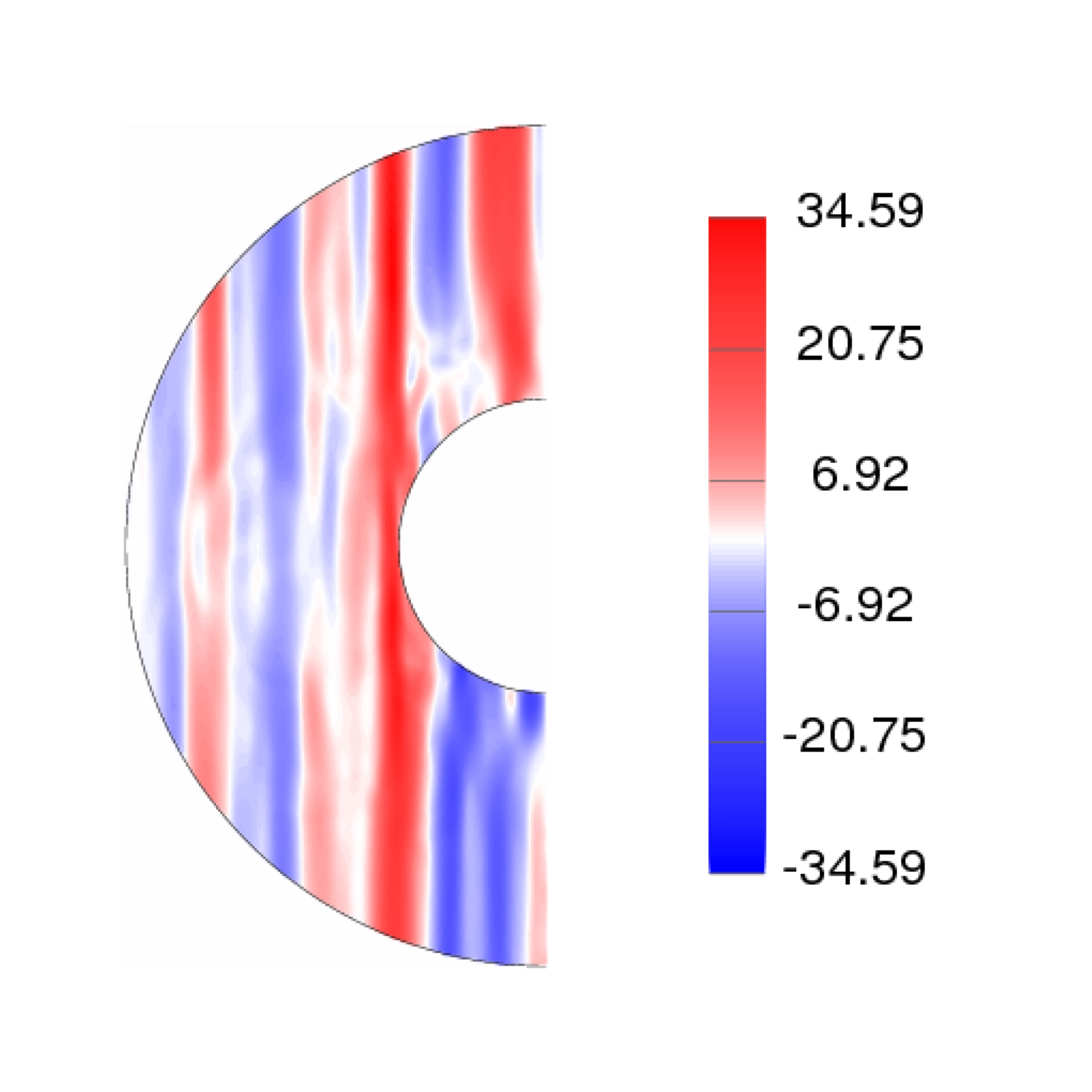}
% \caption{$t=0.0016$}
\label{fig:5R2uphi80}
% \end{minipage}
\caption{Series of snapshots of $\overline{u_\phi}^\prime$ for the run 5R2. Panels from left to right are at the following times: $t=0.0004$, $t=0.0008$, $t=0.0012$, and $t=0.0016$.}
\label{fig:5R2uphisnap}
\end{figure*}

\begin{figure*}
% \begin{minipage}[b]{0.3\linewidth}
\centering
\includegraphics[width=0.25\linewidth]{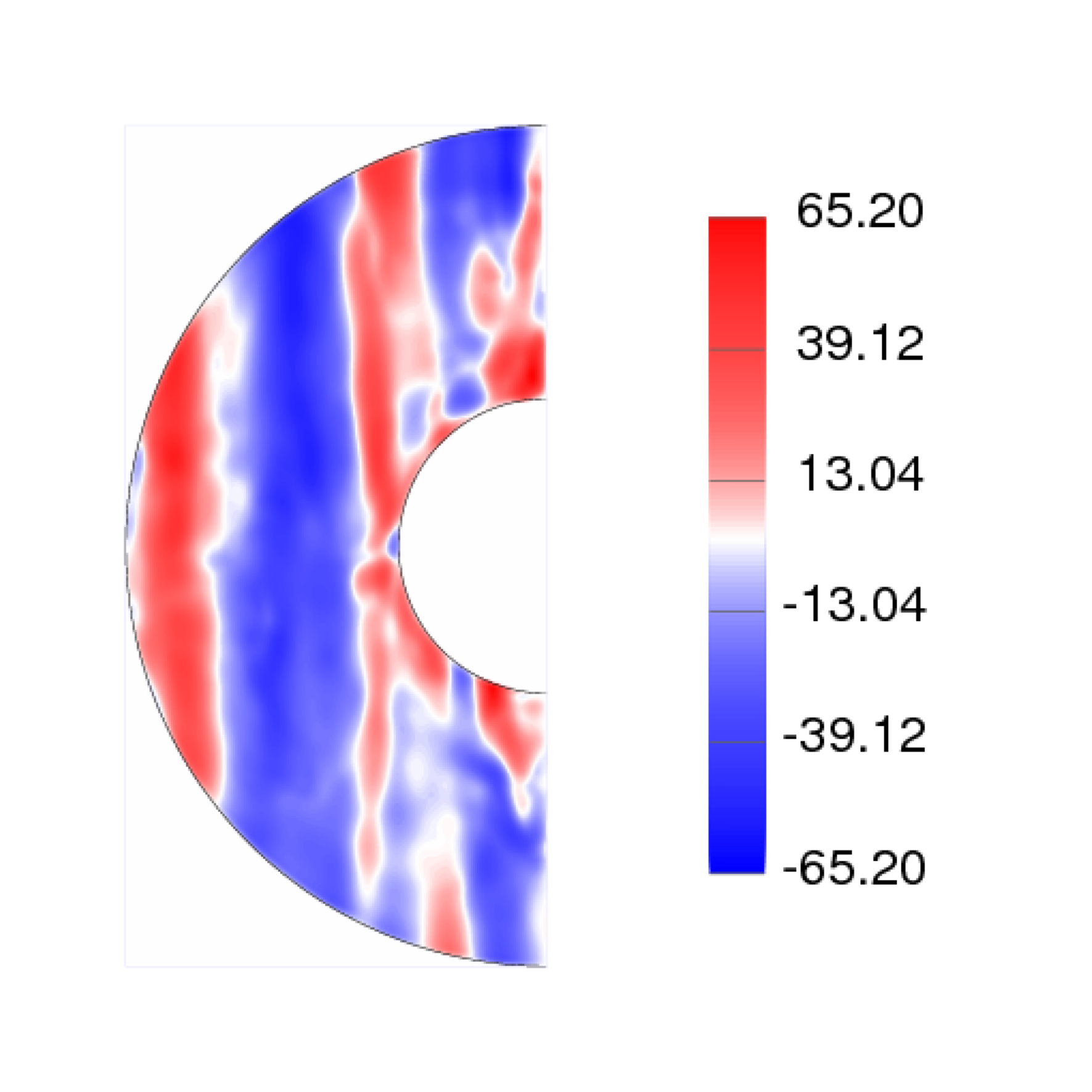}
% \caption*{$t=0.009$}
\label{fig:5R2uphi50}
% \end{minipage}
 \hspace{-1.1cm}
% \begin{minipage}[b]{0.3\linewidth}
\centering
\includegraphics[width=0.25\linewidth]{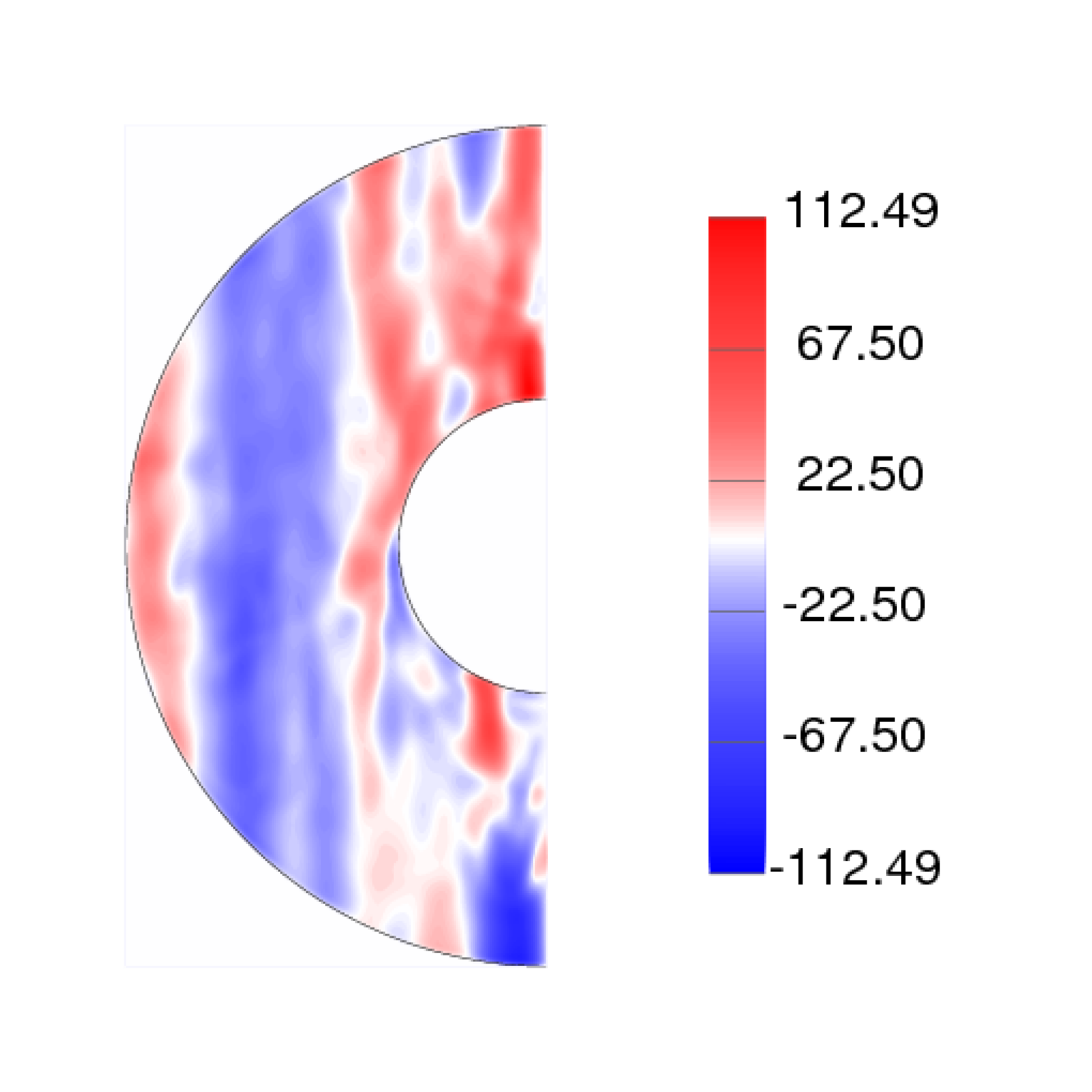}
% \caption*{$t=0.010$}
\label{fig:5R2uphi100}
% \end{minipage}
\hspace{-1.1cm}
% \begin{minipage}[b]{0.3\linewidth}
\centering
\includegraphics[width=0.25\linewidth]{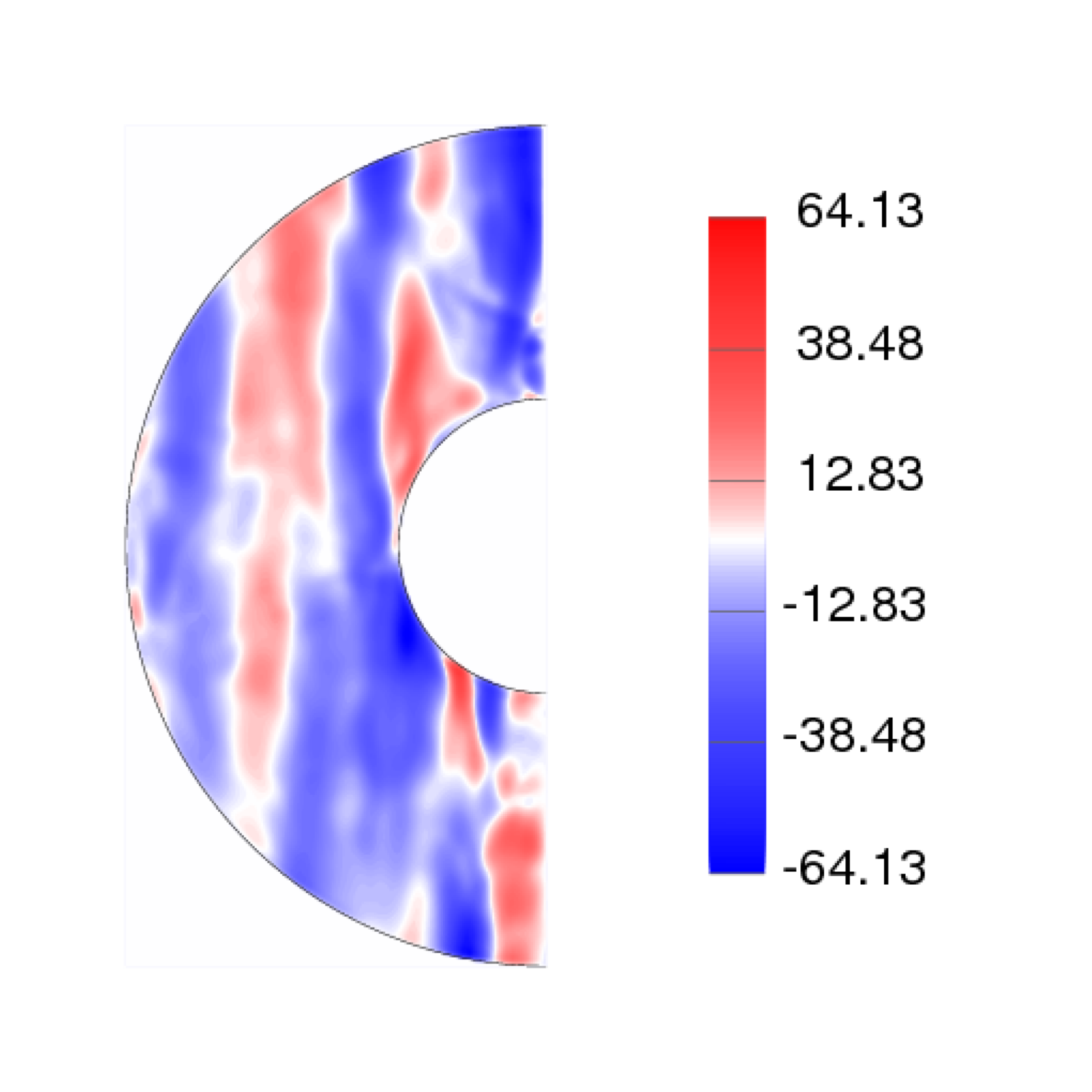}
% \caption*{$t=0.012$}
\label{fig:5R2uphi175}
% \end{minipage}
\hspace{-1.1cm}
% \begin{minipage}[b]{0.3\linewidth}
\centering
\includegraphics[width=0.25\linewidth]{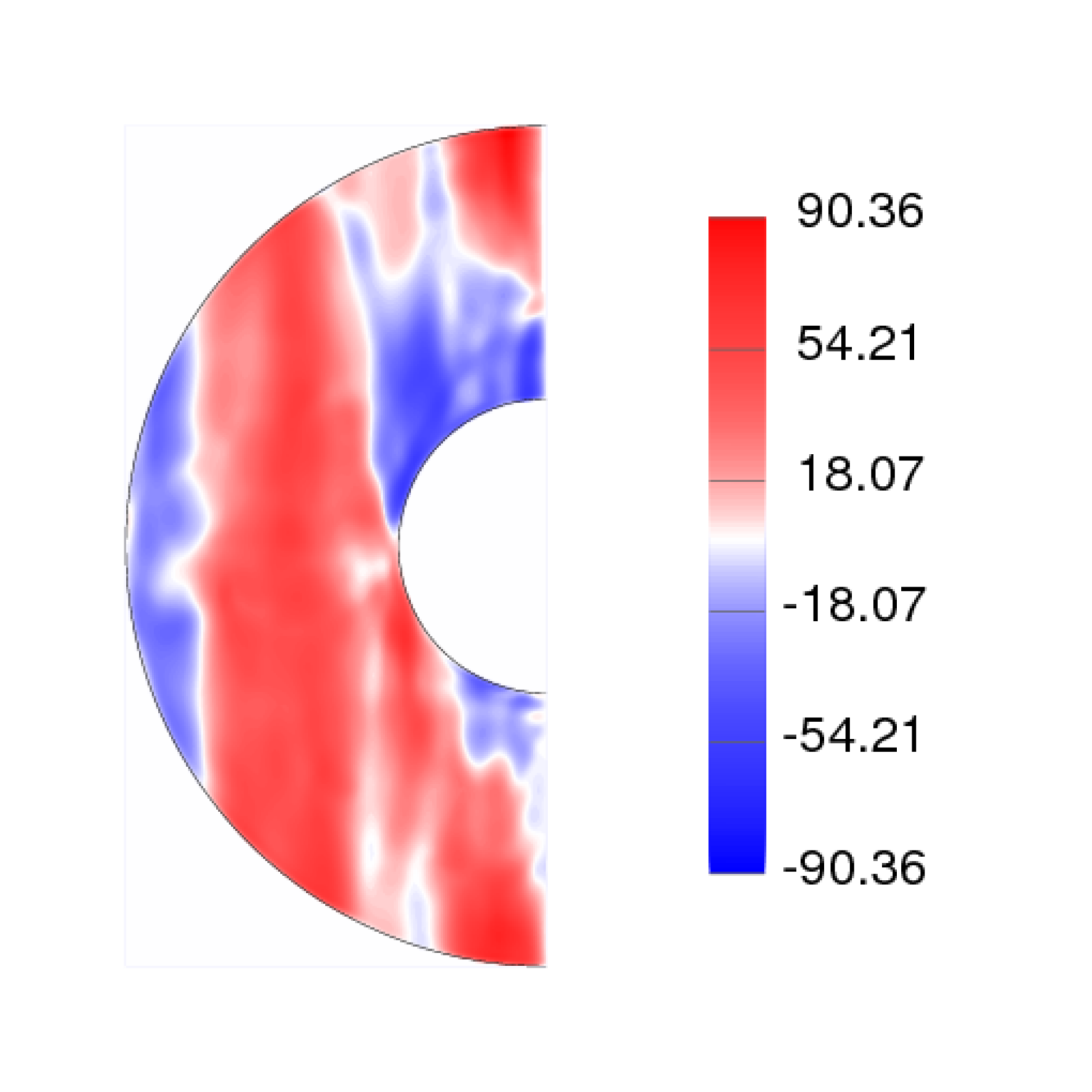}
% \caption*{$t=0.014$}
\label{fig:5R2uphi275}
% \end{minipage}
\caption{Series of snapshots of $\overline{u_\phi}^\prime$ for the run 4R5. Panels from left to right are at the following times: $t=0.009$, $t=0.010$, $t=0.012$, and $t=0.014$.}
\label{fig:4R5uphisnap}
\end{figure*}

Figs \ref{fig:5R2uphisnap} and \ref{fig:4R5uphisnap} show a series of snapshots of $\overline{u_\phi}^\prime$ in a meridional section for two runs. In the first set of snapshots, for run 5R2, we see that the azimuthal velocity is very columnar both inside and outside the TC. However, it proves difficult to see evidence of propagation of these columns either inwards or outwards. Analysis of a movie shows occasional propagation of columns but for the most part the oscillations act as standing waves. This is to be expected because we observed from Fig. \ref{fig:5R2uphi} that this run contains both inwards and outwards moving waves in approximately equal numbers. Therefore it is tricky to distinguish between the two directions of travel.

Although the columnar structure of Fig. \ref{fig:4R5uphisnap}, for run 4R5, is less striking, we are able to observe features moving radially outwards. Between $t=0.009$ and $t=0.010$ a positive (red) structure in $\overline{u_\phi}^\prime$ propagates towards the equator and by $t=0.012$ it has dissipated at the boundary. This is shortly followed by a negative (blue) structure that at $t=0.009$ resides in the centre of the region OTC but by $t=0.014$ has moved to the equator as a newly formed positive structure now dominates OTC. These outwards propagating positive and negative features can be directly matched with those of Fig. \ref{fig:4R5uphi} for the section of time from $t=0.009$ to $t=0.014$.

The plots displayed, and more generally the runs considered, in this subsection are representative of other runs from Table \ref{tab:runs} that are in neighbouring regions of parameter space. The general features observed in the figures can be extrapolated to the runs for which we have not displayed plots. For example, runs with $Pm=1$ are found to have an even more columnar structure with even fewer propagating waves compared with the $Pm=2$ cases. Additionally, we find that repeating runs with stress-free boundary conditions do not appear to alter our findings from the rigid case since various plots of the data for the runs 5F1 to 5F5 broadly match those of runs 5R1 to 5R5. This is, perhaps, not surprising when reflecting on the similarity of the output parameters from these two sets of runs (Table \ref{tab:output}).

One feature of TOs that we have not observed is the possible reflection of waves at the equator. This is true not only for the runs for which we have displayed plots, but, more generally, is the case across all of our simulations. Our results are therefore in agreement with \cite{sch12} who suggest that the observation of wave reflection in dynamo simulations with insulating no-slip BCs is not possible due to a small reflection coefficient. 

%%%%%%%%%%%%%%%%%%%%%%%%%%%%%%%%%%%%%%%%%%%%%%%%%%%%%%%%%%%%%%%%%%%%%%%%%%%%%%%%%%%%%%%%%%%%%%%%%%%%%%%%%%%%%%%%%%%%%%%%%%%%%%%%

\subsection{Core travel times}
\label{sec:time}

We are able to estimate the travel time for our observed waves to cross the outer core. However, such estimates must be treated with a considerable degree of caution since the parameter regimes used to produce these simulations are inconsistent with that of the Earth resulting in a difficulty in identifying the timescale to use when converting back from our nondimensional time to physical time.

Consideration of the diffusion timescale reveals that it is not ideal for conversion in our study of TOs since our fields in these units are often too strong. Therefore we choose to convert by matching the Alfv\'{e}n speed at the CMB. Using $0.7$mT as the magnetic field strength at the CMB \citep{gil10} and $\rho=1\times10^4$kgm$^{-3}$ (as well as $\mu_0=4\pi\times10^{-7}$) this gives an Alfv\'{e}n speed of approximately $6\times10^{-3}$ms$^{-1}$ at the CMB.

We can use the values of $U_A(r_o)$ (the Alfv\'{e}n speed at the equator at the CMB) given in Table \ref{tab:output}, as well as $D\approx2.2\times10^6$km, to calculate the dimensional version of $\tau$ from Table \ref{tab:runs}. Since TOs are approximately operating on the $\tau$ timescale we thus find that the outer core travel time of the TOs in our simulations ranges from months to $\approx$ 6 years. TOs in the core are currently believed \citep{gil10} to operate on a 4 to 6 year timescale and, from our set of simulations, it is runs in the strong field regime that fare best at operating on or near to this timescale. In particular, runs 4R5, 5R3, 5R4, 5R5, 6.5R4 and 6.5R5 have all shown TOs with core crossing travel times in the 4 to 6 year range.

%%%%%%%%%%%%%%%%%%%%%%%%%%%%%%%%%%%%%%%%%%%%%%%%%%%%%%%%%%%%%%%%%%%%%%%%%%%%%%%%%%%%%%%%%%%%%%%%%%%%%%%%%%%%%%%%%%%%%%%%%%%%%%%%

\subsection{Bandpass filtering}
\label{sec:filter}

In order to observe TOs more clearly in our simulation data we perform bandpass filtering on our $ts$-data from section \ref{sec:tos}. Hence we perform a Fourier transform on the data in the $t$-direction and filter frequencies using a step function. This is a similar analysis to that performed by \cite{gil10} albeit on our synthetic data rather than observational data.

\begin{figure}
% \begin{minipage}[b]{0.45\linewidth}
\centering
\includegraphics[width=0.8\linewidth]{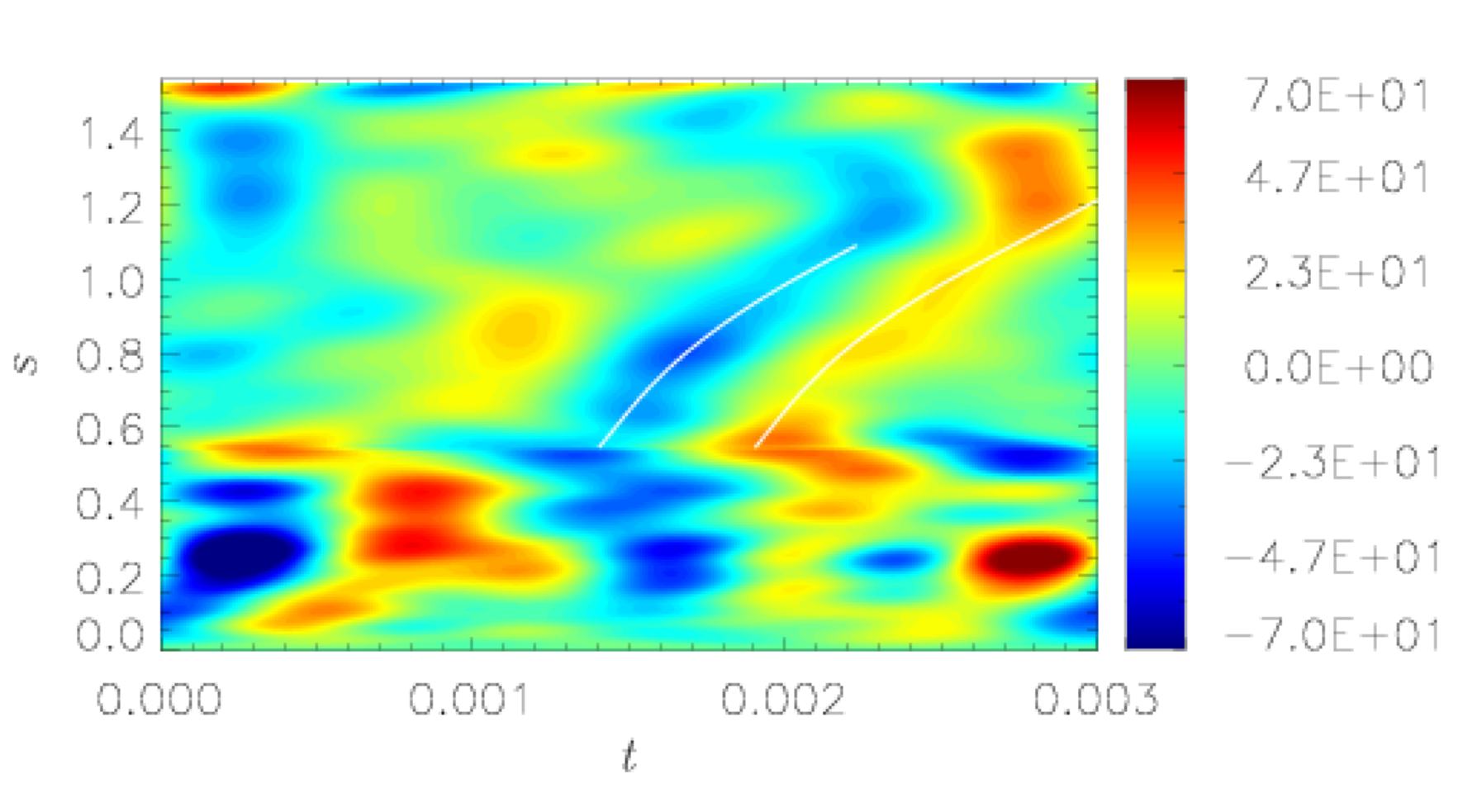}\\
\includegraphics[width=0.8\linewidth]{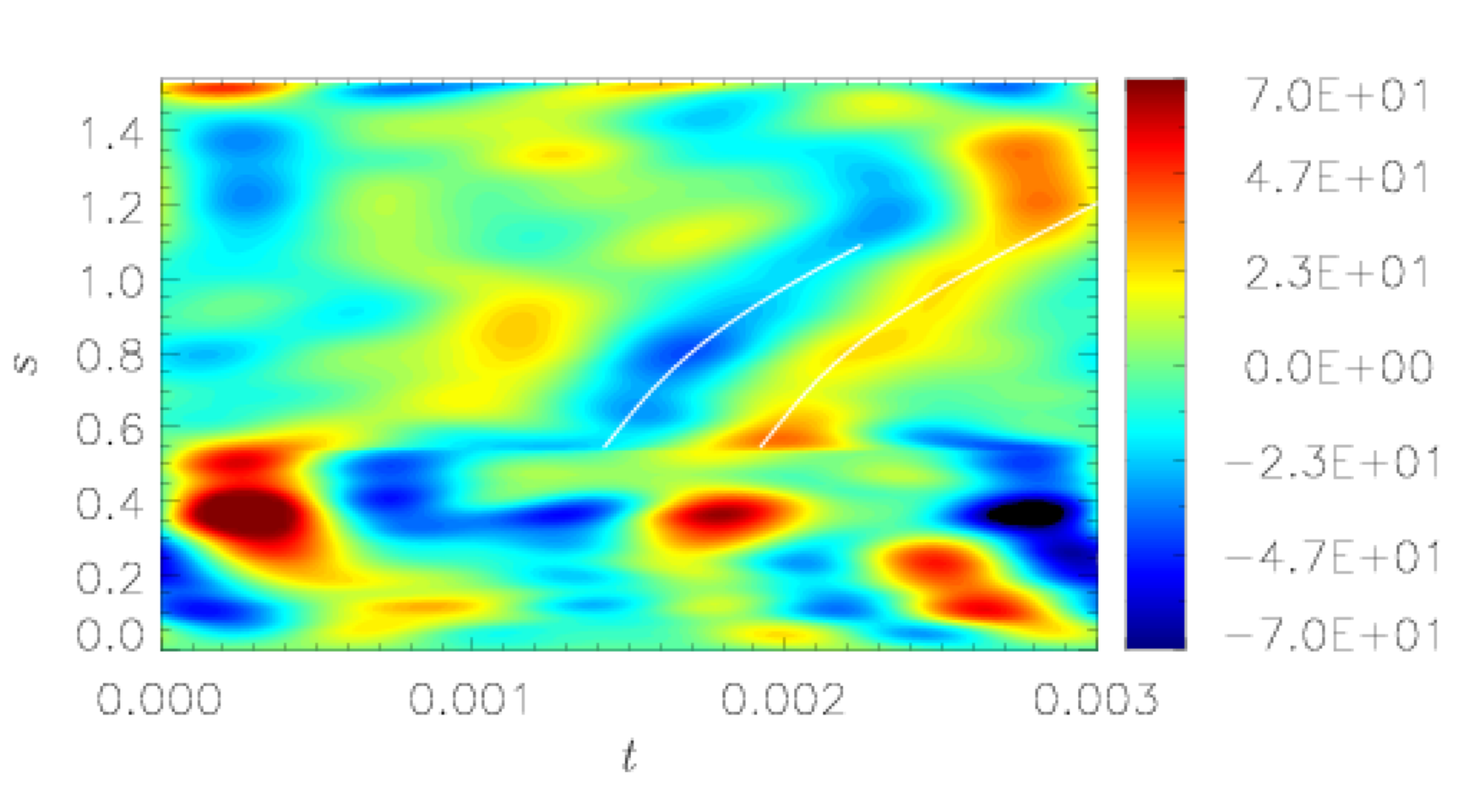}\\
\includegraphics[width=0.8\linewidth]{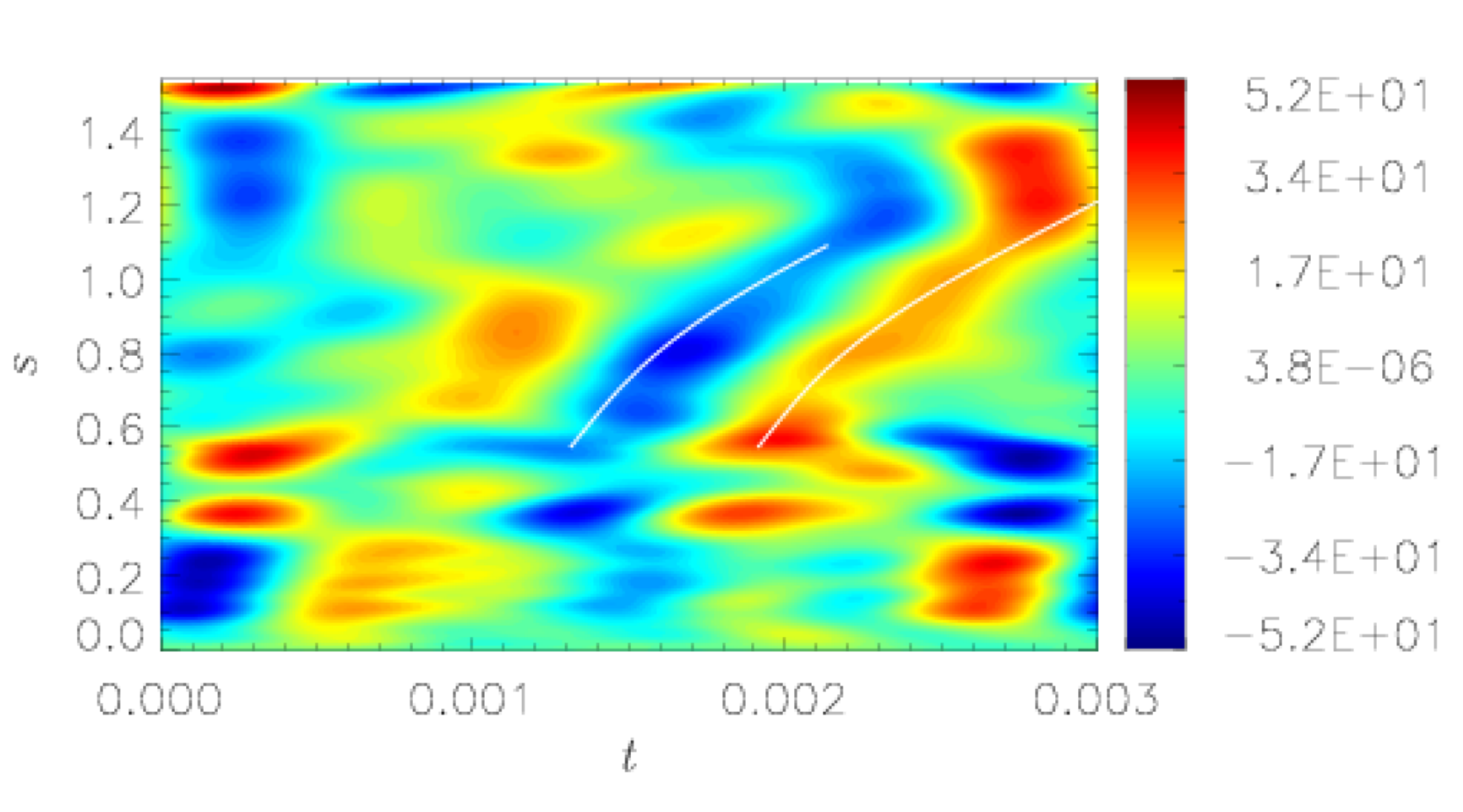}
\caption{$\langle\overline{u_\phi}\rangle$ bandpass filtered over modes 2 to 4, for the run 5R5.}
\label{fig:5R5trunc2to4}
% \end{minipage}
\end{figure}

\begin{figure}
% \begin{minipage}[b]{0.45\linewidth}
\centering
\includegraphics[width=0.8\linewidth]{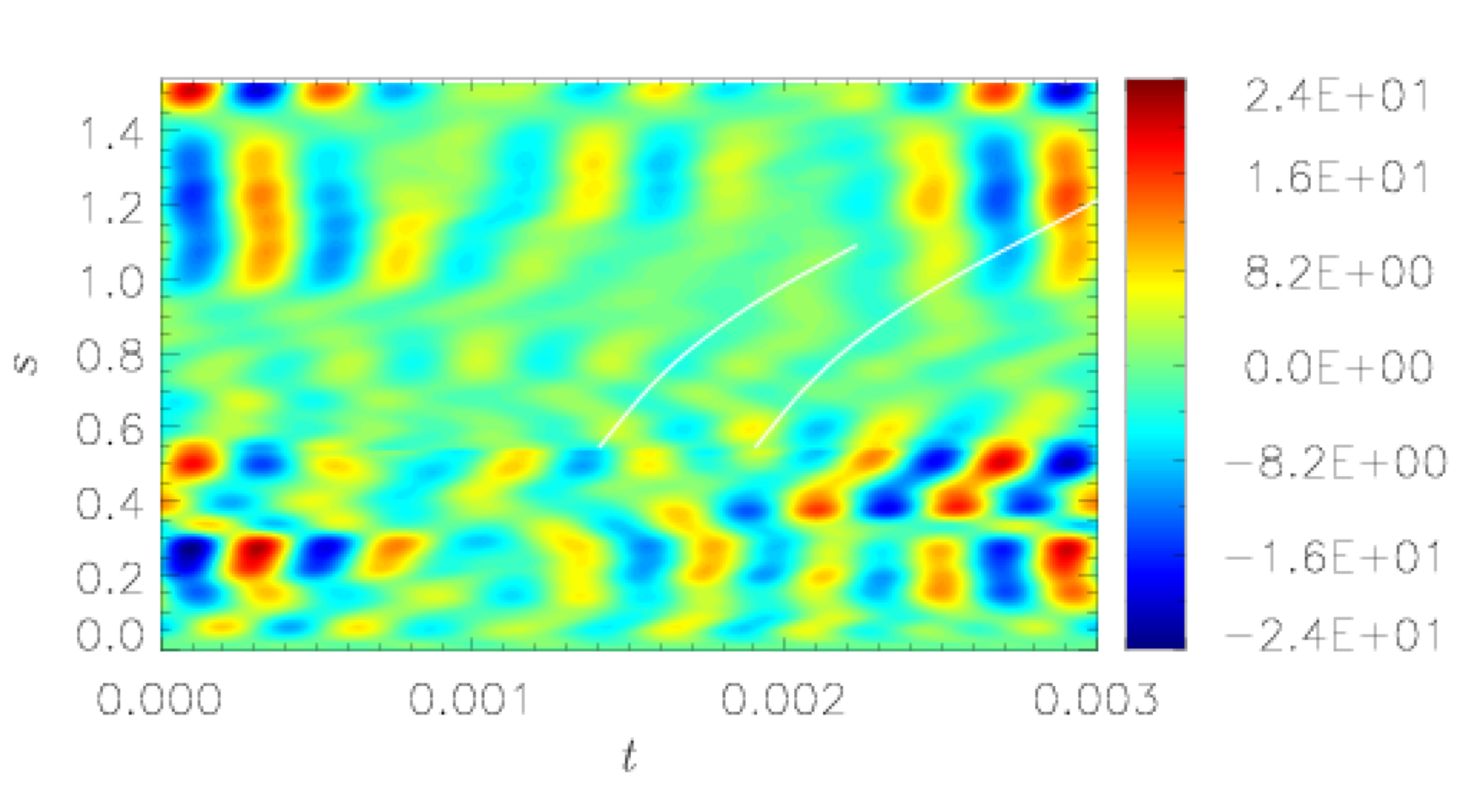} \\
\includegraphics[width=0.8\linewidth]{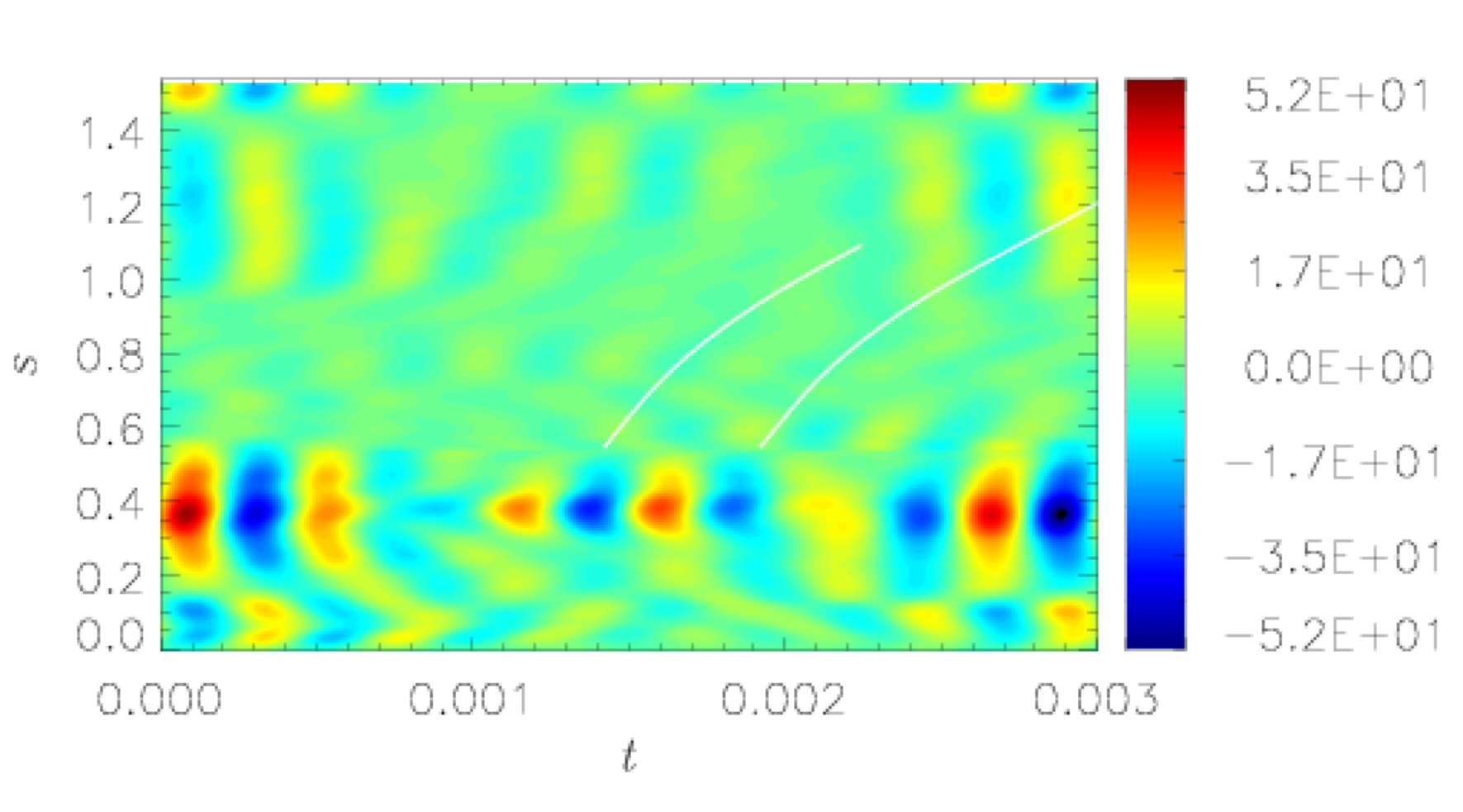} \\
\includegraphics[width=0.8\linewidth]{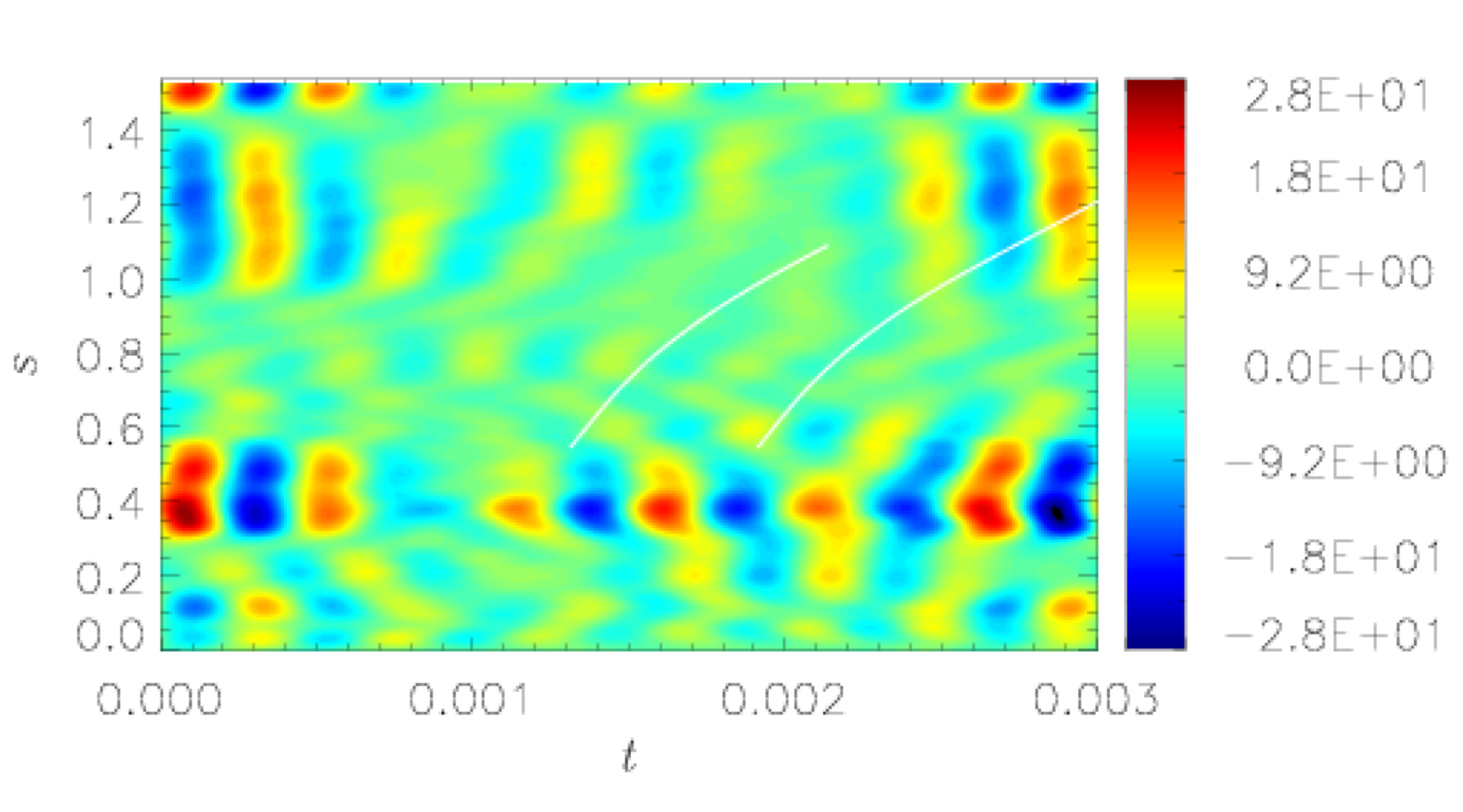}
\caption{$\langle\overline{u_\phi}\rangle$ bandpass filtered over modes 6 to 8, for the run 5R5.}
\label{fig:5R5trunc6to8}
% \end{minipage}
\end{figure}

\begin{figure}
% \begin{minipage}[b]{0.45\linewidth}
\centering
\includegraphics[width=0.8\linewidth]{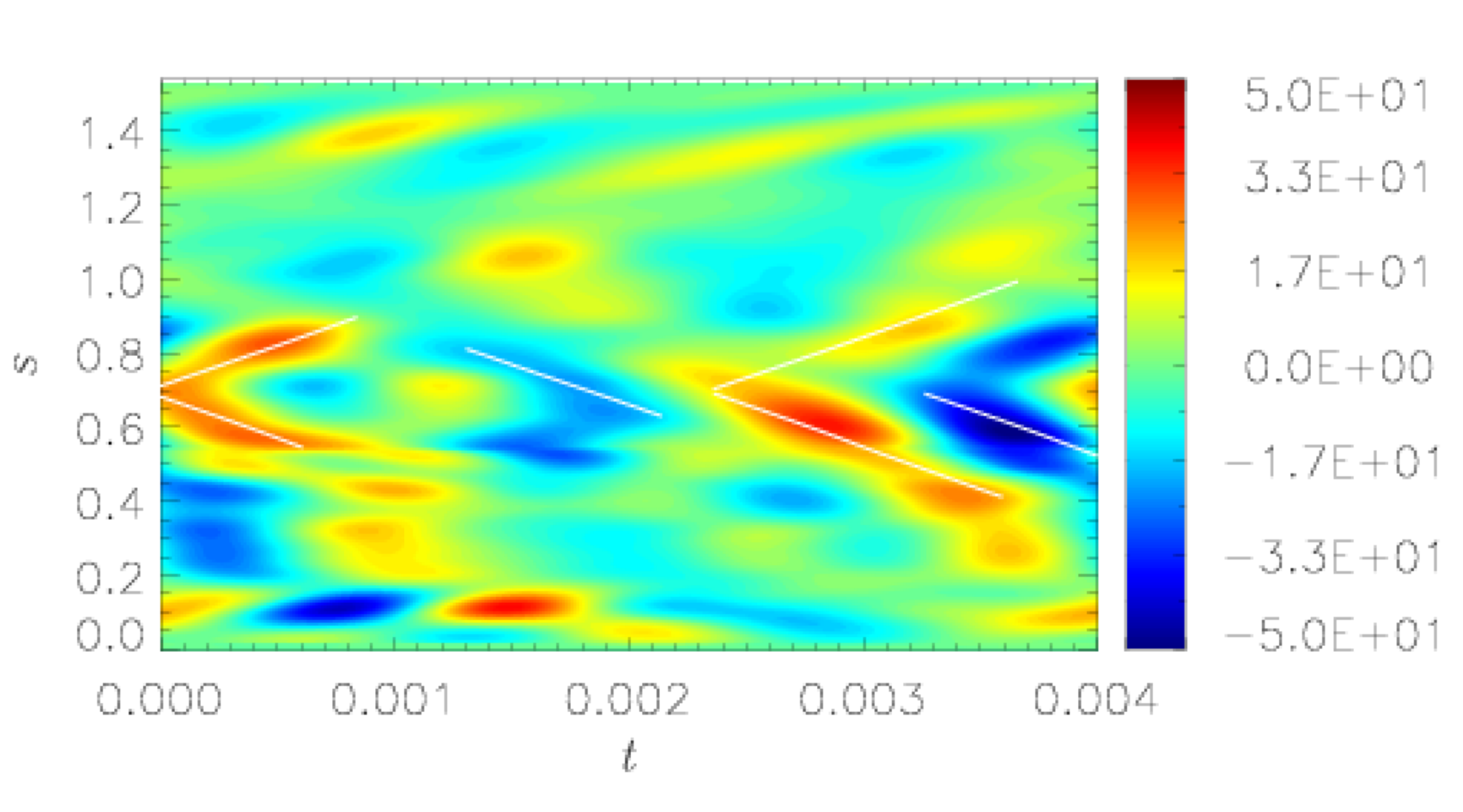}\\
\includegraphics[width=0.8\linewidth]{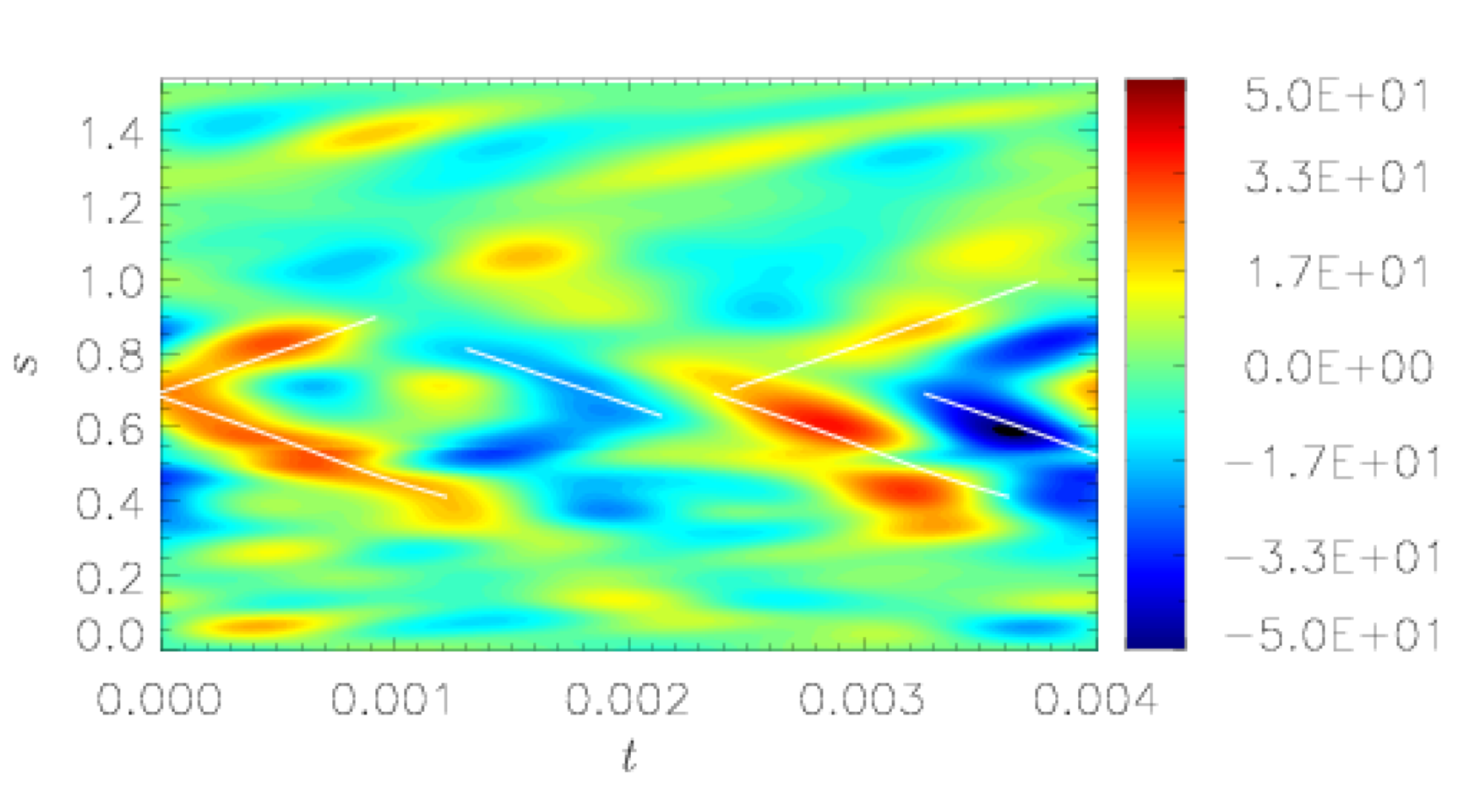}\\
\includegraphics[width=0.8\linewidth]{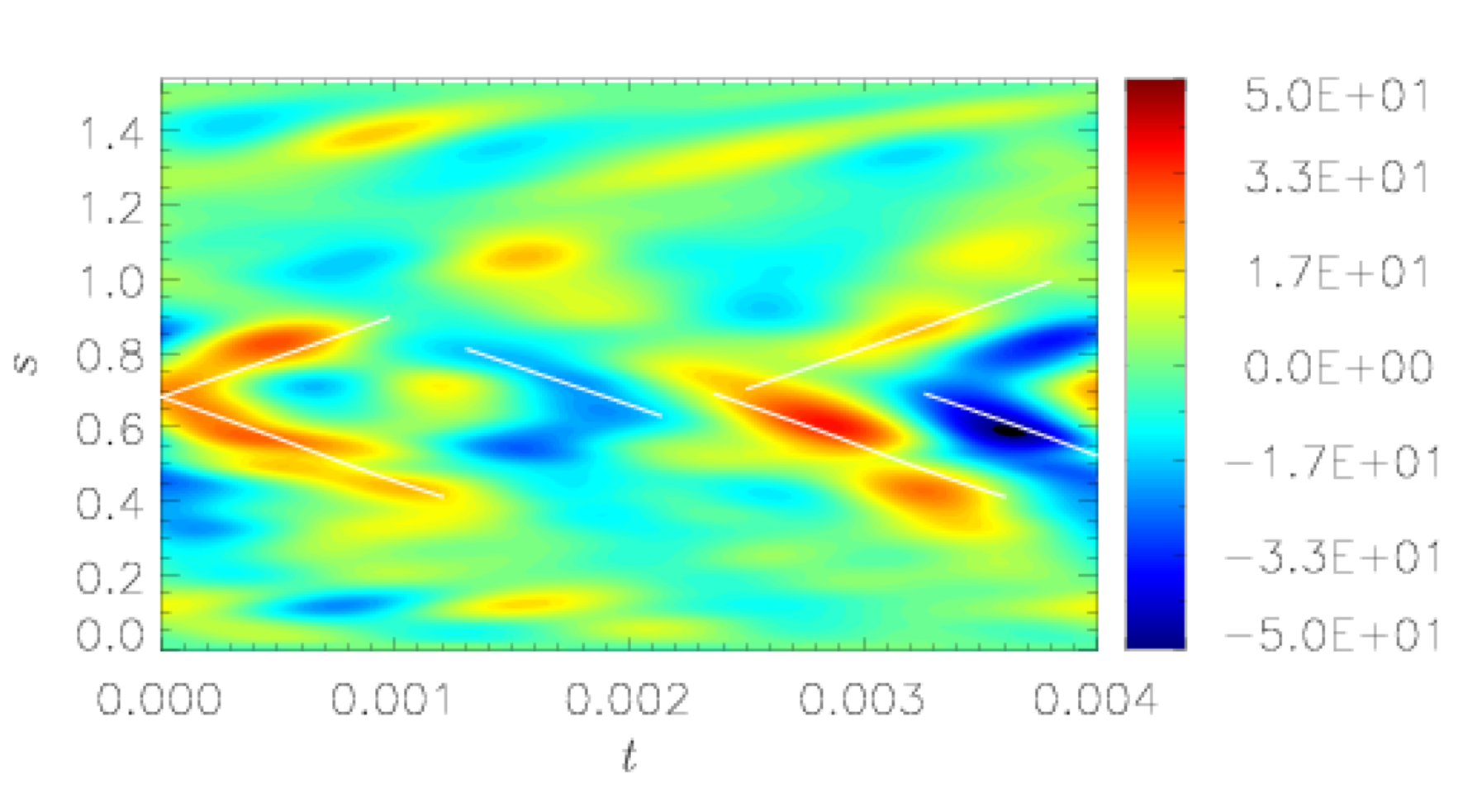}
\caption{$\langle\overline{u_\phi}\rangle$ bandpass filtered over modes 2 to 4, for the run 6.5R2.}
\label{fig:6.5R2trunc2to4}
% \end{minipage}
\end{figure}

\begin{figure}
% \begin{minipage}[b]{0.45\linewidth}
\centering
\includegraphics[width=0.8\linewidth]{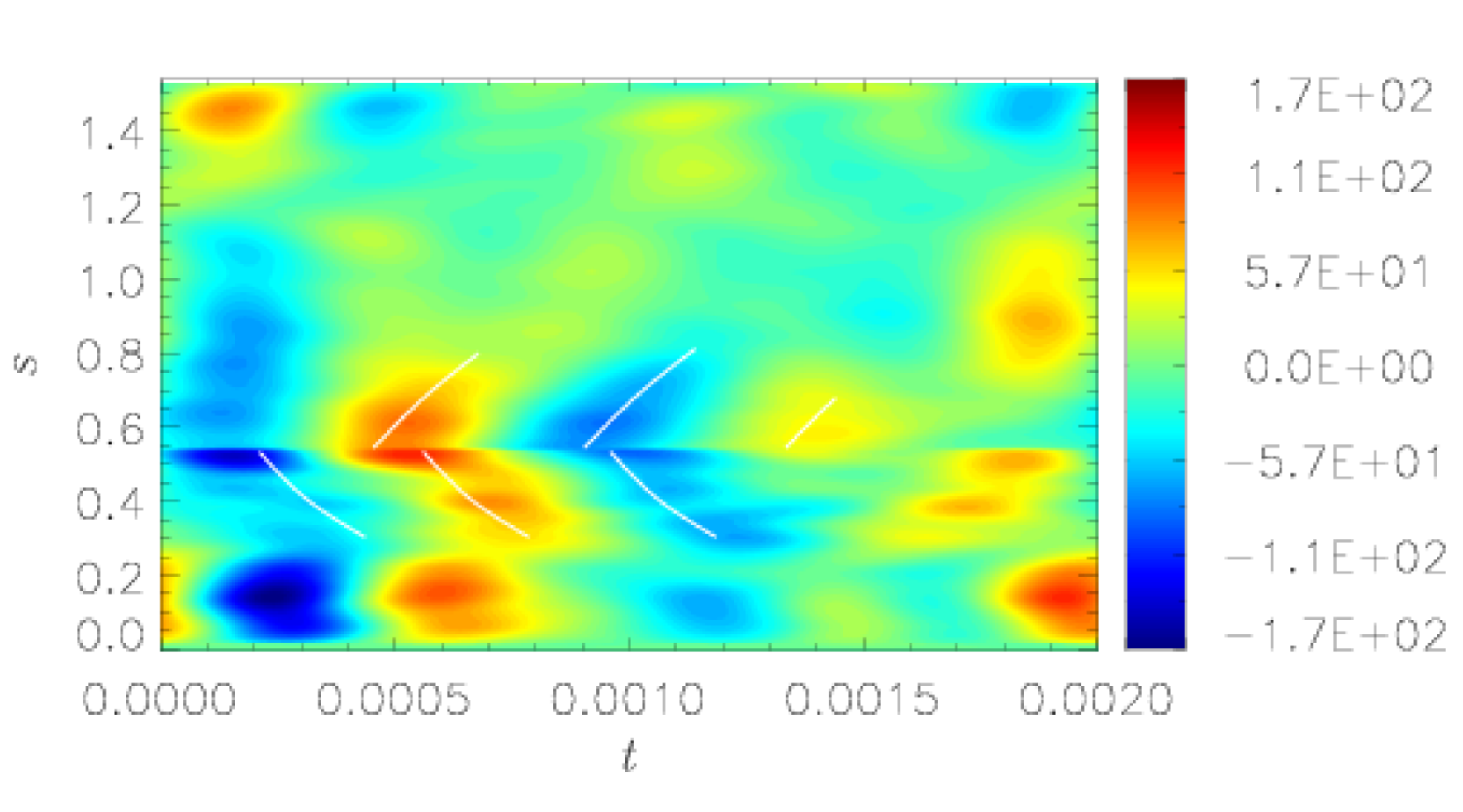}\\
\includegraphics[width=0.8\linewidth]{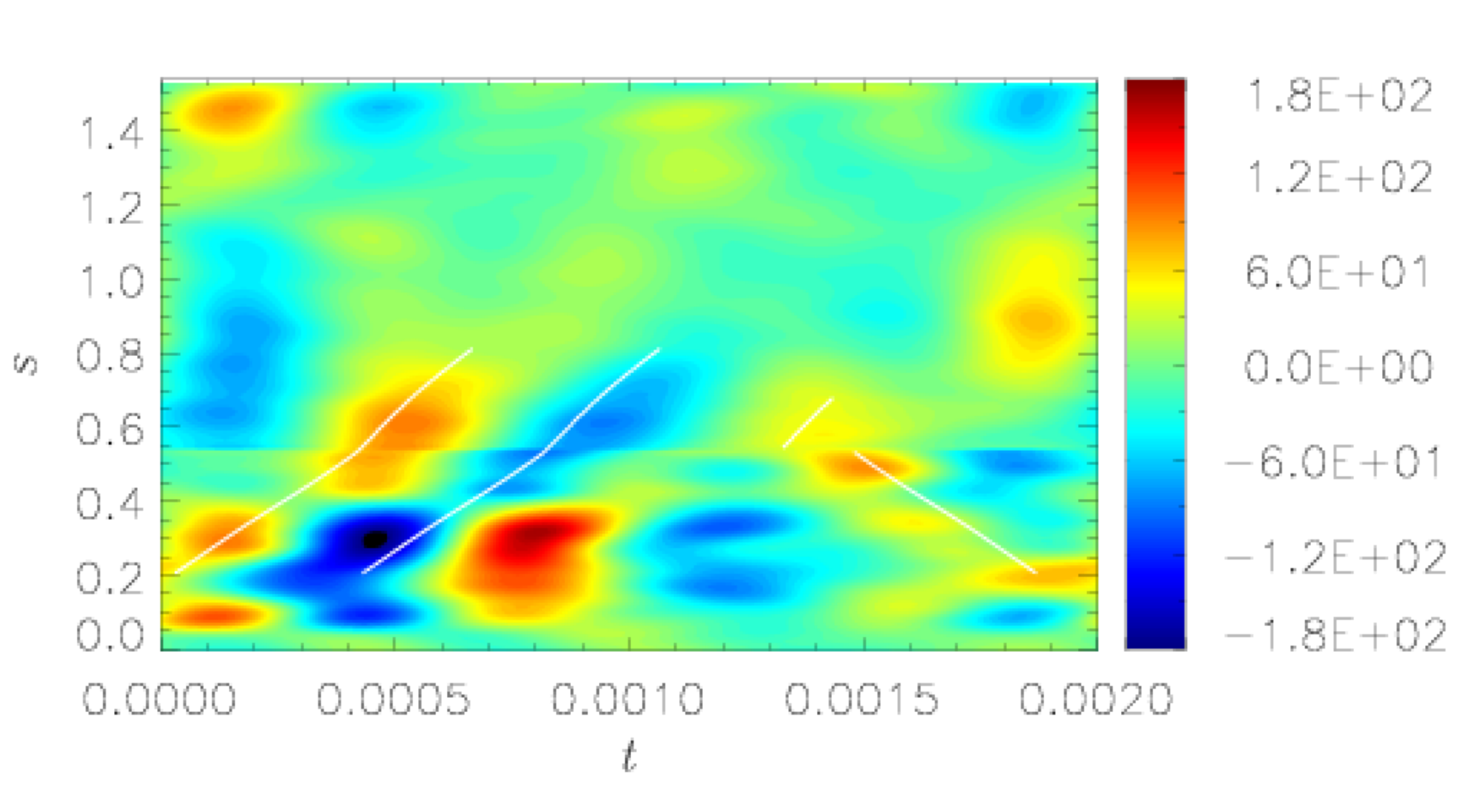}\\
\includegraphics[width=0.8\linewidth]{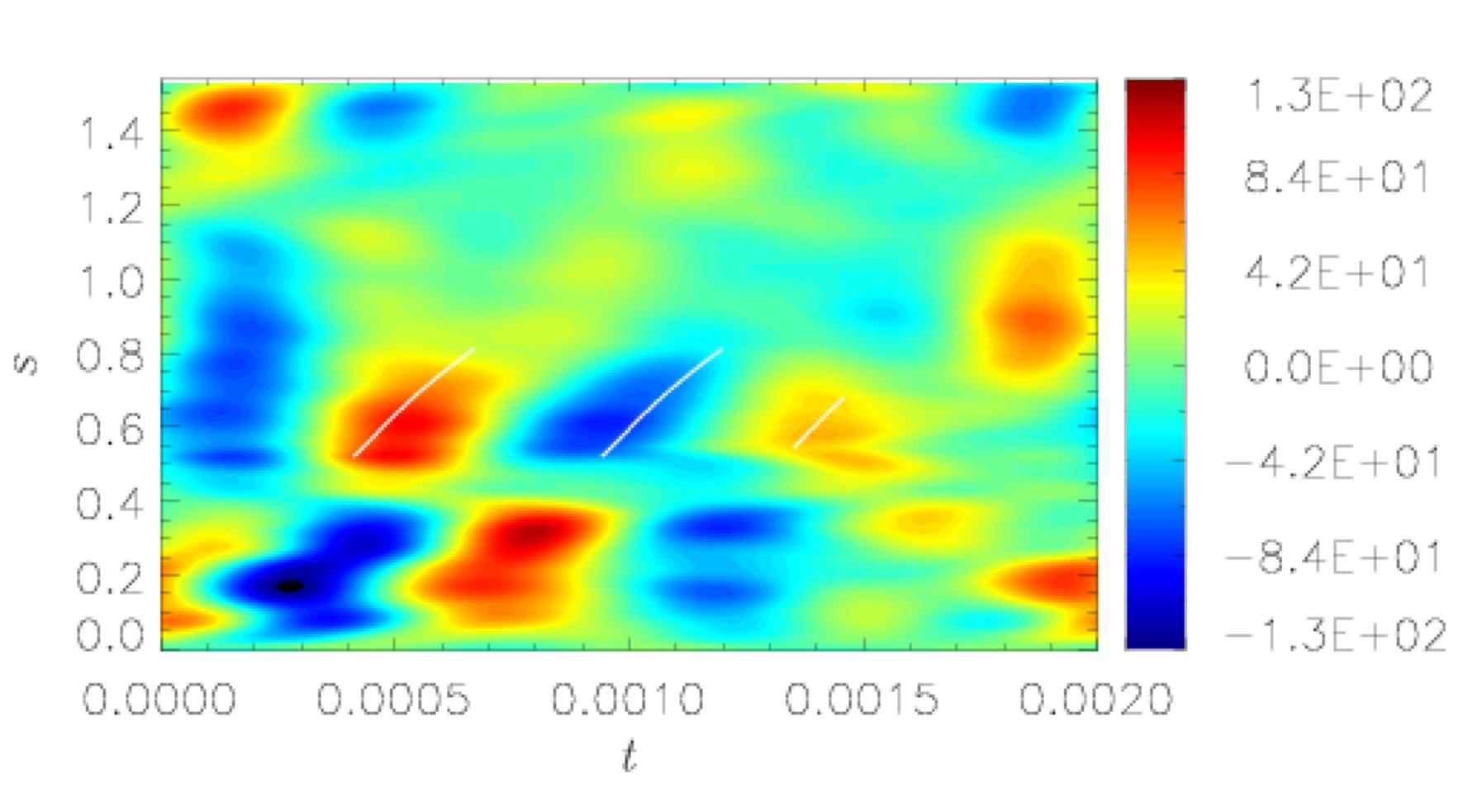}
\caption{$\langle\overline{u_\phi}\rangle$ bandpass filtered over modes 2 to 4, for the run 6.5R5.}
\label{fig:6.5R5trunc2to4}
% \end{minipage}
\end{figure}

Figs \ref{fig:5R5trunc2to4} to \ref{fig:6.5R5trunc2to4} show $ts$-data for several of our simulations that has been filtered of certain frequencies. The plots in each figure follow the same layout as previous figures so from top to bottom: data for ITCN, ITCS and the average over the entire $z$-average, respectively. In all of our runs we find that filtering out higher frequencies allows us to better identify the TOs in our data. Fig. \ref{fig:5R5trunc2to4}, for run 5R5, further highlights the two TOs that were identified in this data previously (cf. Fig. \ref{fig:5R5uphi}). This data has been filtered of frequency modes above 4 (as well as the mean). If we instead filter these low frequency modes out of the data we remove the structures travelling at the correct Alfv\'{e}n speed. We can see this in Fig. \ref{fig:5R5trunc6to8}, again for run 5R5, where all but frequency modes 6 to 8 are filtered. The structures present in $\langle\overline{u_\phi}\rangle$ no longer follow the trajectories given by the white curves and instead move outwards at a faster rate.

Further bandpass filtered plots for $\langle\overline{u_\phi}\rangle$, also over the frequency modes 2 to 4, for runs 6.5R2 and 6.5R5 are presented in Figs \ref{fig:6.5R2trunc2to4} and \ref{fig:6.5R5trunc2to4}, respectively. We have omitted plots filtered of higher frequencies for runs 6.5R2 and 6.5R5 due to their similarity to the plots of Fig. \ref{fig:5R5trunc6to8}. All data filtered over ranges other than approximately modes 2 to 4 only show structures moving at rates inconsistent with the TO Alfv\'{e}n speed.

Fig. \ref{fig:6.5R2trunc2to4} allows us to identify a complicated structure of inwards and outwards propagating waves OTC near to the TC, which was not immediately obvious in the earlier unfiltered plots (cf. Fig. \ref{fig:6.5R2uphi}). It is clear that some inwards moving waves propagate through the TC and often into one hemisphere only. For example the earliest instance of an inwards propagating wave in Fig. \ref{fig:6.5R2trunc2to4} reaches the TC at $t\simeq 0.0006$ and passes through into the region ITCS but not in the northern hemisphere.

Filtering all but low frequency structures again highlights the previously identified TOs in Fig. \ref{fig:6.5R5trunc2to4}, for run 6.5R5 (cf. Fig. \ref{fig:6.5R5uphi}). In fact, several of the features previously identified have only become clear upon filtering.  We can clearly see the structures propagating outwards from the TC, as well as inwards from the TC in the northern hemisphere. Conversely, the structures ITC in the southern hemisphere propagate outwards and through into the region OTC. This run, in particular, highlights the complicated nature of waves incident on the TC.

The sensitivity in the bandpass filtering and preference for low frequency modes draws our attention to two points. Firstly, it validates our choice of $\tau$ for each run since TOs appearing at low frequencies implies that they do indeed operate on the $\tau$ timescale. Secondly, the lack of TOs appearing at higher frequencies also suggests that TOs do not operate on timescale much smaller than $\tau$. This was not immediately obvious from our unfiltered data.
% \hspace{0.5cm}
% \hspace{0.5cm}
% 

%%%%%%%%%%%%%%%%%%%%%%%%%%%%%%%%%%%%%%%%%%%%%%%%%%%%%%%%%%%%%%%%%%%%%%%%%%%%%%%%%%%%%%%%%%%%%%%%%%%%%%%%%%%%%%%%%%%%%%%%%%%%%%%%

\subsection{Excitation mechanisms}
\label{sec:fb}

We now explore the role various forces have in the driving of the torsional waves observed in sections \ref{sec:tos} and \ref{sec:filter}. In section \ref{sec:taylor} we discussed how there were three possible driving forces in our system and hence we plot quantities appearing on the right-hand-side of (\ref{eq:acceqsforce}). Since we aim to find correlation between these forcing terms and the origins of TOs we retain, on our plots throughout this section, the white curves from the associated azimuthal velocity plots of sections \ref{sec:tos} and \ref{sec:filter}. However, in our $ts$-contour plots for $F_R$, $F_V$ and $F_{LD}$, we do not expect features to be travelling along the white curves; rather we expect to find features at the origins of the curves.

\begin{figure}
% \begin{minipage}[b]{0.45\linewidth}
\centering
\includegraphics[width=0.8\linewidth]{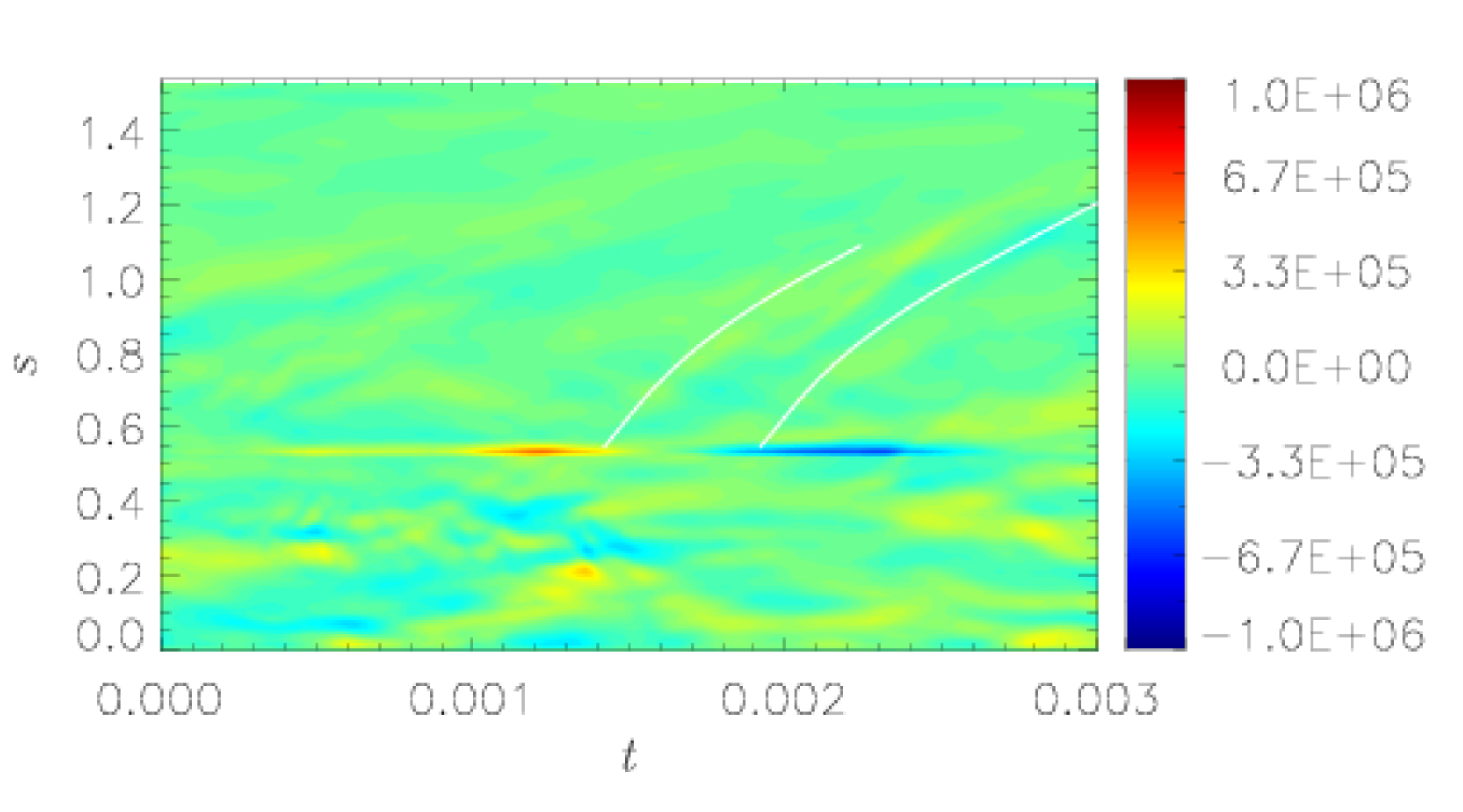}\\
\includegraphics[width=0.8\linewidth]{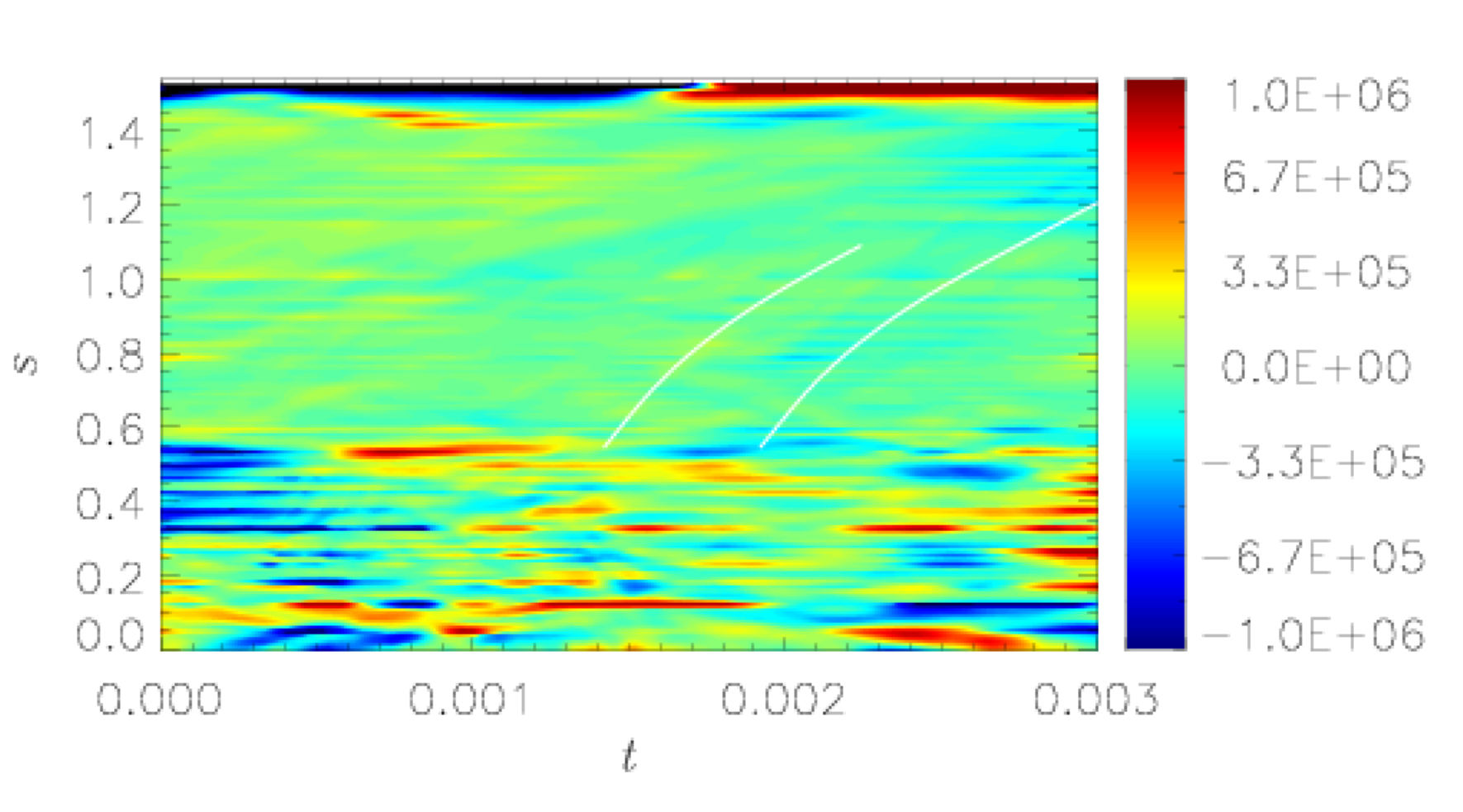}\\
\includegraphics[width=0.8\linewidth]{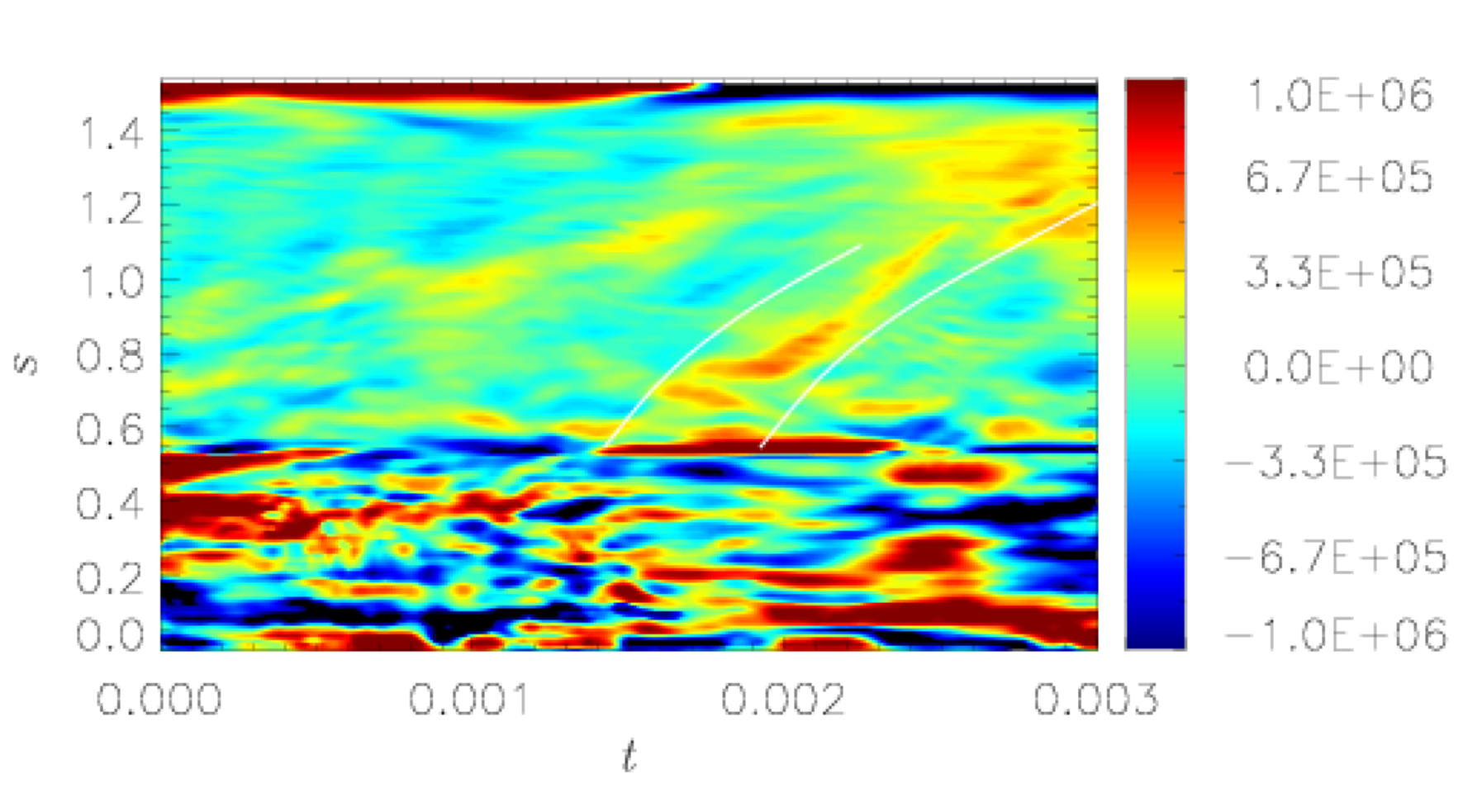}\\
\caption{Forcing terms for ITCS and OTC for the run 5R5. From top to bottom: $F_R$, $F_V$ and $F_{LD}$.}
\label{fig:5R5forcesS}
% \end{minipage}
\end{figure}

From Fig. \ref{fig:5R5forcesS}, displaying forcing terms for run 5R5 (for the regions OTC and ITCS only), we can make several observations. All three forces are weak for most of the region OTC except at the TC itself. The viscous dissipation and the Lorentz forcing are also strong at the equator, where the rapid changes in velocity due to the CMB boundary layer have a significant effect. Within the TC all three forces, but especially $F_V$ and $F_{LD}$, are larger. However, one of the most striking features of these plots in the context of TO driving is the excellent correlation between large Reynolds force at the TC and the excitation of waves represented by the origin of the two curves. Although the Reynolds force is clearly weaker than the Lorentz forcing (by approximately a factor of three), its correlation is superior since there are regions of large Lorentz force that do not coincide with TO initiation. Conversely, whenever the Reynolds force is large at the TC, a TO is produced.

\begin{figure}
% \begin{minipage}[b]{0.45\linewidth}
\centering
\includegraphics[width=0.8\linewidth]{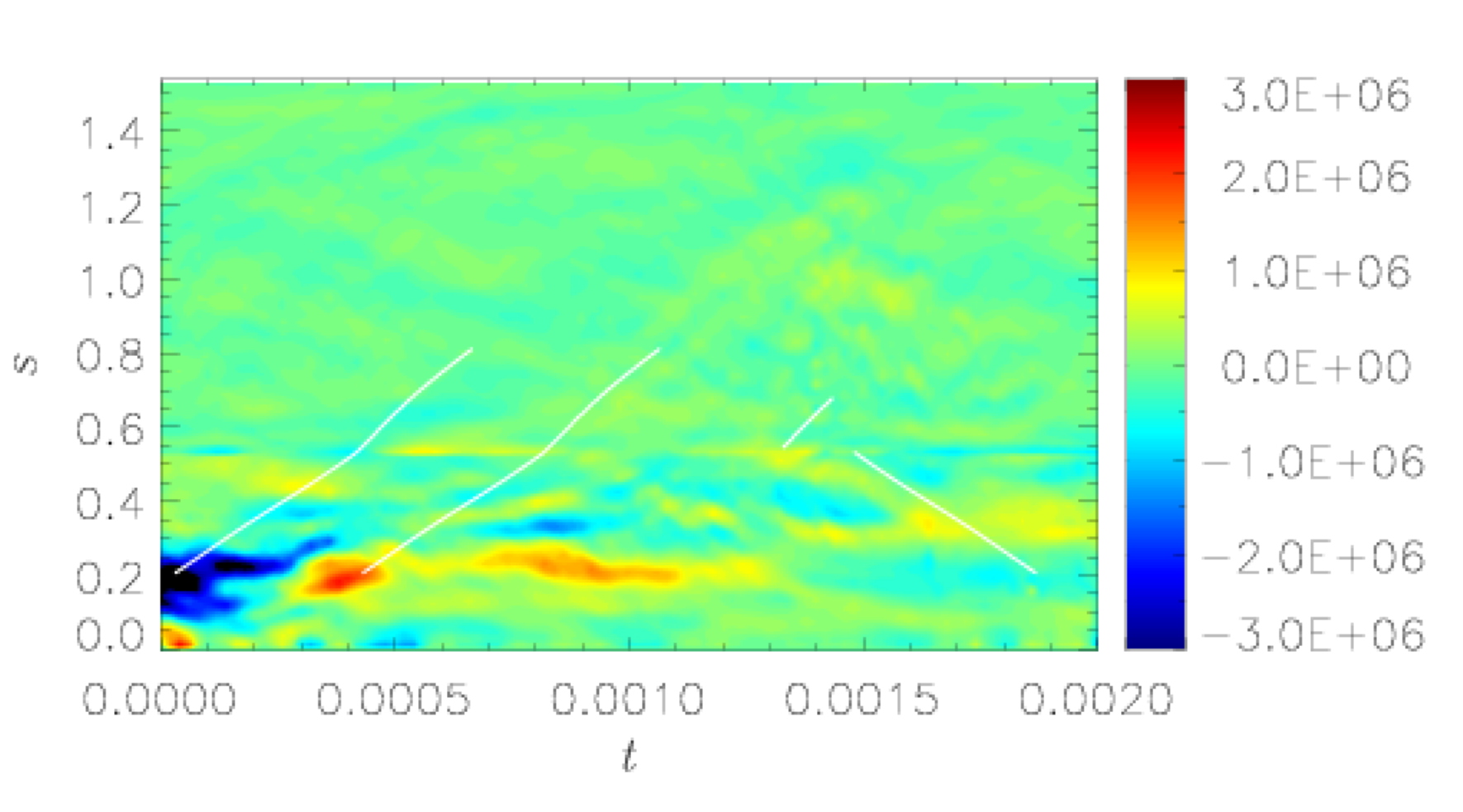}\\
\includegraphics[width=0.8\linewidth]{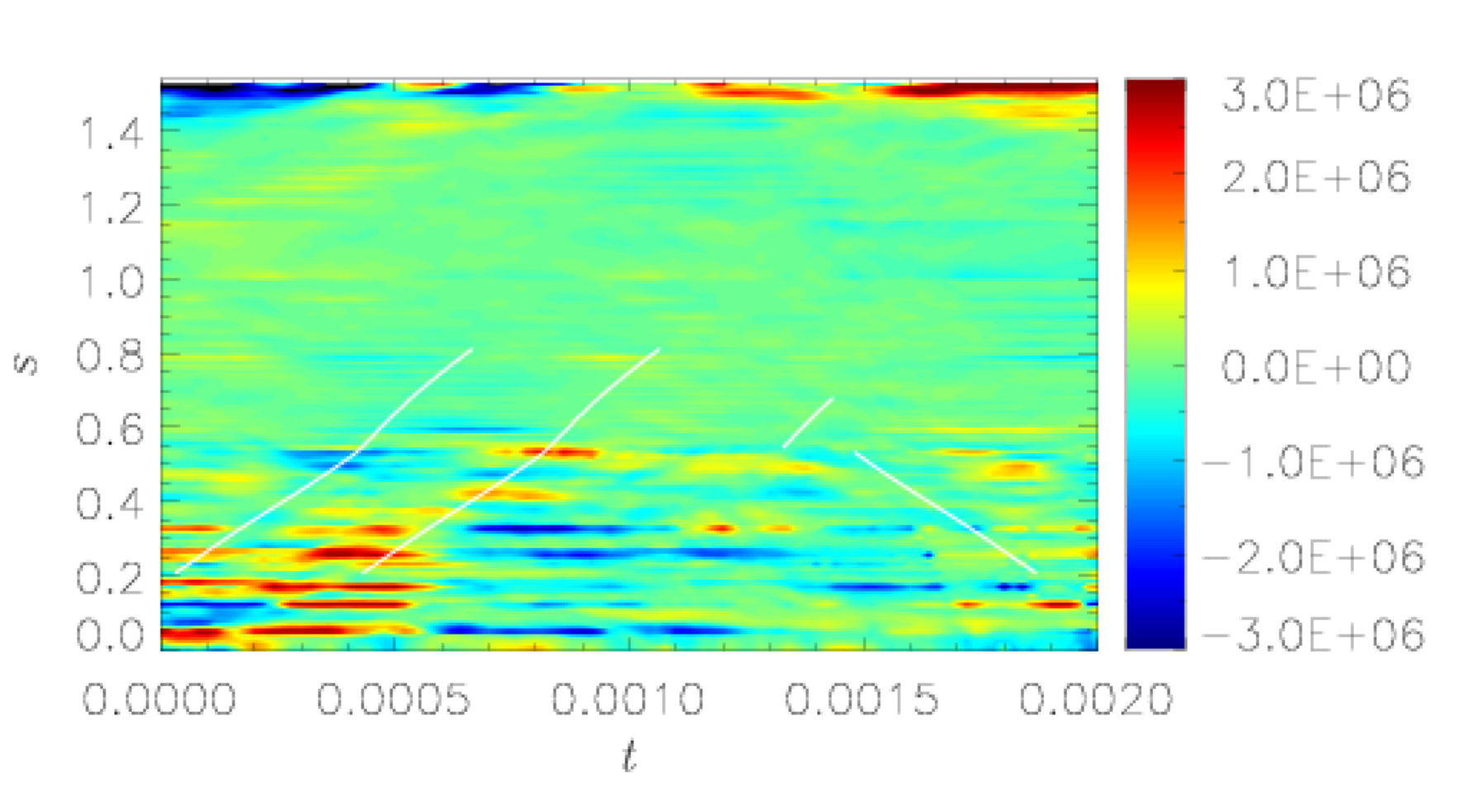}\\
\includegraphics[width=0.8\linewidth]{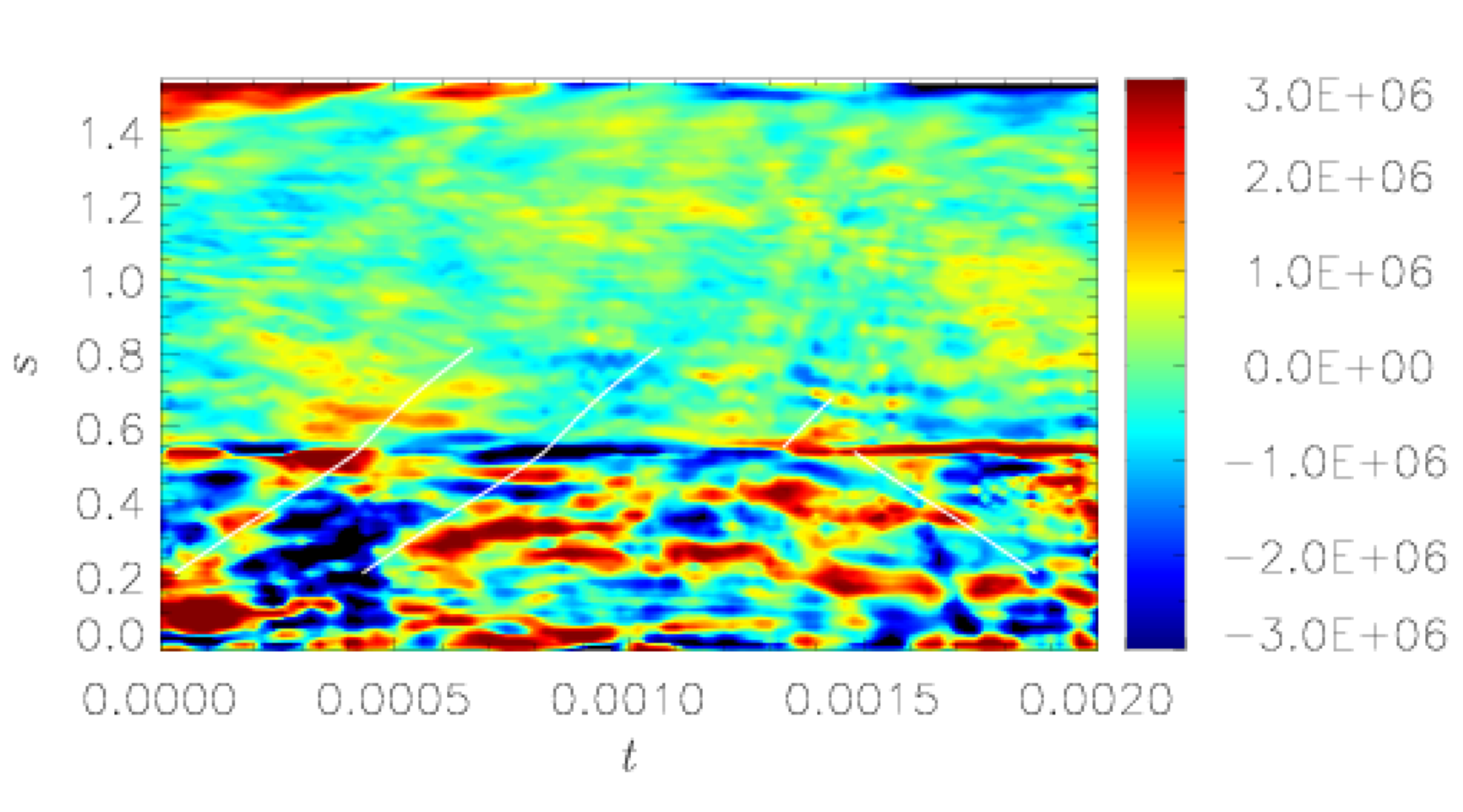}\\
\caption{Forcing terms for ITCS and OTC for the run 6.5R5. From top to bottom: $F_R$, $F_V$ and $F_{LD}$.}
\label{fig:6.5R5forcesS}
% \end{minipage}
\end{figure}

In Fig. \ref{fig:6.5R5forcesS} we again plot forcing terms, this time for run 6.5R5. The plots for the three forces are broadly similar to the 5R5 case OTC. Once again the locations of the origin of identified TOs are well correlated with large regions of Reynolds force, this time ITC. A lack of correlation of large $F_R$ at the TC with the waves propagating outwards there suggests that the waves ITC do indeed traverse the TC and thus do not require an excitation mechanism at the TC in this case. Evidence for correlation between Reynolds forcing and TO excitation comes not only from Figs \ref{fig:5R5forcesS} and \ref{fig:6.5R5forcesS}, but from a series of snapshots from our runs, too numerous to display here.
 
%%%%%%%%%%%%%%%%%%%%%%%%%%%%%%%%%%%%%%%%%%%%%%%%%%%%%%%%%%%%%%%%%%%%%%%%%%%%%%%%%%%%%%%%%%%%%%%%%%%%%%%%%%%%%%%%%%%%%%%%%%%%%%%%

\section{Discussion}
\label{sec:cons}

Through our numerical simulations we have observed torsional oscillations at a range of Ekman numbers including at the relatively large $E=10^{-4}$. These oscillations are able to propagate either inwards or outwards in the cylindrical radial direction. The torsional waves travel fastest under parameter regimes that promote the production of strong magnetic fields. Thus, large magnetic Prandtl number and rapidly rotating regimes produce the quickest oscillations.

Torsional oscillations are often found to propagate from the TC, both inwards and outwards. Hence we have observed waves ITC, a region of the spherical shell not considered in previous work. Although waves are mostly found to originate at the TC, it is possible for excitation to occur at other locations in the shell. This indicates a complicated non-uniform excitation mechanism with various processes likely to excite oscillations at the different locations. 

Within our set of simulations we identified two dynamo regimes for which a given system is able to organise itself. Whether the dynamo is in a weak or strong field regime has implications on the torsional waves observed. Weak field regimes found at $Pm\in[1,3]$ for a range of Ekman numbers are able to produce approximately equal numbers of inward and outward propagating waves. Conversely, strong field regimes found at $Pm\in[3,5]$ are dominated by waves of outwards propagation. Plots (and movies) of meridional sections of $\overline{u_\phi}^\prime$ are able to show the outwards propagation of columns in strong field runs whereas the same graphics show features more reminiscent of standing waves in the weak field runs. The  speed of waves is found to best match that predicted for the Earth in the strong field regime with a core travel time of between 4 and 6 years.

Oscillations observed ITC almost exclusively originate at the TC and thus move radially inwards. This is either via an excitation mechanism at the TC or by a wave propagating across the TC from OTC. Additionally, weak field regimes are more likely to promote torsional oscillations within the TC. If waves are being excited at the TC then the weak field regime, with its greater ability to promote inwards propagation, is naturally preferred for disturbances ITC. Conversely, the preference for outwards movement in the strong field regime leads to disturbances at the TC commonly travelling through the region OTC towards the equator.

One of the most intriguing results from our simulations is the apparent ability of waves to cross the tangent cylinder. Waves can cross in either direction, however waves entering the region ITC often dissipate quickly, probably owing to the large viscous dissipation there. Features propagating from OTC are often absorbed into only one hemisphere ITC suggesting that conditions and flow patterns have to be desirable, in a given hemisphere, for a crossing of the TC to take place in this direction. The crossing of waves in the opposite direction is possible but rarer. The likelihood of movement of oscillations into the region OTC is increased if waves are found to be approaching the TC in each hemisphere approximately concurrently. Since the regions north and south of the inner core effectively act independently, propagation from ITC to OTC is a random and often infrequent phenomenon resulting in the scarcity of such events. One of our most studied simulations (6.5R5) was one of the few to display propagation of waves from ITC to OTC. 

We have been able to investigate the excitation mechanisms of torsional waves within our simulations. We split these into three categories, the damping due to viscous forces, the Reynolds forces, and the Lorentz forces. We have shown that the Lorentz force can be usefully divided into that part which gives the restoring force of the torsional oscillation itself, and the part that comes from the ageostrophic convection. Although the convection is relatively small-scale, the Lorentz force it produces does not vanish when averaged over the Taylor cylinder, and may be an important excitation mechanism for TOs.

Despite the Reynolds force consistently being the weakest of the three forces, correlation with TO propagation from the TC leads us to conclude that it is also an important excitation mechanism in our simulations. At lower, more Earth-like, Ekman numbers the Reynolds forcing will inevitably become small relative to the Lorentz force and may play a diminished role. However, the thin region near the TC may well become thinner at low Ekman number, so the velocity gradients driving the Reynolds force might be sufficient to have an effect even though the velocity itself is small in magnitude. The scaling of the relative size of the Reynolds and Lorentz contribution with Ekman number needs to be explored further, but this will require a new approach, as reducing the Ekman number is notoriously expensive in full geodynamo simulations. The Lorentz force excited by ageostrophic convection, which seems particularly strong inside the TC, is currently the preferred explanation of TO excitation in the Earth's core.

Viscous forces were found to be significant near the CMB equator and inside the TC in our models, though we expect their impact to be much reduced at the very low Ekman numbers of the Earth's core. Their damping effect may be replaced by electromagnetic coupling with the mantle and the inner core, which has not yet been included in our model.

Several of the observations from our results highlight a common problem in numerical geodynamo simulations: we are restricted by limited computing resources when attempting to reach a parameter regime that can quantitatively replicate many of the geodynamo's features, including torsional oscillations. A reduction of geometric complexity by considering, for example, magnetoconvection in an annulus would help to alleviate this problem by allowing one to perform simulations at more realistic Ekman numbers. Alternatively, spherical geometry could be retained and a lower Ekman number achieved by performing simulations of magnetoconvection where the requirement of a long period of time integration to ensure a dynamo state is found is not necessary. These topics are the subject of future work.

  \def\newblock{\hskip .11em plus .33em minus .07em} 
  \bibliographystyle{gji}
  \bibliography{paper}

\end{document}